\documentclass{elsart}
\usepackage{amsmath,amssymb,graphicx}
\usepackage[colorlinks=true,pdftex]{hyperref}

\begin{document}
\begin{frontmatter}

 \title{Numerical study of blow-up in solutions to generalized Korteweg-de Vries equations}

\author{C.~Klein}
\address{Institut de Math\'ematiques de Bourgogne,
		Universit\'e de Bourgogne, 9 avenue Alain Savary, 21078 Dijon
		Cedex, France,
Tel: 00333803 95858, Fax: 00333803 95869, christian.klein@u-bourgogne.fr}

\author{R.~Peter}
\address{Institut de Math\'ematiques de Bourgogne,
		Universit\'e de Bourgogne, 9 avenue Alain Savary, 21078 Dijon
		Cedex, France}
\date{\today}    

\begin{abstract}

We present a detailed numerical study of solutions to 
general Korteweg-de Vries equations with critical and 
supercritical nonlinearity. We study the stability of solitons and 
show that they are unstable 
against being radiated away and blow-up. In the $L_{2}$ critical 
case, the blow-up mechanism by Martel, Merle 
and Rapha\"el can be numerically identified. In the limit of small 
dispersion, it is shown that a dispersive shock always appears before 
an eventual blow-up. In the latter case, always the first soliton to 
appear will blow up. It is shown that the same type of blow-up as for 
the perturbations of the soliton can be observed which indicates that 
the theory by Martel, Merle 
and Rapha\"el is also applicable to initial data with a mass much 
larger than the soliton mass. We study the 
scaling of the blow-up time $t^{*}$ in dependence of the small dispersion 
parameter $\epsilon$ and find an exponential dependence 
$t^{*}(\epsilon)$ and that there is a minimal blow-up time 
$t^{*}_{0}$ greater than the critical time of the corresponding Hopf 
solution for $\epsilon\to0$. To study the cases  with blow-up in 
detail, we apply the first dynamic
rescaling for generalized Korteweg-de Vries equations. This allows 
to identify the type of the singularity. 

\end{abstract}


\end{frontmatter}

\thanks{We thank Y.~Martel and J.C.~Saut for useful discussions and 
hints. 
This work has been supported by the project FroM-PDE funded by the European
Research Council through the Advanced Investigator Grant Scheme,  and the ANR via the program ANR-09-BLAN-0117-01. }


\section{Introduction}
\subsection{Background and motivation}

The celebrated Korteweg-de Vries (KdV) equation provides an asymptotic 
description of one-dimensional waves in shallow water in the long 
wave-length limit.  For shorter wavelengths the dispersion in the KdV 
equation is in general too strong compared to what is observed in applications. A possible approach 
to address this short-coming of the model equation KdV is to tilt the 
balance between nonlinearity and dispersion towards 
nonlinearity which leads to generalized KdV (gKdV) equations, 
\begin{equation}
    u_{t}+u^{n}u_{x}+\epsilon^{2}u_{xxx}=0;
    \label{gKdV}
\end{equation}
here the parameter 
$n$ in the nonlinearity is taken to be a positive integer, KdV 
corresponds to   $n=1$, the modified KdV (mKdV) equation to $n=2$; the parameter $\epsilon$ is a small 
dispersion parameter.
The 
gKdV equations have three conserved quantities, the $L_{1}$ and the 
$L_{2}$ norm of $u$ (or the \textit{mass}    $M[u] = ||u||_2^{2}$) and the energy,
\begin{equation}
            E[u] = \int_{-\infty}^{\infty}\left(\frac{\epsilon^2}{2}u_x^2
           - \frac{1}{(n+1)(n+2)}
	   u^{n+2}\right)dx
    \label{energy}.
\end{equation}
The KdV and mKdV equation are completely integrable which implies 
that they have an infinite number of conserved quantities. For 
general $n$, the gKdV equations have just the three conserved 
quantities mentioned above. 
 
The introduction of a stronger nonlinearity in KdV equations 
increases its importance with respect to dispersion, but has the well-known
effect that solutions to gKdV equations may have finite time blow-up for $n\geq 
4$, i.e., a loss of regularity with respect to the initial data in 
the form of diverging norms of the solution. The global 
well-posedness of the Cauchy problem in $H^{1}$ was proven in 
\cite{Kato83,KPV93}.  The case $n=4$ is 
critical in the sense that solutions to gKdV for $n<4$ for 
sufficiently smooth initial data $u(x,0)=u_{0}(x)$
are globally regular in time. The case $n=4$ is also distinguished by the invariance of the 
mass
under rescalings of the form $x\to x/\lambda$, $t\to 
t/\lambda^{3}$ and $u\to \lambda^{2/n}u$ with $\lambda=const$ which 
leave equation (\ref{gKdV}) invariant. 

\subsection{Basic mathematical properties of generalized Korteweg-de 
Vries equations}

The gKdV equations have localized travelling wave solutions 
of the form $u=Q_{c}(x-x_{0}-ct)$ with $x_{0},c=const$ and with
\begin{equation}
    Q_{c}(z) = \left( 
    \frac{(n+1)(n+2)c}{2}\,\mbox{sech}^{2}\frac{\sqrt{c}n}{2\epsilon}z\right)^{1/n}
    \label{soliton}.
\end{equation}
Obviously we have $Q_{c}(z)=c^{1/n}Q(\sqrt{c}z)$, where we have put 
$Q=Q_{1}$. This simple scaling property of the \emph{solitons} allows to 
concentrate on the case $c=1$. If we refer in the following to 
the gKdV soliton, it is always implied that $c=1$. It was shown in \cite{BSW1987} that these solitons are linearly 
unstable for $n\geq 4$. Note that the energy of the soliton vanishes for $n=4$.

The initial value problem for gKdV for the $L_{2}$ critical case $n=4$ has been studied in 
detail in a series of papers 
\cite{MarMer2000,MarMer2001,MarMer2002_1,MarMer2002_2,MarMer2002_3,Merle2001}. In 
these papers, it was shown that initial data $u_{0}$ with negative 
energy subject to the 
condition $u_{0}\in H_{1}$, $||Q||_{2}\leq 
||u_{0}||_{2}<||Q||_{2}+\alpha_{0}$ with $\alpha_{0}\ll 1$ lead to a blow-up of the solution in finite or infinite time. 
For asymptotically decreasing data such that 
$\int_{x>1}^{}x^{6}u_{0}^{2}dx<\infty$, it was proven that there is 
blow-up in finite time. In the cases with blow-up, the 
\textit{universal blow-up profile} of the self-similar blow-up 
is given by the soliton $Q$ 
(\ref{soliton}).  A variational argument \cite{weinstein} implies 
that  $H^{1}$ initial data with subcritical mass $M[u_{0}]<M[Q]$ 
generate global in time solutions in $H^{1}$.

The work on $L_{2}$ critical blow-up was recently continued in 
three articles by Martel, Merle and Rapha\"el
\cite{MMR2012_I,MMR2012_II,MMR2012_III} to which we refer the reader 
for details. 
The stability of the soliton was addressed in \cite{MMR2012_I}, where 
it was shown that the soliton is both unstable against blow-up and against being 
radiated away towards infinity. More precisely the authors gave 
(see Theorem 1.1 of \cite{MMR2012_I}) the 
following
\begin{thm}[Martel, Merle, Rapha\"el (2013)]\label{MMR}
    Let $\mathcal{T}_{\alpha^{*}}$ be the set given by
    $$\mathcal{T}_{\alpha^{*}}=\left\{
    u\in H^{1} \mbox{ with } \inf_{\lambda_{0}>0, x_{0}\in 
    \mathbb{R}}\left|\left|u-\frac{1}{\lambda_{0}}Q\left(\frac{\cdot-x_{0}}{
    \lambda_{0}}\right)\right|\right|_{2}<\alpha^{*}\right\}$$ and let 
    $\mathcal{A}$ be the set of initial data $u(x,0)=u_{0}$ given by
    \begin{equation}
        \mathcal{A}=\left\{u_{0}=Q+\epsilon_{0} \text{ with } 
    ||\epsilon_{0}||_{H^{1}}<\alpha_{0}\ll 1 \text{ and } 
    \int_{x>1}^{}x^{10}\epsilon_{0}^{2}dx<\infty\right\},
        \label{A}
    \end{equation}
    where $0<\alpha_{0}\ll\alpha^{*}\ll1$ are universal constants, 
    and let 
    $u_{0}\in\mathcal{A}$. If $E[u_{0}]<0$ and if $u(x,t)$ is not a 
    soliton, then $u(x,t)$ blows up at the finite time $t^{*}$ and 
    $u(t)\in \mathcal{T}_{\alpha^{*}}$ for $t<t^{*}$. Then there 
    exist a constant (with respect to $t$) $l_{0}(u_{0})> 0$ and 
    functions $L(t)$ and $x_{m}(t)$ such 
    that for     $t\to t^{*}$
    \begin{equation}
    u(x,t)- 
    \frac{1}{\sqrt{L(t)}}\,Q\left(\frac{x-x_{m}(t)}{L(t)}\right)\to
    \tilde{u}\in L_{2}
    \label{selfs},
\end{equation}
and 
\begin{equation}
    ||u_{x}||_{2}\sim \frac{||Q'||_{2}}{l_{0}(t^{*}-t)} 
    \label{ux2}
\end{equation}
with  $Q$ from (\ref{soliton}), where
\begin{equation}
    L(t)\sim l_{0}(t^{*}-t),\quad x_{m}(t)\sim 
    \frac{1}{l_{0}^{2}(t^{*}-t)}.
    \label{ll0}
\end{equation}

\end{thm}

In addition it is shown in Theorem 1.2 of \cite{MMR2012_I} that there are 
just three scenarios possible for initial data $u_{0}\in 
\mathcal{A}$ (\ref{A}): blow-up in finite time as detailed in theorem 
\ref{MMR}, the solution is identical to the soliton $Q$, or it leaves 
the set $\mathcal{T}_{\alpha^{*}}$ in finite time.  In addition 
it is proven that the minimal mass for initial data to lead to a 
blow-up is the mass of the soliton. Note, 
however, that the theorem does not exclude the possibility that the solution $u$ 
for initial data with smaller mass than the soliton 
leaves $\mathcal{T}_{\alpha^{*}}$ at some time $t_{1}>0$ and reenters 
it after some finite time $t_{2}>t_{1}$, 
which means that it becomes 
almost solitonic again. In addition there is no 
statement in the theorem whether the blow-up profile (\ref{selfs}) 
can be also observed for localized initial data with negative energy 
and a mass much larger than the soliton, i.e., for $u_{0}\notin 
\mathcal{A}$. It is one of the motivations 
of this paper to address numerically these questions which have 
not yet been investigated analytically.

The picture is much less clear for the supercritical cases $n>5$. 
It is just known that blow-up is possible in this case, but not for which 
initial data, and the precise mechanism of the blow-up is unknown. It was also shown in \cite{BSW1987} 
that the soliton is unstable in this case, but the mechanism of the 
instability or of a blow-up is not rigorously established. Therefore 
numerical experiments have been carried out in 
\cite{BDK1986,BDKMcK1992,BDKMcK1995} to address this issue. The basic 
idea of the numerical approach in these papers 
was to use cubic (or higher) splines for the spatial dependence and an 
implicit fourth order Runge-Kutta method for the time dependence 
(which we will also use for some code), 
essentially fourth order methods in time and space. The studied 
examples were periodic on $[0,1]$, and the maximum of the initial 
data was localized at $x=0.5$. To address the 
blow-up, an adaptive mesh refinement was used. The maximum of the 
solution was shifted manually every few time steps back to $x=0.5$. 
When $\sqrt{h}||u||_{\infty}/||u||_{2}$ dropped 
below a prescribed tolerance, the mesh size $h$ of the spatial grid was refined close to 
the maximum of the solution. The resolution in time was increased to 
keep some conserved quantity, essentially the numerically computed 
energy, at a prescribed accuracy, which was not easy to achieve. This 
technique allowed in \cite{BDKMcK1995} to trace the $L_{\infty}$ norm and 
certain $L_{p}$ norms of 
the solution in cases with blow-up.  It was 
possible with this numerical approach to study blow-up 
in the supercritical cases $n>4$, but the identification of the 
asymptotic behavior for the critical case $n=4$ was not conclusive. 
Similar experiments were carried out in the presence of dissipation 
in the equations. The adaptive mesh refinement was very effective in 
tracing blow-up, but the lack of resolution in the rest of the 
computational domain, due to a numerical approach of finite order and 
due to less powerful computers than accessible today, made it impossible to 
decide whether the observed small oscillations were numerical 
artifacts or true effects in the solution. 
This was for instance 
noted in  \cite{DixMcKinney} where the oscillations were seen as 
an artifact. In \cite{DixMcKinney} the authors studied the gKdV equation for the 
case $n=5$ with numerical (the same approach as in \cite{BDKMcK1995}) and asymptotic methods. This allowed to 
conclude that a self-similar solution acts as an attractor for 
blow-up as in the critical case, but that this is not a rescaled soliton 
(\ref{soliton}). Instead the blow-up profile is given by a solution 
vanishing for $|\xi|\to\infty$ of the ordinary differential 
equation (ODE)
\begin{equation}
    -a_{\infty}\left(\frac{2}{n}U_{\infty}+\xi 
    U_{\infty,\xi}\right)-v_{\infty}U_{\infty,\xi}
    +U^{n}_{\infty}U_{\infty,\xi}+\epsilon^{2}U_{\infty,\xi\xi\xi}=0
    \label{ODE},
\end{equation}
where $U^{\infty}=U^{\infty}(\xi)$ is independent of $t$ and where 
$a_{\infty}$ and $v_{\infty}$ are constants. The ODE  (\ref{ODE}) 
was studied in \cite{DixMcKinney} and in more 
detail in \cite{Koch}. 
Since it contains two free parameters, a best fit of 
the numerical gKdV solution to a numerically found asymptotically 
decreasing solution of (\ref{ODE}) was performed in \cite{DixMcKinney} confirming the 
validity of the approach. Note that this is in contrast to the 
$L_{2}$ critical case $n=4$ for which theorem \ref{MMR} was proven. 
There the blow-up profile is asymptotically given by the soliton 
$Q_{c}$ (\ref{soliton})
which only contains one free parameter $c$ which can be fitted by the 
rescaling $Q_{c}(z)=c^{1/n}Q(\sqrt{c}z)$. 

\subsection{Small dispersion limit}
The parameter $\epsilon$ in (\ref{gKdV}) appears in the 
study of the small dispersion limit $\epsilon\ll1$ as in \cite{DGK2011}. It can be 
seen as being introduced in the gKdV equation (\ref{gKdV}) with 
$\epsilon=1$ by some initial data, which vary on large 
scales of order $1/\epsilon$. Studying the solution of gKdV for such 
initial data on large time scales of order $1/\epsilon$ can be 
achieved by a transformation $x\to x\epsilon$, $t\to t\epsilon$ 
which takes gKdV with $\epsilon=1$ to the form (\ref{gKdV}). Note that this rescaling 
changes the mass of $u$ by multiplication with a factor $1/\epsilon$, 
which will of course be taken into 
account when addressing conditions on the norm of the initial data in 
the context of blow-up studies formulated for $\epsilon=1$. 

It is known that 
solutions to the formal limit of (\ref{gKdV}) for $\epsilon=0$, 
generalized Hopf equations, have for generic localized initial data a 
point of gradient catastrophe for some finite time $t_{c}$. At this 
point, also denoted as \emph{hyperbolic blow-up}, the $L_{\infty}$ 
norm of the solution stays finite in contrast to the $L_{\infty}$ blow-up in 
solutions to gKdV for finite $\epsilon$.  An important feature of the small 
dispersion limit of nonlinear dispersive PDEs as gKdV is the 
appearance of a \emph{dispersive shock}, a zone of rapid modulated 
oscillations in the vicinity of the point of gradient catastrophe of 
the solution to the dispersionless equation for the same initial 
data. This small dispersion limit was studied  in \cite{DGK2011} also 
numerically. It was found in \cite{DGK2011} that for $t>t_{c}$ a 
dispersive shock is observed.  But due to an 
instability in the time integration scheme (an \emph{exponential time 
differencing method} (ETD)) in the numerical approach, which will be 
addressed in more detail elsewhere, it was not possible to decide 
whether there will be an $L_{\infty}$ blow-up of the solution in 
finite time. It is one of the goals of the present paper to answer 
this question (which is equivalent to blow-up for initial data of 
large mass) and to show, that there is finite time blow-up at some 
time $t^{*}$ which is always greater than $t_{c}$. Moreover there 
will be always a dispersive shock in contrast to the corresponding 
situation of nonlinear Schr\"odinger equations, see \cite{DGK2013}. 
With the numerical methods presented in this paper, it is possible to 
address both the dispersive shock and a blow-up with the necessary 
resolution.  We show that for $n=4$, the asymptotic description of 
theorem \ref{MMR} still applies though the studied initial data are 
very far from the soliton. 

In the small dispersion limit of generalized NLS equations, it was 
conjectured in \cite{DGK2013} that the 
NLS solution for small $\epsilon$ is locally given near the critical time of the semiclassical 
solution by the \emph{tritronqu\'ee} solution of the Painlev\'e I 
equation where $x$ and $t$ have to be rescaled by a factor $\epsilon^{4/5}$. It was 
found numerically in \cite{DGK2013} that the same scaling in 
$\epsilon$ applies also for the blow-up time in the $L_{2}$ critical 
case. No dispersive shock is observed in this case. 
For gKdV equations is was conjectured in \cite{DGK2011} that 
the  gKdV solution  near the break-up point of the generalized 
Hopf solution is given by a special solution to the so-called 
Painlev\'e I2 equation where $t$ has to be rescaled by a factor 
$\epsilon^{4/7}$. An important question in the context of gKdV is  
whether a dispersive shock can be observed before an eventual blow-up 
and whether the Painlev\'e asymptotics near the critical time of the 
generalized Hopf solution plays a role up to blow-up. 

As in \cite{DGK2011} we discuss the initial data 
$u_{0}=\beta\, \mbox{sech}^{2}x$ in the semiclassical limit $\epsilon\ll 1$. 
These data are motivated by the KdV soliton and have the advantage 
to be more rapidly decreasing than the gKdV soliton 
(\ref{soliton}). Thus it is possible to analytically continue the 
solution (within numerical precision) as a periodic function on a 
smaller computational domain. This allows to use effectively higher resolutions 
in Fourier space. The critical time for solutions  
to the generalized Hopf equation (gKdV (\ref{gKdV}) for 
$\epsilon=0$) for these initial data can be given in closed form:
\begin{equation}
    t_{c}=\frac{(1+2n)^{n+1/2}}{(2n)^{n+1}\beta^{n}}
    \label{tch}.
\end{equation}
Note that we use a different 
scaling of the gKdV equation here with respect to \cite{DGK2011} (a 
factor 6 is missing in front of the nonlinearity), which leads to a 
different critical time. 
The energy (\ref{energy}) is obviously always 
negative  for fixed $\beta$ if $\epsilon$ is small enough. 

\subsection{Structure of the present work and main results}

In this paper, we present a careful numerical study 
of soliton stability in the critical and supercritical cases, of the 
small dispersion limit $\epsilon\ll 1$, 
and of the asymptotic behavior in the cases with blow-up. To this end 
we employ a dual approach: a direct integration of the gKdV equation 
(\ref{gKdV}), and a dynamically rescaled gKdV equation (\ref{gKP5}) 
as detailed in the next section for cases with blow-up. 
Since these numerical 
approaches are independent, they provide a mutual check of their respective 
accuracy. In all cases, we use a Fourier 
spectral method for the spatial dependence which is optimal for the 
smooth initial data we discuss here. For the direct 
integration of gKdV we apply an implicit fourth order Runge-Kutta  method for the time 
dependence as 
in \cite{BDKMcK1995},  for  the dynamically 
rescaled gKdV equation (\ref{gKP5}) an ETD
scheme of fourth order. This implies the use of two high order methods for space 
and time allowing for an efficient numerical solution of high 
accuracy. For the dynamical rescaling, we follow the approach presented in 
\cite{McLPSS1986,LeMPSS1987,LPSS1988,PSSW1993}, as well as in the 
monograph \cite{SulemSulem1999}, for numerical studies of the blow-up 
of solutions to the nonlinear Schr\"odinger equations (NLS). These 
numerical methods have proven instructive in the analytic description 
of blow-up in NLS systems. A similar approach was 
applied in \cite{KPSZ1988,ZakharovShvets1988,KSZ1991}, but there 
only the independent variables were rescaled. The rescaling allows a 
high resolution computation up to times very close to       
blow-up. Since the critical case $n=4$ is expected to show the 
behavior (\ref{selfs}) near blow-up, this implies that the blow-up is 
expected for $x_{m}\to \infty$. To address this problem, we solve the 
equation in a frame commoving with the maximum of the solution. In 
addition, the blow-up profile  for $n=4$ is 
reached algebraically in the rescaled time whereas an exponential dependence is 
expected for $n>4$. Thus it is numerically extremely challenging to get 
close to the critical time $t^{*}$. But as will be shown 
in the paper, we get close enough to allow for some fitting to the 
theoretical predictions. To the 
best of our knowledge, we present here the first implementation of a 
fully dynamical rescaling for gKdV equations. In the case of very large 
mass of the initial data, the direct integration of the dynamically 
rescaled equation (\ref{gKP5}) suffers from numerical instabilities 
at the boundaries of the computational domains. As in 
\cite{KP14,KS2014b,KS2014,KSM14} the quantities of the dynamical rescaling 
are obtained  
in these cases by postprocessing the data from the direct 
integration of the gKdV equation (\ref{gKdV}).

Our numerical simulations are compatible with the following conjectures  
about the analytic behaviour of the solution 
in the critical and supercritical case. For $n=4$ we have
\begin{conj}\label{conj4}
   Let $Q$ be the soliton (\ref{soliton}) and $u(x,t)$ be the 
   solution of (\ref{gKdV}) for $n=4$ for 
   smooth initial data $u_0(x)\in L_{2}(\mathbb{R})$ of sufficiently 
   large mass with a single 
   maximum.
   
   \begin{itemize}
       \item If  $E[u_{0}]>E[Q]$ and $||u_0||_2 
   < ||Q||_2$, the solution is radiated away to infinity.
   
   \item If 
      $E[u_{0}]<E[Q]$ and $||u_0||_2 
   > ||Q||_2$, 
   then there is a a blow-up at the finite time $t^{*}$ as detailed in 
   theorem \ref{MMR}, even for initial data not in the class 
   $\mathcal{A}$ (\ref{A}).
      \end{itemize}
\end{conj}
This means that perturbations of the soliton with smaller mass than 
the soliton will be radiated away to infinity without coming close 
again to the soliton state (there is no refocusing of the mass). In 
the case of a blow-up, the theory by Martel, Merle and Rapha\"el \cite{MMR2012_I} 
gives the asymptotic description of the blow-up even for localized 
initial data with much larger mass than the soliton. 

For the supercritical case $n>4$, we confirm the 
numerical results by 
Bona et al.~\cite{BDKMcK1995} and Dix and McKinney 
\cite{DixMcKinney} for the appearance of blow-up, but can in addition reliably trace 
dispersive oscillations and dispersive shocks. We 
give via a dynamical rescaling an asymptotic description of the 
blow-up profile. The study can be summarized in the following
\begin{conj}\label{conj1}
   Let $P$ be the solution to the ODE 
      (\ref{ODE}) discussed in 
   \cite{Koch} and $u(x,t)$ be the solution of (\ref{gKdV}) for $n=5$ for 
   smooth initial data $u_0(x)\in L_{2}(\mathbb{R})$ of sufficiently large mass
   with a single 
   maximum, 
   then there is a a blow-up at the finite coordinates $(x^*,t^*)$ such 
   that
   \begin{itemize}
       \item   \begin{equation}\label{selfssc}  u(x,t)- 
    \frac{1}{L(t)^{2/n}}\, P\left(\frac{x-x_{m}(t)}{L(t)}\right)\to 
    \tilde{u}\in L_{2}
    \end{equation}
    where the scaling factor $L$ has for $t\sim t^{*}$ the form
           $ L(t) = C(t^* - t)^{1/3}$ 
         with $C = C(u_0)$ a constant depending on the initial data $u_0$
      \item  the position $x_m$ of the maximum has the asymptotic 
      behavior
         \begin{equation}\label{eq:conj_L_xm}
            x_m(t) = \gamma L(t) + x^* \quad \text{as} \quad t\nearrow t^*
         \end{equation}
         where $\gamma = \gamma(u_0)$ is a constant depending on the initial 
         data $u_0$. 
   \end{itemize}
\end{conj}

In the small dispersion limit $\epsilon\ll 1$, we show 
that there is always a dispersive shock in contrast to the NLS case, and that the scaling of the 
blow-up time follows an exponential law in $\epsilon$. More precisely 
we find 
\begin{conj}\label{sdconj}
Let $u_{0}\in L_{2}(\mathbb{R})$ be smooth
initial data with a single maximum 
and $\epsilon\ll1$. 
Then the solution to the gKdV equation (\ref{gKdV}) for $n\geq 4$ has 
a dispersive shock for $t$ between $t\sim t_{c}$, the break-up time of the 
solution to the generalized Hopf equation, and the blow-up time 
$t^{*}$. The $\epsilon$ dependence of the blow-up time is given by 
\begin{equation}
    t^{*}(\epsilon)=t^{*}_{0}\exp(\alpha \epsilon)
    \label{tstar},
\end{equation}
where $t^{*}_{0}>t_{c}$ and 
$\alpha$  are constant with respect to $\epsilon$.
\end{conj}

The paper is organized as follows: In section 2 we present the used 
numerical methods and perform tests. The $L_{2}$ critical case $n=4$ is discussed in 
section 3. We study perturbations of the soliton, the small 
dispersion limit
 and blow-up. The same examples are studied in section 4 for 
the supercritical case $n=5$. In both these 
sections we test the numerical approaches in the presence of blow-up 
and identify viable strategies to characterize the mechanism of 
the blow-up. This allows in 
section 5 to study initial data with larger mass for the example of 
the small dispersion limit with $\epsilon\ll1$ and to determine the 
$\epsilon$ dependence of the blow-up time $t^{*}$. 

\section{Numerical methods}
In this section we present the used numerical methods. Two approaches 
will be applied which complement each other: On one hand we directly 
integrate the gKdV equation for given initial data. 
With this code we study soliton stability and the small 
dispersion limit of gKdV. In the cases with blow-up, 
we use in addition for the first time a dynamical rescaling for the gKdV 
equation to study the blow-up in more detail and to identify the type 
of formed singularity. This approach can also be used a posteriori on 
the data obtained with the direct integration to identify scaling 
laws for certain norms near blow-up. The spatial 
dependence will be treated with a Fourier spectral method in all 
cases, the time integration with fourth order methods. 
Both codes will be tested at the example of the explicitly 
known soliton solutions (\ref{soliton}). Since the approaches are 
independent, the consistency of results with both codes provides an 
additional test of the 
accuracy.

\subsection{Direct numerical integration}

The choice to use Fourier methods is based on the excellent 
approximation properties spectral methods have for smooth functions 
as the ones studied here. We will always consider initial data which 
are rapidly decreasing, and which can thus be analytically continued 
as periodic within the finite numerical precision if the 
computational domain is chosen large enough. Since we study here 
dispersive equations, it is also important that spectral methods 
minimize the introduction of numerical dissipation that might 
suppress dispersive effects. The 
\emph{method of lines} approach implies that the partial differential 
equation (PDE) (\ref{gKdV}) 
will be approximated via a system of 
ODEs for the Fourier coefficients. The latter system is 
\emph{stiff}\footnote{The term stiffness is used here to indicate that there 
are different timescales in the studied problem which make the use of 
explicit methods inefficient for stability reasons.} because of the third derivative in $x$, but has the 
advantage that the stiffness appears in the linear term of the 
equation, 
\begin{equation}
    \hat{u}_{t}=\mathcal{F}(u)+\mathcal{L}\hat{u},
    \label{gKdVfourier}
\end{equation}
where $\mathcal{L}=\epsilon^{2}ik^{3}$ and $\mathcal{F}(u)=-ik\widehat{u^{n+1}}/(n+1)$. For equations of 
the form (\ref{gKdVfourier}), there are many efficient high-order time 
integrators, see e.g.~\cite{CoxMatthews2002,KassamTrefethen2005,HO,Klein2008,KleinRoidot2011}, especially for 
diagonal $\mathcal{L}$ as in the Fourier case. Thus the Fourier approach is not 
only very convenient for the spatial dependence, but allows also 
efficient time integration of high accuracy. 

In \cite{KleinRoidot2011}, most of the studied integrators as in \cite{KassamTrefethen2005} are 
explicit and thus in general much more efficient than implicit 
integrators. It was shown, however, that an implicit Runge-Kutta of 
fourth order (IRK4), a two-stage Gauss scheme, could be competitive if the 
iteration is optimized. This method has also been used in 
\cite{BDKMcK1995,DixMcKinney}.
Similar to \cite{KleinRoidot2011} we 
apply a simplified Newton scheme to solve the nonlinear equations for 
the IRK4 scheme.
In this form the iteration converges rapidly 
(at early times in 3 to 4 iterations) because the  stiffness in this 
example appears in the linear part $\mathcal{L}$ which is addressed explicitly. 
Note that in \cite{BDKMcK1995,DixMcKinney}, too, a simplified Newton 
iteration was applied, which in this case was efficient, but 
approximate since the differentiation matrices were not diagonal. 
We apply this approach since it is robust up to an eventual 
blow-up of the solution whereas the explicit schemes used in 
\cite{Klein2008} would require  prohibitively small time steps 
to address stability issues to be discussed elsewhere. This is the 
reason why one could not get close enough to actual blow-up in 
\cite{DGK2011} to uniquely identify it. 

Accuracy of the numerical solution is controlled as discussed for 
instance in 
\cite{Klein2008,KleinRoidot2011} via the numerically computed energy (\ref{energy}) 
which will depend on time due to unavoidable numerical errors. We use 
the quantity
\begin{equation}
    \Delta=|E(t)/E(0)-1|
    \label{Delta}
\end{equation}
as an indicator of the numerical 
accuracy (if $E(0)=0$ we just consider the difference $|E(t)-E(0)|$). It was shown in \cite{Klein2008,KleinRoidot2011} that the numerical accuracy 
of this quantity overestimates the $L_{\infty}$ norm of the 
difference between numerical and exact solution by two to three 
orders of magnitude. A precondition for the usability of this 
quantity is sufficient resolution in Fourier space, i.e., a large 
enough number of Fourier modes. 

We generally choose the number $N$ of Fourier modes high enough 
such that they decrease to machine precision ($10^{-16}$ here) for times 
much smaller than the time of blow-up. The number $N$ thus depends   
on the size of the computational domain, $x\in[-\pi,\pi]D$ where the 
real constant $D$ is chosen large enough to ensure `periodicity' of 
the initial data in the sense discussed above.
An occurrence of blow-up in the code is indicated by an increase of 
the Fourier modes for the high wave numbers which implies eventually a 
failure of the iterative solution of IRK4 relations to converge. We 
always choose the time step sufficiently small  to ensure that 
the solution is obtained with machine precision for times much 
smaller than the blow-up time. 

\begin{rem}
The chosen time step in the case of a blow-up
is always small enough that a lack of resolution in Fourier space 
constrains the accuracy. In other words, the presented resolution in 
time is always high enough that a further reduction of the time 
step does not change the presented result. The code also does not get 
decisively faster, if an adaptive scheme is used, since the iteration 
converges more rapidly for smaller time steps. Therefore instead of 
an adaptive scheme, with which we were experimenting,  we reduce the 
size of the time step until the final recorded time (controlled via 
energy conservation) does not change anymore. 
\end{rem}

\subsection{Dynamic rescaling}
The results in \cite{MMR2012_I} for the critical case $n=4$ and the 
numerical studies \cite{BDKMcK1995,DixMcKinney} suggest that an $L_{\infty}$ blow-up 
of solutions to gKdV is to be expected for initial data with negative 
energy and mass larger than the soliton mass. This can be numerically 
addressed by an adaptive approach, a rescaling of the coordinates 
and of the solution $u$ by some time dependent factor $L$, which
is supposed to vanish at the blow-up time. In addition it is 
convenient to address the fact that the maximum which eventually 
develops into the blow-up is moving in contrast to NLS cases where 
such techniques have been applied  for self similar blow-up 
(for the description of 
the method and the results in the NLS case, we refer to 
\cite{SulemSulem1999} and references therein). Thus we also shift the 
$x$ coordinate by  $x_{m}(t)$, the location of the maximum, 
in order to keep it at 
a constant position in the transformed coordinates.  In the case 
$n=4$, the quantity $x_{m}$ is expected to tend to infinity at 
blow-up. With the 
coordinate change 
\begin{equation}
    \xi = \frac{x-x_{m}}{L},\quad 
    \frac{d\tau}{dt}=\frac{1}{L^{3}},\quad U = L^{2/n}u
    \label{gKP4},
\end{equation}
which leaves gKdV invariant for constant $L$ and $x_{m}$, we get for (\ref{gKdV})
\begin{equation}
    U_{\tau}-a\left(\frac{2}{n}U+\xi U_{\xi}\right)-vU_{\xi}
    +U^{n}U_{\xi}+\epsilon^{2}U_{\xi\xi\xi}=0
    \label{gKP5},
\end{equation}
with 
\begin{equation}
    a=(\ln L)_{\tau}, \quad v = \frac{x_{m,\tau}}{L},
    \label{a}
\end{equation}
where $x_{m,\tau}$ denotes the derivative of $x_m$ with respect to 
$\tau$. The coordinate transformation (\ref{gKP4}) implies in the 
cases studied here that a blow-up 
occurs at $\tau=\infty$. Thus the space and time scales are changed 
adaptively around blow-up, exactly in the way predicted by 
\cite{MMR2012_I} for the critical case.

Equation (\ref{gKP5}) is also important for theoretical purposes and 
has been used in \cite{MarMer2000} and later works. In the limit 
$\tau\to\infty$, i.e., at blow-up, the functions $U$, $v$ and $a$ are expected to 
become independent of $\tau$ which is denoted by an index $\infty$. 
For $v$ this can be seen from the fact that $x_{m}\propto 1/L$ for 
$n=4$ and that we find numerically that $x_{m}\sim \gamma L+x^{*}$ 
for $n=5$ which implies that $v_{\infty}\propto a_{\infty}$ in both 
cases.
Thus (\ref{gKP5}) is expected to reduce  in the limit $\tau\to\infty$ to the ODE 
(\ref{ODE}).

For the numerical implementation, the scaling factor $L$ and the 
\emph{speed} $v$ have to be chosen in a convenient way. 
We choose $\xi$ such that the maximum of $|U|$ is located at $\xi=\xi_0$ which 
implies  $U_{\xi}(\xi_0,\tau)=0$.  
This condition implies by differentiating (\ref{gKP5}) 
with respect to $\xi$, 
\begin{equation}
    v= - a\xi_0 + 
    U(\xi_0,\tau)^{n}+\frac{\epsilon^{2}U_{\xi\xi\xi\xi}(\xi_0,\tau)}{U_{\xi\xi}(\xi_0,\tau)}
    \label{gKP8}.
\end{equation}

The scaling function $L$ on the other hand can be chosen to keep certain norms constant, for 
instance the $L_{\infty}$ norm, 
\begin{equation}
    L^{2/n}=\frac{||U||_{\infty}}{||u||_{\infty}}.
    \label{gKP6}
\end{equation}
We get from (\ref{gKP5}) in this case
\begin{equation}
    a=\frac{n\epsilon^{2}}{2}\frac{U_{\xi\xi\xi}(\xi_0,\tau)}{U(\xi_0,\tau)}
    \label{gKP7}.
\end{equation}
Another possibility is to choose $L$ such that the $L_{2}$ norm 
of $U$, which reads 
\begin{equation}
   L^{1/2-2/n}= \frac{||U||_2}{||u||_2}
    \label{L2},
\end{equation}
 is constant.
However, this is not convenient here since we 
also want to study the $L_{2}$ critical case $n=4$ for which the 
$L_{2}$ norm is invariant under a rescaling of the form (\ref{gKP4}) 
(thus there is no condition on $L$ in this case by imposing a 
constant $L_{2}$ norm of $U$).
Alternatively one can impose that the $L_{2}$ norm of $U_{\xi}$ is 
constant which is attractive since the blow-up theorems in 
\cite{MMR2012_I} are 
formulated for this norm. In this case we have the relation
\begin{equation}
    \label{u2x}
    L^{2/n + 1/2} = \frac{||U_{\xi}||_2}{||u_x||_2}
\end{equation}
and we get by differentiating the norm with respect to $t$ and by using 
equation (\ref{gKP5}) to eliminate derivatives with respect to $t$ 
after some partial integrations
\begin{equation}
    a=\frac{2n}{(n+1)(n+4)||U_{\xi}||_{2}^{2}}\int_{\mathbb{R}}^{}U^{n+1} U_{\xi\xi\xi}\,d\xi
    \label{L2x}.
\end{equation}
For NLS computations with dynamic rescaling this approach proved to 
be numerically more stable than fixing the $L_{\infty}$ norm to be 
constant, see \cite{SulemSulem1999} and references therein. We 
observe the same here. This appears to be due to the fact that 
condition (\ref{L2x}) involves an integral over the whole 
computational domain, whereas condition (\ref{gKP7}) is local. In 
addition the  appearance 
of the integral in (\ref{L2x}) provides a global control of how well 
the numerical solution solves the rescaled equation (\ref{gKP5}). 
Therefore we will always apply the choice (\ref{L2x}) in the following
Thus all quantities in (\ref{gKP5}) can be expressed in terms of $U$ 
alone. 

The spatial dependence in equation (\ref{gKP5}) will be again 
treated with Fourier spectral methods for the reasons outlined above. 
In addition the appearance of a fourth derivative in (\ref{gKP5}) 
implies that spectral accuracy is needed in the computation of the 
derivatives. 
Since the coordinate $\xi$ is dynamically rescaled with respect to 
$x$, this implies that the computational domain has to be chosen 
large enough to avoid that the imposed periodicity of the 
solution affects the blow-up profile. As will be shown, this is possible 
for small enough masses except for the perturbed solitons where spurious 
oscillations in the quantity $a(\tau)$ appear which can be clearly 
attributed to the periodic boundary conditions. But due to the 
comparatively slow decrease of the soliton solution towards infinity, 
this case is problematic with any approach. Nevertheless we  recover the 
predicted behavior of Theorem \ref{MMR} \cite{MMR2012_I} for $n=4$. Note that the high spatial 
resolution also allows to avoid spurious oscillations as observed in 
\cite{DixMcKinney} for
the numerical study of blow-up in the supercritical case.

For the 
Fourier approach $\xi U_{\xi}$ in (\ref{gKP5}) is the numerically problematic term\footnote{It is exactly this term 
which requires a special fall-off condition for the initial data in 
(\ref{A}).}. Since we have to choose large 
domains to address the `zooming in' effect ($L\to0$) at blow-up and since the 
solutions to (\ref{gKP5}) except for exact 
solitons have dispersive oscillations with slow decrease towards 
infinity, numerical errors at the boundaries for these \emph{dispersive 
tails} lead to a pollution of the Fourier coefficients at the high 
frequencies. A damping of the oscillations at the boundaries is 
equivalent to imposing incorrect boundary conditions. These lead to 
reflections of the oscillations at the boundary which will eventually 
destroy the solution. A
filtering in Fourier space to suppress an increase of the 
coefficients for high wave numbers has a similar effect. Thus the only 
way to address this problem without affecting the numerical solution 
appears to be the use of sufficient resolution in Fourier space and high time 
resolution, thus effectively propagating the solution with machine 
precision for as long as possible. As will be shown, this can be done 
for  masses close to the soliton mass, but the needed resolution in 
time will become prohibitive once the mass is several times the 
soliton mass.  

High time resolution can be achieved because 
equation (\ref{gKP5}) is in Fourier space again of the form 
(\ref{gKdVfourier}). Since it is dynamically rescaled,
there is no blow-up of the solution for finite $\tau$. Thus we can use an 
explicit integrator. It was shown in \cite{Klein2008} that 
ETD schemes perform best for KdV 
in the small dispersion limit, and 
in \cite{KleinRoidot2011} that the performance is similar for different ETD 
schemes. The basic idea of these methods is to use a constant time 
step $h=t_{m+1}-t_{m}$ and to integrate (\ref{gKdVfourier}) with an 
exponential factor,
$$\hat{u}(t_{m+1})=e^{\mathcal{L}h}\hat{u}(t_{m})+\int_{0}^{h}d\theta\, 
e^{\mathcal{L}(h-\theta)}\mathcal{F}(\hat{u}(\theta+t_{m}),\theta+t_{m}).$$
The different ETD schemes differ in the approximation of the integral.
We 
use here the method by Cox and Matthews \cite{CoxMatthews2002} which 
will be implemented as discussed  in \cite{Klein2008}.

Note that the used ETD scheme is of classical order four (for a 
detailed discussion see \cite{HO,HO09}). But the information on the 
quantity $a$ in (\ref{a}) at the stages of the scheme is not of the 
same accuracy as at the full steps. Thus we only use $a$ at the time 
steps $t_{n}$ and obtain $\ln L$ via the trapezoidal rule which is of 
second order. Due to the high number of time steps we use in 
practice, $L$ is obtained with more than sufficient accuracy. In the 
same way we 
obtain the time $t$ from (\ref{gKP4}) and $x_{m}$ from (\ref{gKP8}).
To control the accuracy of both the numerical solution $U$ and the 
computed factor $L$, we use that equation (\ref{gKP5}) has the 
conserved energy
\begin{equation}
            E[U] = \frac{1}{L^{4/n+1}}\int_{-\infty}^{\infty}\left(\frac{\epsilon^2}{2}U_\xi^2
           - \frac{1}{(n+1)(n+2)}
	   U^{n+2}\right)d\xi.
    \label{energyL}
\end{equation}
Since $L(\tau)$ explicitly enters this quantity, $E[U]$ controls both $U$ 
and $L$ at the same time. 

 \begin{rem}
     It would be possible to use a transformation of the form $x\to 
     x-x_{m}(t)$ also for the direct integration of gKdV to take care 
     of the propagation of the maximum of the solution and to choose 
     a commoving frame of reference. 
     In this article we will always study positive initial data, 
     i.e., the case which includes solitons. These propagate to the 
     right. But except for the exact soliton solution, there will be 
     dispersive oscillations also propagating to the left of the initial hump 
     localized at 0. Thus resolution for a not commoving frame is 
     needed in practice, or for one where the soliton is fixed close 
     to the right boundary. A commoving frame is mainly beneficial in 
     the context of a dynamical rescaling and will not be used for 
     the direct integration of gKdV. In the latter case, we will sometimes 
     place the initial hump closer to the right boundary at 
     $\xi=\xi_{0}$ in order to leave 
     maximal space to the dispersive oscillations propagating to the 
     left. 
 \end{rem}

In cases of large mass where the solution of the rescaled equation 
(\ref{gKP5}) cannot be obtained close enough to the blow-up, 
we will directly integrate (\ref{gKdV}) as described in 
\cite{KP14}, and then postprocess the data to obtain the function $L$ 
to identify blow-up. There are in principle two different cases important in this 
context,  an algebraic or an exponential decay of the scaling factor 
$L$ with $\tau$. In the algebraic case observed for $n=4$ we have $L(\tau) = 
C_1/\tau$  and thus $a_{\infty}=0$. In 
this case the ODE (\ref{ODE}) reduces to the ODE for travelling wave 
solutions of the gKdV equation in a commoving frame which has the 
unique localized solution $Q$ (\ref{soliton}) (recall that the 
soliton speed $c$ can always be rescaled to 1). 
For $n>4$ exponential decay $L(\tau) = C_2 e^{a_{\infty}\tau}$ with 
$C_{2}=const$ is numerically
observed which leads to $a_{\infty}<0$ and thus to the ODE (\ref{ODE}). 
Since we solve in this paper gKdV for small mass in the 
rescaled form (\ref{gKP5}) and since we observe the expected behavior 
of the scaling factor $L(\tau)$, we solve for large $\tau$ also (\ref{ODE}). 
For a given solution $u(x,t)$, the quantity $L(t)$ is determined either via 
(\ref{u2x}) and the condition that 
$||U_{\xi}(\cdot,0)||_{2}=||u_{x}(\cdot,0)||_{2}$ 
or via (\ref{gKP4}) and the condition that 
$||U||_{\infty}=||u_{0}||_{\infty}$. For an 
exponential dependence of $L$ on $\tau$, we have for $\tau\to\infty$
\begin{equation}
 L\propto (t^{*}-t)^{1/3},\quad  ||u_{x}||_{2}^{2}\propto \frac{1}{(t^{*}-t)^{3/5}},\quad 
   ||u||_{\infty}\propto \frac{1}{(t^{*}-t)^{2/15}}
    \label{exp}
\end{equation}
and for an algebraic one
\begin{equation}
 L\propto (t^{*}-t)^{1/2},\quad    ||u_{x}||_{2}^{2}\propto \frac{1}{t^{*}-t},\quad 
   ||u||_{\infty}\propto \frac{1}{(t^{*}-t)^{1/4}}
     \label{alg}.
\end{equation}
This allows to fit the $L$ obtained from the norms of the solution to the 
expected asymptotic behavior. The $L_{2}$ norm of 
$u_{x}$ involves an integral and thus takes into account the solution 
on the whole computational domain, whereas the $L_{\infty}$ 
norm provides a local criterion. Consistency of the  fitting results 
provides a test of the approach. In practice we run the 
codes until the conservation of the numerically computed energy drops 
below $10^{-3}$ which means the solution is no longer reliable. For 
the resulting data, we fit the last 1000 time steps to the expected 
asymptotic behavior (\ref{exp}) or (\ref{alg}). In the case of a 
linear dependence to be fitted, we use standard linear regression. 
For a nonlinear dependence as on $t^{*}$ in (\ref{exp}) and 
(\ref{alg}), we apply the optimization algorithm
\cite{fminsearch} distributed with Matlab as \emph{fminsearch}.

\subsection{Tests of the numerics}
Before using the above codes to study stability of the soliton 
solution (\ref{soliton}) and the small dispersion limit of gKdV, we 
will first test with which accuracy the code can propagate the 
explicitly known soliton solution. It will be shown that this can be 
done essentially with machine precision which is important in view of 
the fact that the soliton is unstable against perturbations, which 
will be studied in the following sections. The fact that no 
instability is observed for the used methods on the studied timescales for the exact 
solution makes clear that a perturbation has to be considerably 
larger than the numerical error to lead to a visible effect for the 
considered times.

For the direct integration of gKdV for $\epsilon=1$ with the IRK4 code, we use the 
initial data $Q(x+3)$ (\ref{soliton}) and 
compute the solution until $t=6$ for $x\in 10[-\pi,\pi]$. 
In this way we assure that the modulus of the 
Fourier coefficients for $N=2^{10}$ Fourier modes
decreases to $10^{-14}$ for $n=4$ and $10^{-12}$ 
for $n=5$ during the whole 
computation. If one wants to trace the solution for larger times with the same 
resolution, one has to use larger computational domains and a larger 
value of $N$. We use in both cases $N_{t}=10^{4}$ time steps and find  
$\Delta\sim 6.46*10^{-15}$ for $n=4$ as well as 
$\Delta\sim1.07*10^{-13}$ (\ref{Delta}) for 
the computation with $n=5$. The 
difference between numerical and exact solution is 
$1.24*10^{-12}$ for 
$n=4$ and $1.25*10^{-12}$ for $n=5$. Note that the energy cannot be computed to higher precision 
than the resolution in Fourier space. This is why it is almost of the 
same order as the $L_{\infty}$ norm of the difference between 
numerical and exact solution here.

We also test the code with dynamical rescaling for the soliton though 
in this case the function $L(t)$ will be equal to one within numerical 
precision. Nonetheless this test checks whether the code can propagate 
the soliton for finite times with machine precision in a commoving 
frame (due to the condition that the maximum is localized at 
$\xi=\xi_{0}$ with $\xi_{0}=0$, 
the soliton is stationary in this setting). The code is run up to 
$\tau=t=100$ in this case, and the error is still of the order of 
$10^{-10}$ at the final time. There is no indication of an instability due to the 
numerical error on this timescale. The computation is carried out 
with      $N = 2^{14}$ Fourier modes and  $N_t = 2.5* 10^5$  time steps for  
   $x\in 100[-\pi,\pi]$. The computed energy is in this case of the 
   order of $10^{-10}$ in accordance with the $L_{\infty}$ norm of 
   the difference between numerical and exact solution. The $L_{2}$ 
   norm of $U_{\xi}$ is constant to the order of $10^{-12}$. 
  
\begin{figure}[htb!]
 \includegraphics[width=0.49\textwidth]{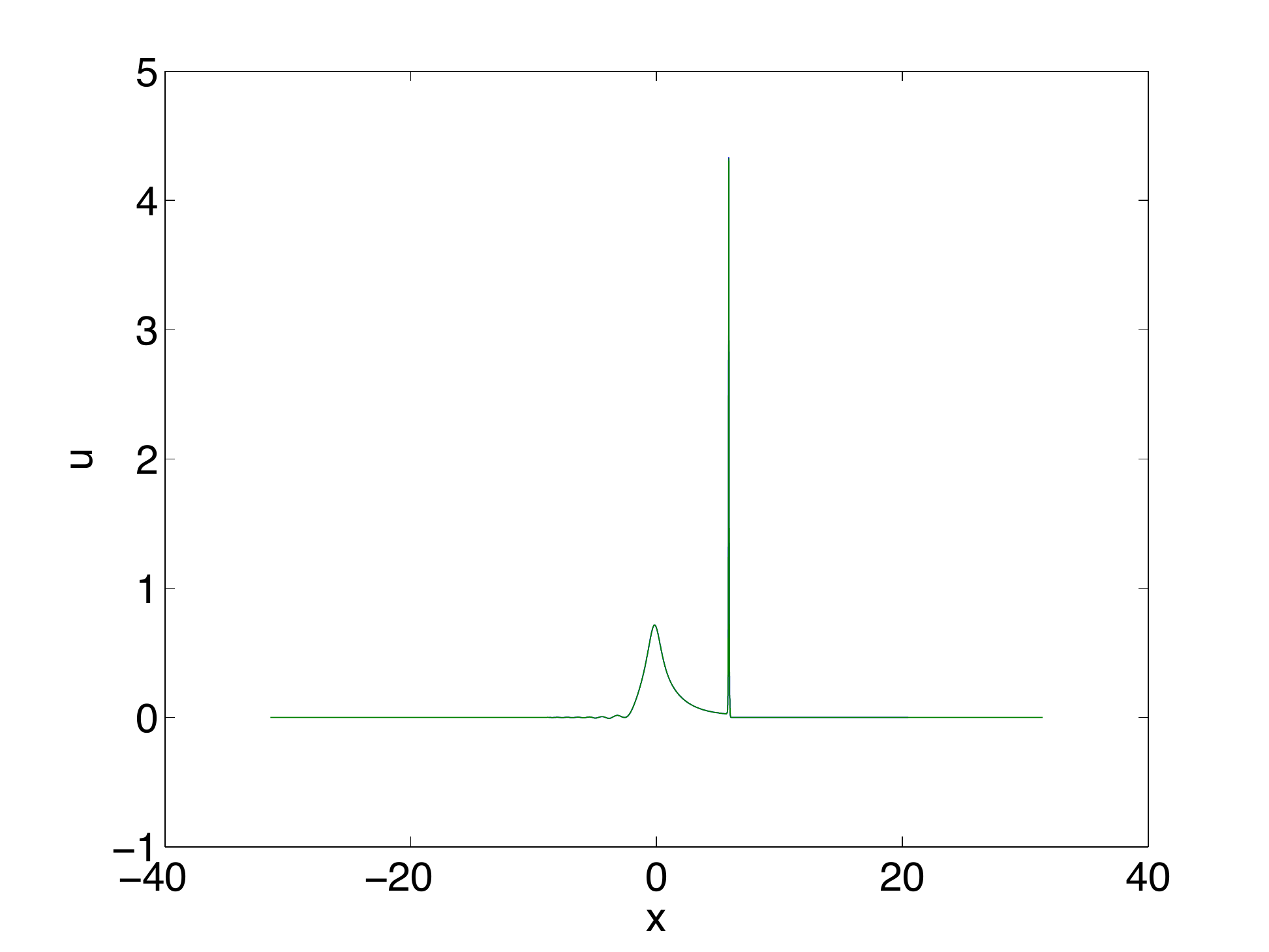}
 \includegraphics[width=0.49\textwidth]{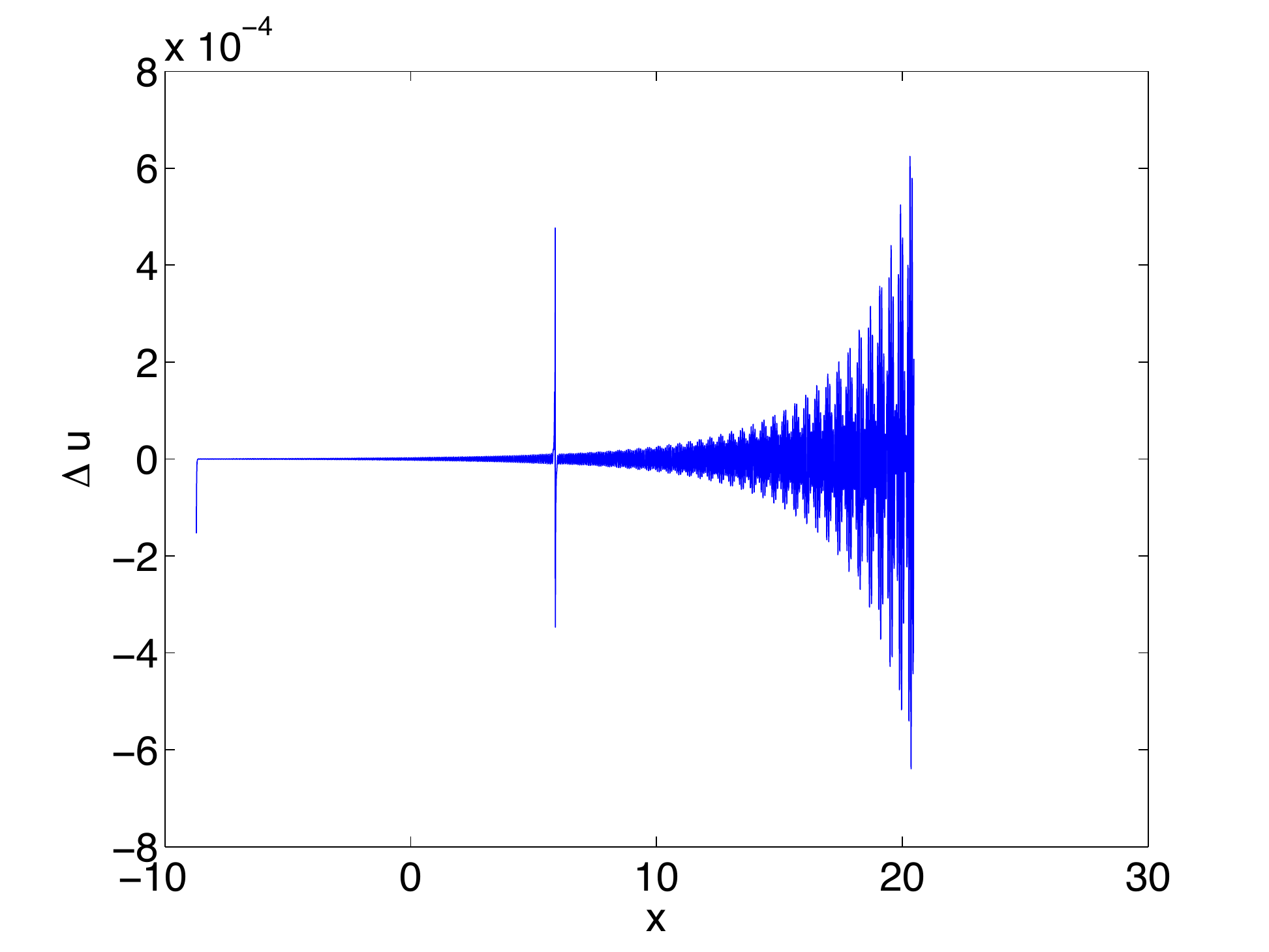}
 \caption{Solution to the gKdV equation (\ref{gKdV}) for $n=4$, 
 $\epsilon=0.1$ and the initial data $u_{0}(x)=\mbox{sech}^{2}x$ for 
 $t=4.1786\ldots$ in blue, and the solution to the rescaled gKdV 
 equation (\ref{gKP5}) for $\tau=1700$ after inverting the scaling 
 (\ref{gKP4}) in green on the right; on the left we show the 
 difference between both solutions.}
\label{deltau}
\end{figure}

Since the numerical approaches for the direct integration of gKdV 
(\ref{gKdV}) and the dynamical rescaled equation (\ref{gKP5}) are 
independent, they can be used to test each other's quality. To 
illustrate this at a concrete example, we consider the case $n=4$ for 
the initial data $u_{0}=\mbox{sech}^{2}x$ in the small dispersion 
limit $\epsilon=0.1$, the example discussed in detail in section 3.2. 
There in Fig.~\ref{fig:SechSquare_n4_U_new}, the solution obtained with the 
dynamically rescaled code is shown for $\tau=1700$ which corresponds 
to $t=4.1786\ldots$. We then use the code for the direct numerical 
integration of gKdV for the same initial data until exactly the same 
time (the parameters for the computation are given in section 3.2).  
In the latter case, the relative computed energy $\Delta$ is of 
the order of $10^{-7}$, and the modulus of the Fourier coefficients 
decreases to $10^{-5}$. Thus we estimate the solution to be accurate 
at least to the order of  $10^{-4}$. In the former case the relative 
computed energy is of the order of $10^{-4}$, and the modulus of the 
Fourier coefficients decreases to the order of $10^{-8}$. Thus we 
expect the solution to be accurate at least to the order of $10^{-3}$. In fact 
if we plot both in dependence of $x$ on the right in Fig.~\ref{deltau} by using 
the scaling (\ref{gKP4}), we see that both solutions cannot be 
really distinguished in the plot. Therefore we use cubic splines to 
interpolate from the grid in $\xi$ to the one in $x$ and present the 
difference of both solutions on the left in Fig.~\ref{deltau}. 
It can be seen that 
this difference is smaller than $10^{-3}$ as expected. This provides 
as already mentioned a strong test of the numerical accuracy of both 
codes since we even have to interpolate between two grids. Thus we are 
convinced that all shown solutions in this paper are correct to at 
least plotting accuracy. 

\section{The $L_{2}$ critical case $n=4$ for small masses}
In this section we study numerically solutions to the gKdV equation 
(\ref{gKdV}) in the $L_{2}$ critical case $n=4$, where we concentrate 
on masses close to the soliton mass (up to three times the soliton 
mass). Cases with larger mass will be 
considered in section \ref{smalldisp}. We consider 
perturbations of the soliton and the limit of small dispersion. The  results 
can be  summarized as follows (see  conjecture 
\ref{conj4}):

\begin{itemize}
    \item  Localized smooth initial data with a single maximum and a 
    mass smaller than the soliton mass are radiated away to infinity 
    (there is no refocusing of the solution).

    \item  Initial data of this class with a mass larger than the 
    soliton mass show self similar blow-up according to Theorem \ref{MMR}. 
\end{itemize}

The limitation to 
masses of the same order as the soliton mass allows also to use two complementary numerical approaches which provides a strong 
test for the numerical methods, a direct integration of gKdV and the dynamical 
rescaling.  In addition this study indicates how to 
address the case of larger masses in which the dynamical rescaled 
code is subject to numerical instabilities. 


\subsection{Perturbations of the soliton}
We consider as initial data perturbations of the soliton (\ref{soliton}) of the form 
$u(x,0)=\sigma\, Q(x-x_{0})$ with $\sigma$ a real constant slightly smaller 
or bigger than 1, and $x_{0}$ the initial position of the soliton. 
This means that the initial data are in the class 
$\mathcal{A}$ (\ref{A}). 
In this subsection, we put  $\epsilon=1$. The mass of  
the soliton is $M[Q] = 6.0837\ldots$ and its energy vanishes, $E[Q] = 0$.

\begin{figure}[htb!]
  \includegraphics[width=\textwidth]{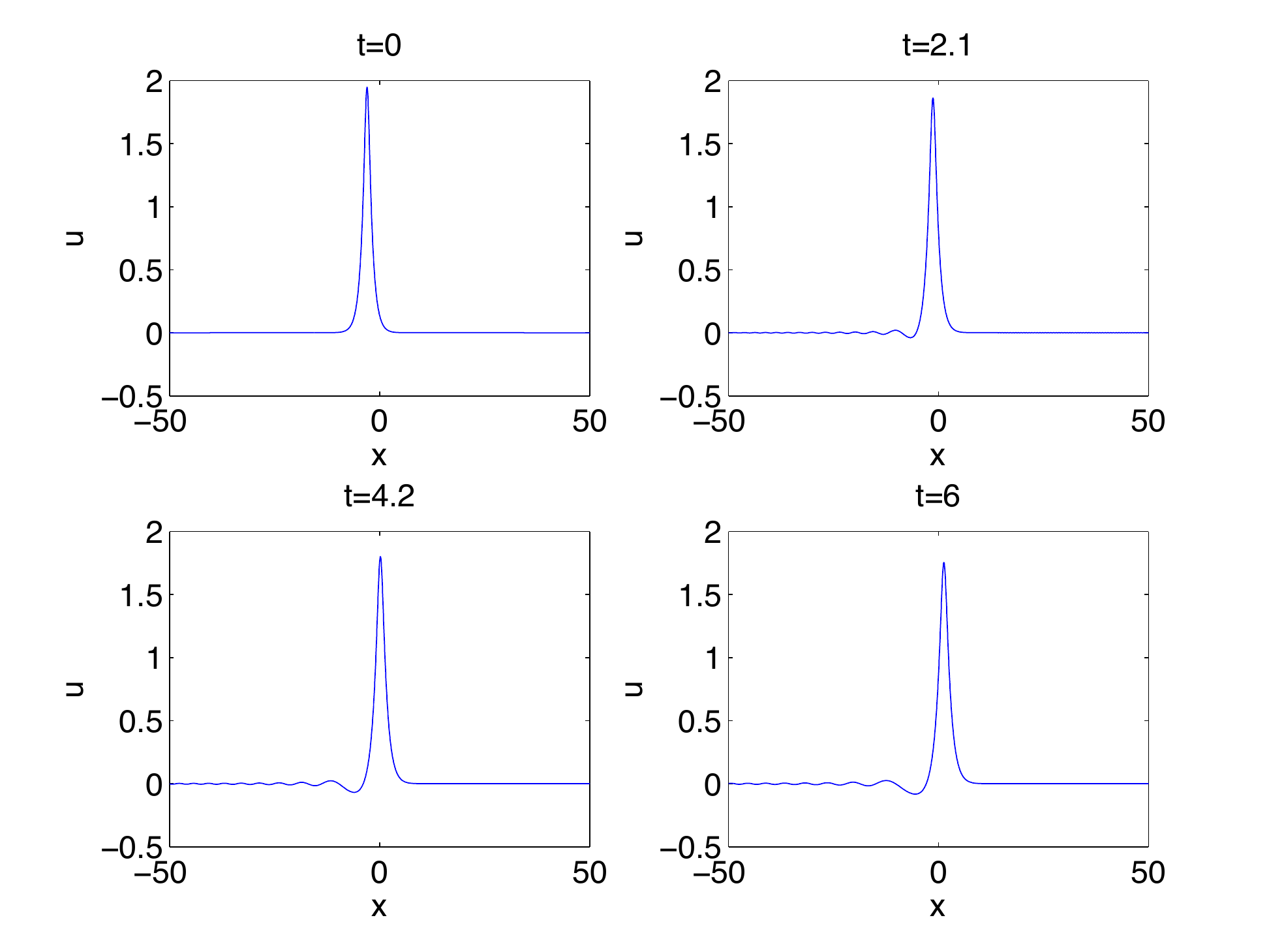}
 \caption{Solution to the gKdV equation (\ref{gKdV}) with 
 $\epsilon=1$ for $n=4$ 
 and the perturbed soliton initial data $0.99\,Q(x+3)$ (\ref{soliton}) 
 for several values of $t$.}
 \label{gKdVn4sol0994t}
\end{figure}

We first consider the initial data $u(x,0)=\sigma\, Q(x+3)$, with $\sigma = 0.99$, on a large domain, 
$x\in 100[-\pi,\pi]$ with $N=2^{14}$ Fourier modes and $N_{t}=10^{4}$ 
time steps. In this situation the mass is obviously smaller than the mass of the soliton and the energy is positive.   It can be seen in 
Fig.~\ref{gKdVn4sol0994t} that dispersive oscillations propagating to 
the left form immediately. The amplitude of these oscillations 
decreases very slowly which is why we choose a large computational 
domain. Only part of it is shown in the figure. 
It can be seen in Fig.~\ref{gKdVn4sol0994t} 
that the $L_{\infty}$ norm of the solution 
for the perturbed soliton decreases in this case monotonically. Thus 
it appears that the soliton will be just radiated away, there is no 
indication of a refocusing of the mass which would not be excluded by 
theorem \ref{MMR}. The Fourier 
coefficients at the final time show that the solution is fully 
resolved in Fourier space. The energy is conserved in the computation 
to the order of $10^{-12}$. 

\begin{figure}[htb!]
  \includegraphics[width=\textwidth]{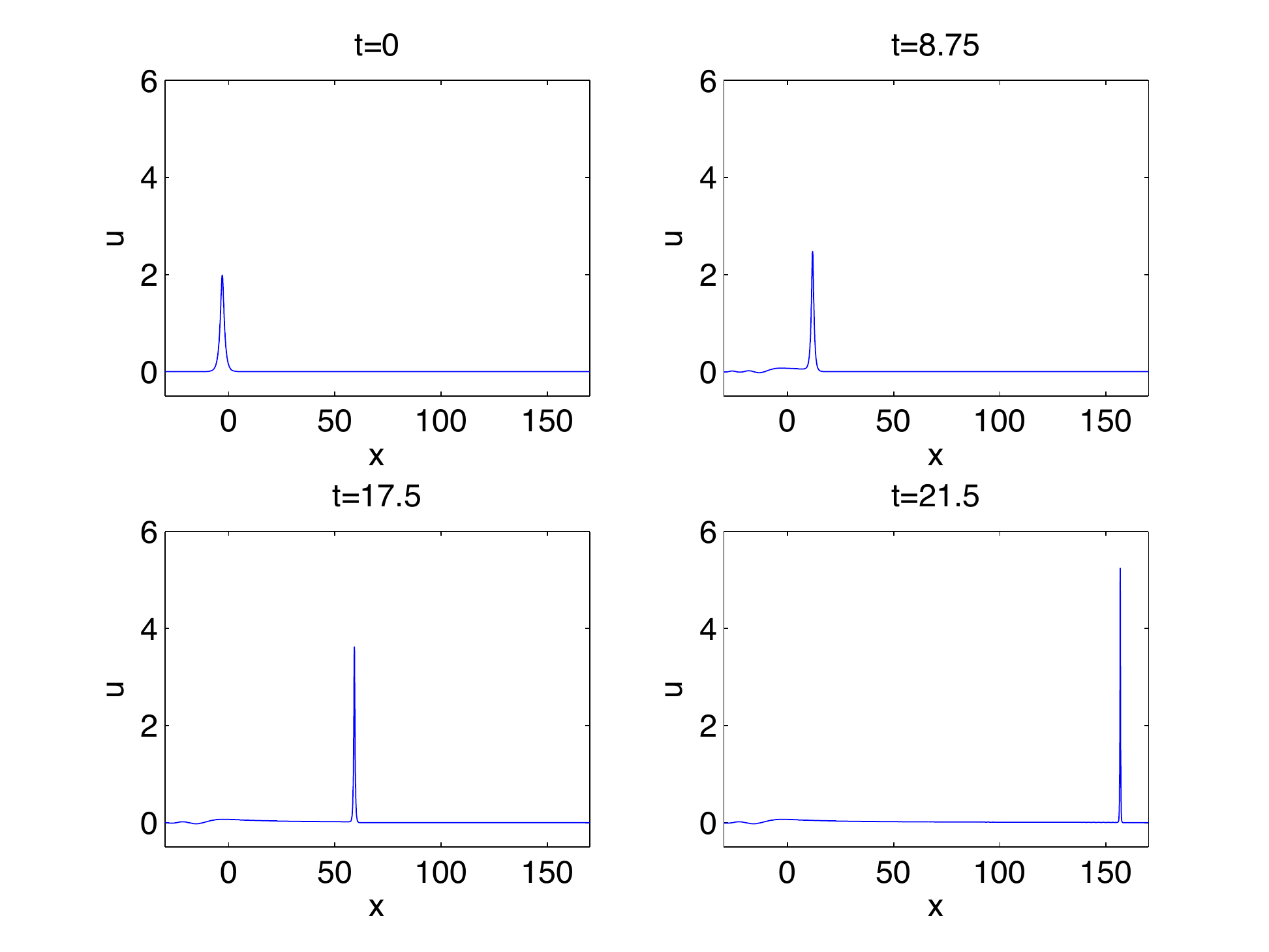}
 \caption{Solution to the gKdV equation (\ref{gKdV}) with 
 $\epsilon=1$ for $n=4$ 
 and the perturbed soliton initial data $1.01\,Q(x+3)$ (\ref{soliton}) 
 for several values of $t$.}
 \label{gKdVn4sol1014t}
\end{figure}

The situation is completely different for initial data of the form 
$\sigma\, Q(x+3)$ with $\sigma$ slightly larger than 1. For 
$\sigma=1.01$ we use $N_{t}=2*10^{5}$ time steps for $t<25$ whereas 
all other parameters are as above. Now the mass is larger than the soliton mass and the energy is negative. The code breaks at $t\approx 22.1814$ 
since the iteration no longer converges. But already at $t=22.15$ the 
energy conservation used to check numerical accuracy 
drops below $10^{-3}$. Thus we suppress the solution for larger times 
due to a lack of reliability. Note that this is not our definition of the 
blow-up time, which will be determined below, it just delimits the 
set of 
reliable data generated with this set of parameters. The solution 
is shown for several times in 
Fig.~\ref{gKdVn4sol1014t}. It can be seen that there will be again 
some dispersive oscillations propagating to the left. But more 
importantly the soliton becomes more and more peaked and thus travels at a 
higher and higher speed. At the same time it gets laterally 
compressed. Thus the solution behaves as predicted in \cite{MMR2012_I} as 
a dynamically rescaled soliton.

\begin{figure}[htb!]
  \includegraphics[width=0.49\textwidth]{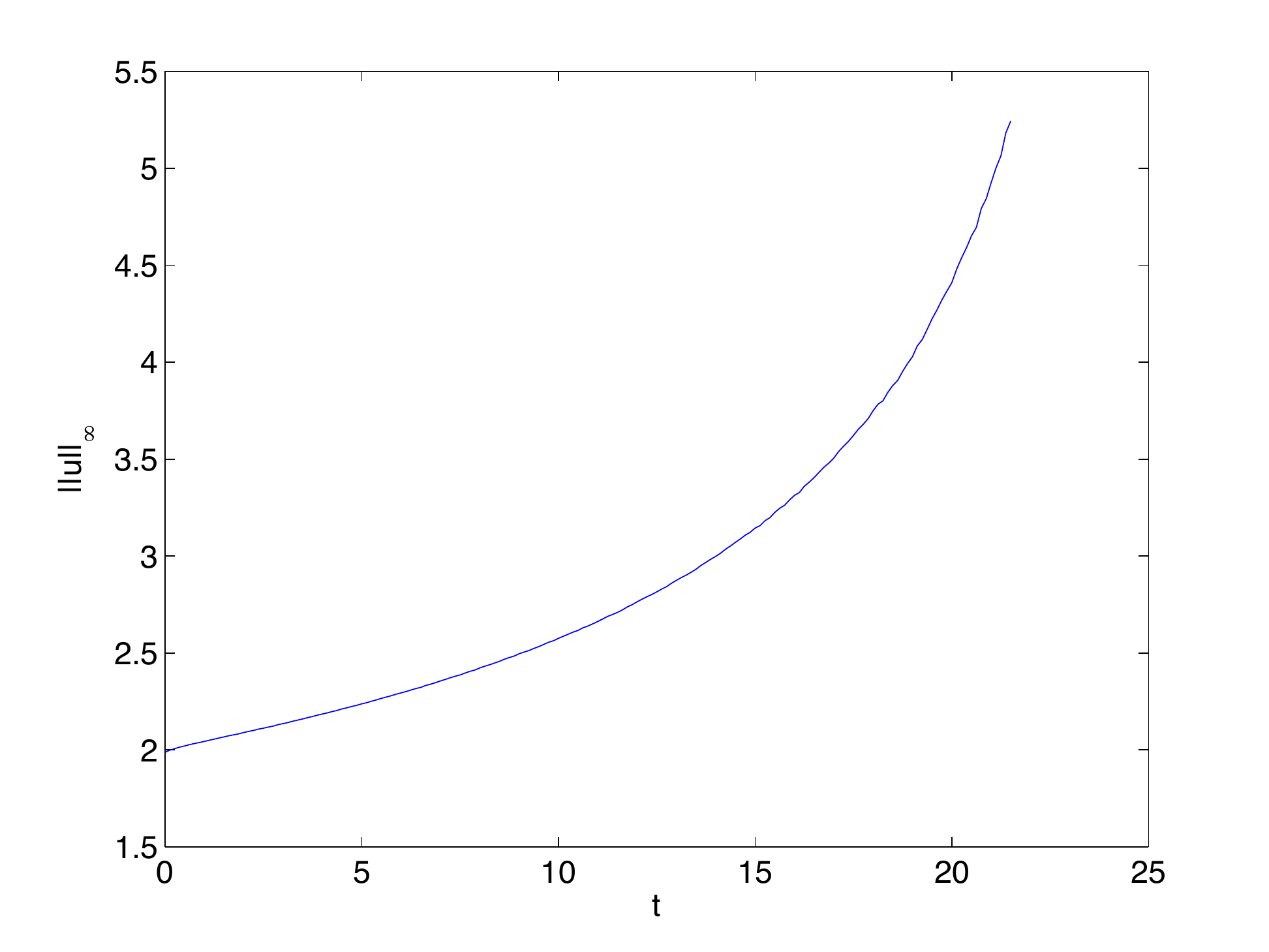}
  \includegraphics[width=0.49\textwidth]{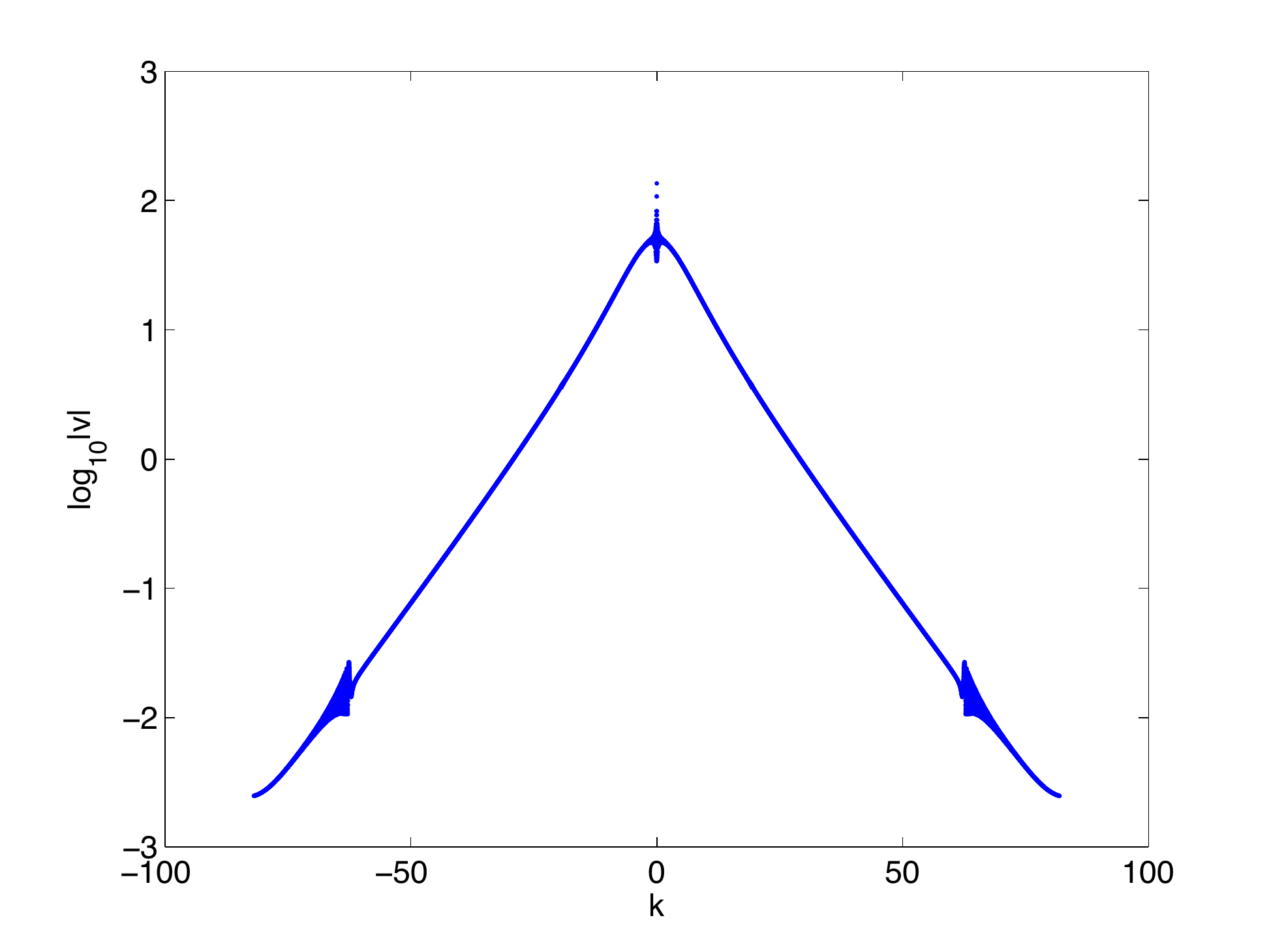}
 \caption{$L_{\infty}$ norm of the solution to the gKdV equation (\ref{gKdV}) 
 with 
 $\epsilon=1$ for $n=4$ 
 and the perturbed soliton initial data $1.01\,Q(x+3)$ (\ref{soliton}) 
 in dependence of time on the left, and the modulus of the Fourier 
 coefficients of the solution for $t=21.5$ on the right.}
 \label{gKdVn4sol101}
\end{figure}

The increase of the $L_{\infty}$ norm in Fig.~\ref{gKdVn4sol1014t} 
is even more obvious  
in Fig.~\ref{gKdVn4sol101}. However the code with the used resolution 
is not able to get arbitrarily close to the blow-up time. As the Fourier 
coefficients at the final time in Fig.~\ref{gKdVn4sol101} show, 
this is not due to a lack of 
resolution in time but in space. 

The slow increase of the $L_{\infty}$ norm of the solution implies 
that it will not be possible to get much closer to the blow-up even with 
considerably higher resolutions in both time and space. But we can 
numerically check whether the analytical prediction that the 
$L_{\infty}$ norm of the solution behaves close to blow-up as 
$||u||_{\infty}\sim (t^{*}-t)^{\alpha}$ with negative constant $\alpha$. By 
fitting $\ln ||u||_{\infty}$ to $\alpha\ln (t^{*}-t)+\kappa$, we find 
$\alpha=-0.4923$,    $\kappa=2.2803$ and   $t^{*}=25.0302$. The value 
of $\alpha$ is compatible with the theoretically predicted $-1/2$ 
(\ref{alg}). The 
difference between $\ln ||u||_{\infty}$ and $\alpha\ln 
(t^{*}-t)+\kappa$ is below $1\%$ and 
largest at the early times, but the values of the fitting 
parameters do not change much if the fitting is done only for larger 
times. This shows that the 
algebraic increase in time of the $L_{\infty}$ norm of the solution 
is followed already for $t\ll t^{*}$ in good approximation. The 
fitting indicates a blow-up roughly for $t=25$, whereas the code was 
stopped at $t=22.15$.

\begin{figure}[ht]
   \centering
   \includegraphics[scale=0.5]{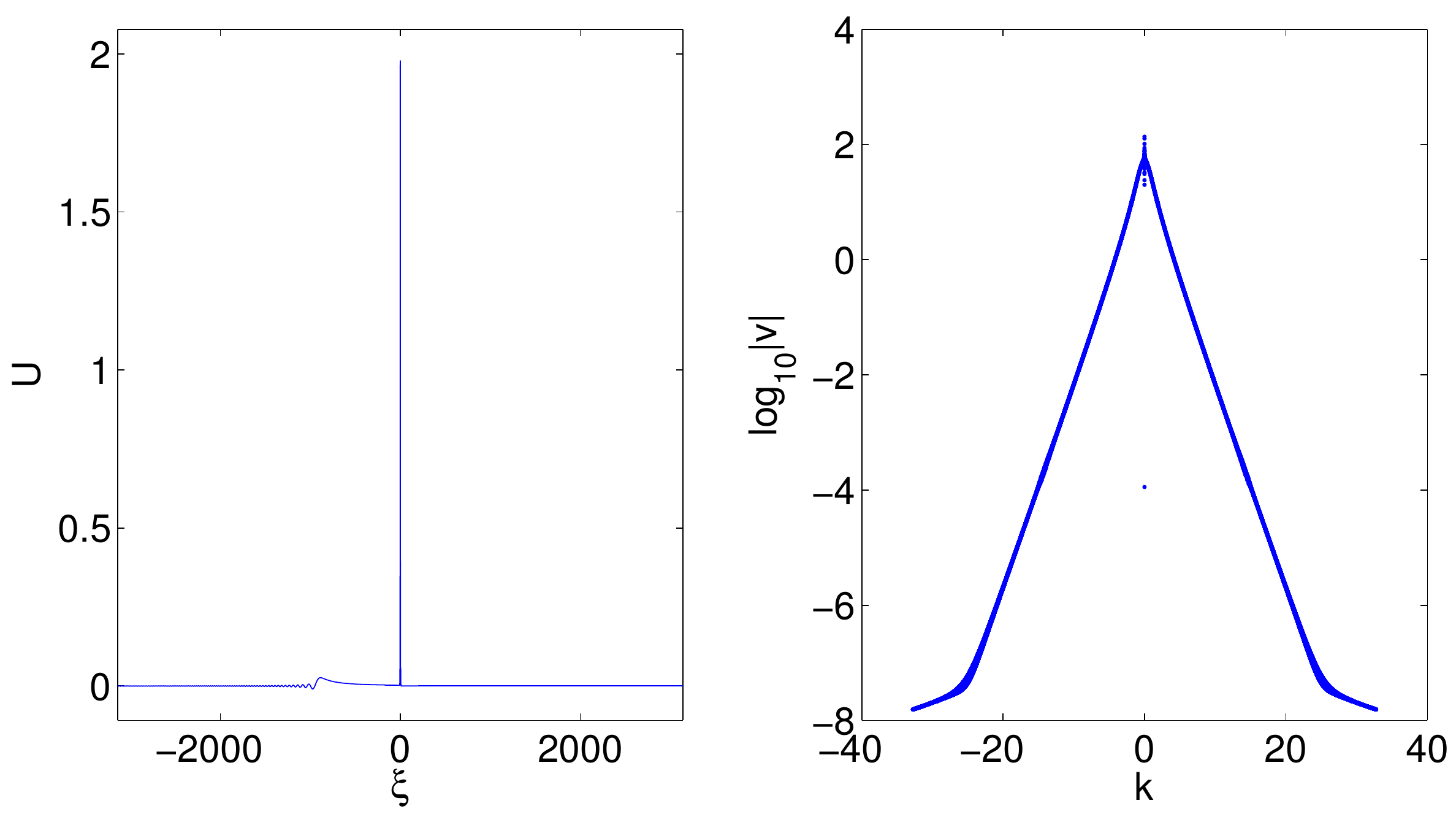}
   \caption{The solution $U(\xi,\tau)$ of the equation (\ref{gKP5}) 
   for the initial data $U(\xi,0) = 1.01\,Q(\xi)$ (\ref{soliton}) for 
    $\tau=500$ (physical time: $t=21.0787$) on the left and the corresponding Fourier 
    coefficients on the right.}
    \label{fig:SolitonPerturb_n4_U_and_FC}
\end{figure}

The above results indicate as expected that the solution near blow-up 
is close to a rescaled soliton. Thus we solve the dynamically 
rescaled equation (\ref{gKP5}) for the same $\sigma=1.01$ for 
$\tau\leq 500$ with $N_{t}=2*10^{6}$ time steps and $N=2^{16}$ 
Fourier modes for $\xi\in1000[-\pi,\pi]$. The solution can be seen in 
Fig.~\ref{fig:SolitonPerturb_n4_U_and_FC}. Since the maximum is fixed at 
$\xi=0$, the dispersive oscillations propagate more rapidly to the 
left than is the case in Fig.~\ref{gKdVn4sol1014t}. The plot suggests 
that the solution will be the rescaled soliton for $\tau\to\infty$, 
and that the remainder of the initial data will be radiated to 
infinity. 
The Fourier coefficients of the solution at the final time are also shown 
in Fig.~\ref{fig:SolitonPerturb_n4_U_and_FC} indicating that there is 
sufficient resolution in Fourier space. The relative computed mass is 
conserved to better than $10^{-4}$. However, the 
energy conservation is only of the order of $10^{-2}$. The comparatively low value of 
the latter is due to the dispersive 
oscillations reentering the computational domain and leading to a 
slight Gibbs phenomenon. Since the mass in contrast to the energy 
does not contain  a derivative, it is less sensitive to this effect. 
Note that a 
slight increase of the computational domain does not change this 
decisively since the factor $a$ tends only slowly to zero, and thus a 
reentering of the dispersive radiation cannot be suppressed in 
practice. In addition a larger domain increases the mentioned 
problems due to the term $\xi U_{\xi}$ in (\ref{gKP5}) which 
eventually lead to numerical instabilities even for prohibitively 
small time steps if the domain is too large.  

\begin{figure}[ht]
   \centering
      \includegraphics[scale=0.5]{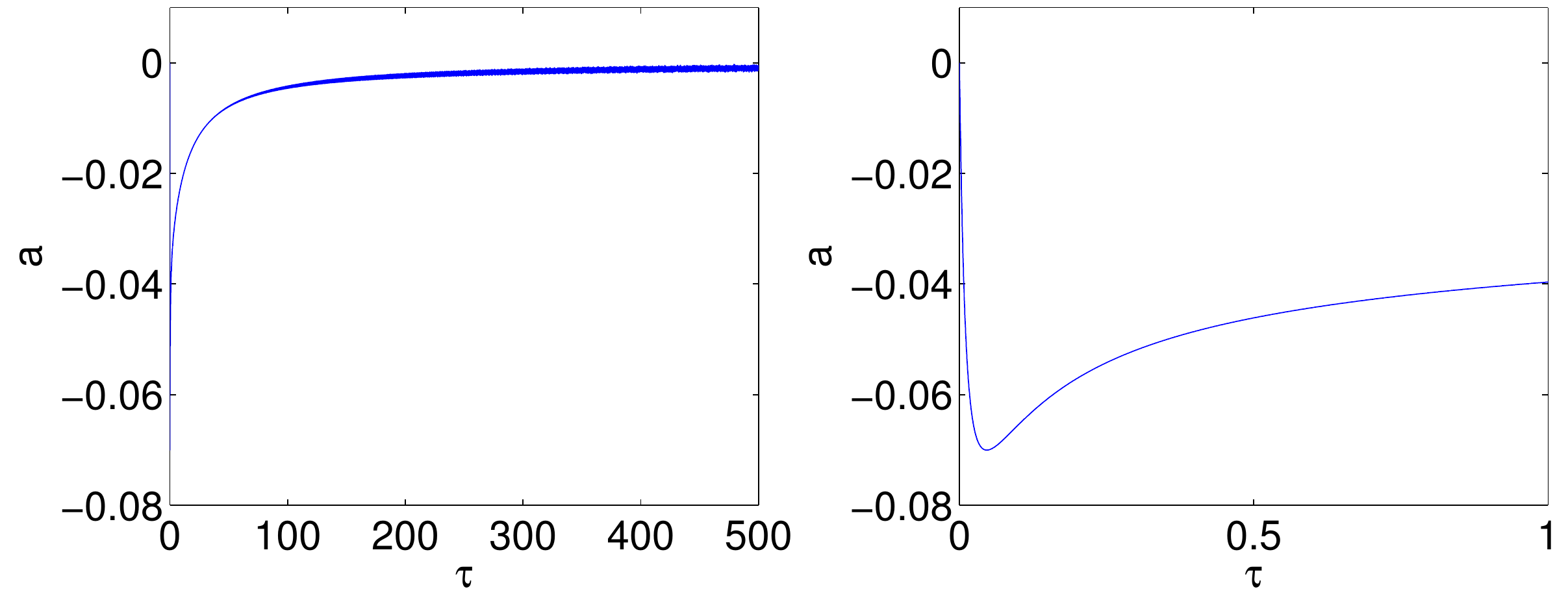}
   \caption{The function $a(\tau)$ on the whole computed $\tau$-range 
   (left) and a close-up of $a(\tau)$ in the interval $\tau\in[0,1]$ (right) for the solution of Fig.~\ref{fig:SolitonPerturb_n4_U_and_FC}.}
   \label{fig:SolitonPerturb_n4_atau}
\end{figure}

\begin{figure}[ht]
   \centering
   \includegraphics[scale=0.5]{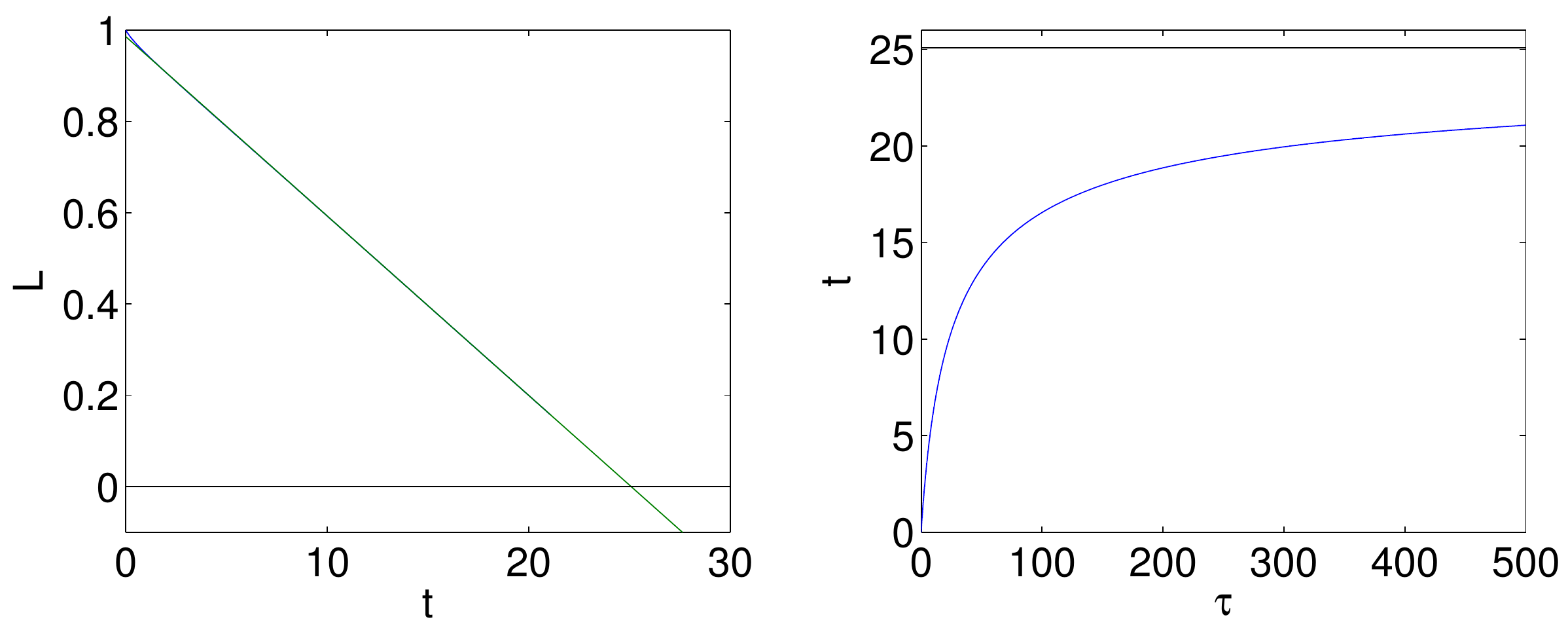}
   \caption{The scaling factor $L$ as a function of the physical time $t$ and its fit for $t>2$ to a straight line (left). The right figure shows the physical time as a function of the rescaled time $\tau$. The black horizontal line in the right figure is 
    the blow up time $t^*$ determined from the fitting of $L$. Both figures correspond to the situation shown in 
   Fig.~\ref{fig:SolitonPerturb_n4_U_and_FC}.}\label{fig:SolitonPerturb_n4_t_L_and_tau_t}
\end{figure}

The reentering of the computational domain of the dispersive 
oscillations due to the imposed periodicity has also an effect for the 
quantity $a(\tau)$ in Fig.~\ref{fig:SolitonPerturb_n4_atau}. The 
slow increase of $a(\tau)$ to zero is superposed by oscillations due 
to the radiation emanating from the perturbed soliton. We only 
observe such spurious oscillations for perturbed soliton initial data 
since the latter and the generated dispersive oscillations decrease 
much slower than the $\mbox{sech}^{2}x$ initial data. This is also 
the reason why we use a large domain and higher resolution in Fourier 
space to treat this case. 
Integrating $a(\tau)$ we obtain $L(\tau)$ and $t(\tau)$. Note that the 
spurious oscillations in $a(\tau)$ appear in both $L(\tau)$ and 
$t(\tau)$, but not if $L$ is shown as a function of $t$. For larger values of 
$t$, the scaling factor $L$ is as expected linear in $t$. Fitting  as 
shown in  Fig.~\ref{fig:SolitonPerturb_n4_t_L_and_tau_t}
$L(t) =  mt + L_0$ led to the values $m = -0.03931$ and $L_0= 0.9861$ 
and thus to the blow-up time $t^* = 25.0836$, in 
accordance with the result from the fitting of the $L_{\infty}$ norm 
of $u$ from the direct solution of gKdV above. The $L_{\infty}$ norm of the difference 
between $L$ and its fit is of the order $10^{-4}$. 


In Fig.~\ref{fig:SolitonPerturb_n4_t_L_and_tau_t} we also present the physical 
time $t$ in dependence of the rescaled time $\tau$. The plot shows 
that we do not get arbitrarily close to the blow-up time in this example due to the 
fact that $a(\tau)$ tends slowly to 0, but close enough to 
extrapolate to the blow-up. This phenomenon was also 
observed in the case of the critical NLS equation \cite{McLPSS1986,PSSW1993,SulemSulem1999}.

\begin{figure}[ht]
   \centering
   \includegraphics[scale=0.5]{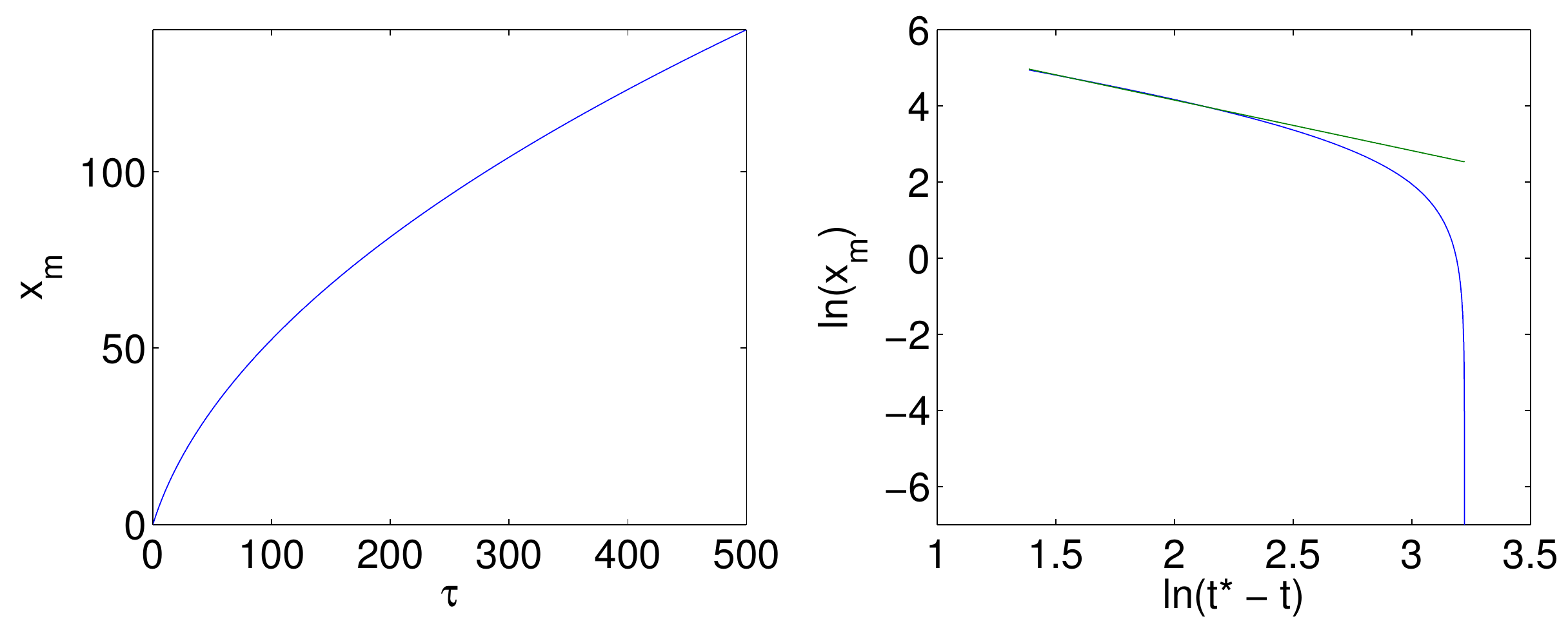}
   \caption{The position $x_m$ of the maximum of the solution of Fig.~\ref{fig:SolitonPerturb_n4_U_and_FC} 
   as a function of the rescaled time $\tau$ on the left and  as a 
   function of $t$ and its fit for small values of $\ln(t^* - t)$ on the right.}\label{fig:SolitonPerturb_n4_xm}
\end{figure}

The position of the maximum $x_{m}(\tau)$ and thus ultimately the 
location of the blow-up is shown in 
Fig.~\ref{fig:SolitonPerturb_n4_xm}. As theoretically predicted, it 
tends to infinity. In  the 
same figure we also present $x_{m}$ in dependence of the physical time. 
A fit  for 
small $t^{*}-t$ ($\ln(t^* - t) < 2.4$) for $\ln x_{m}\sim 
\gamma_{2}\ln(t^{*}-t)+\ln(C_{2})$ gives 
$\gamma_{2}=-1.3282$ and $C_{2}=905.8$. It can be seen 
that the fitting for $x_{m}$ is not as good as for the scaling factor 
$L$ which is not surprising since the distance between the location 
of the blow-up and the dispersive tail is 
infinite. But it is compatible with the theoretical prediction 
$\gamma_{2}=-1$ and the $L_{\infty}$ norm of the difference between 
$x_m$ and its fit is still of order $10^{-2}$. A similar procedure 
 for the $L_2$ norm of $u_x$  leads to $\gamma_3 = -1.0004$ and $C_3 = 13.884$ and 
also match very well with the theoretical prediction. The two control quantities, the $L_2$ norm of $U_{\xi}$ and the value of $U_{\xi}$ at $\xi=0$ are preserved to the order $10^{-6}$ and $10^{-4}$ respectively.

\subsection{Small dispersion limit}
\begin{figure}[htb!]
  \includegraphics[width=\textwidth]{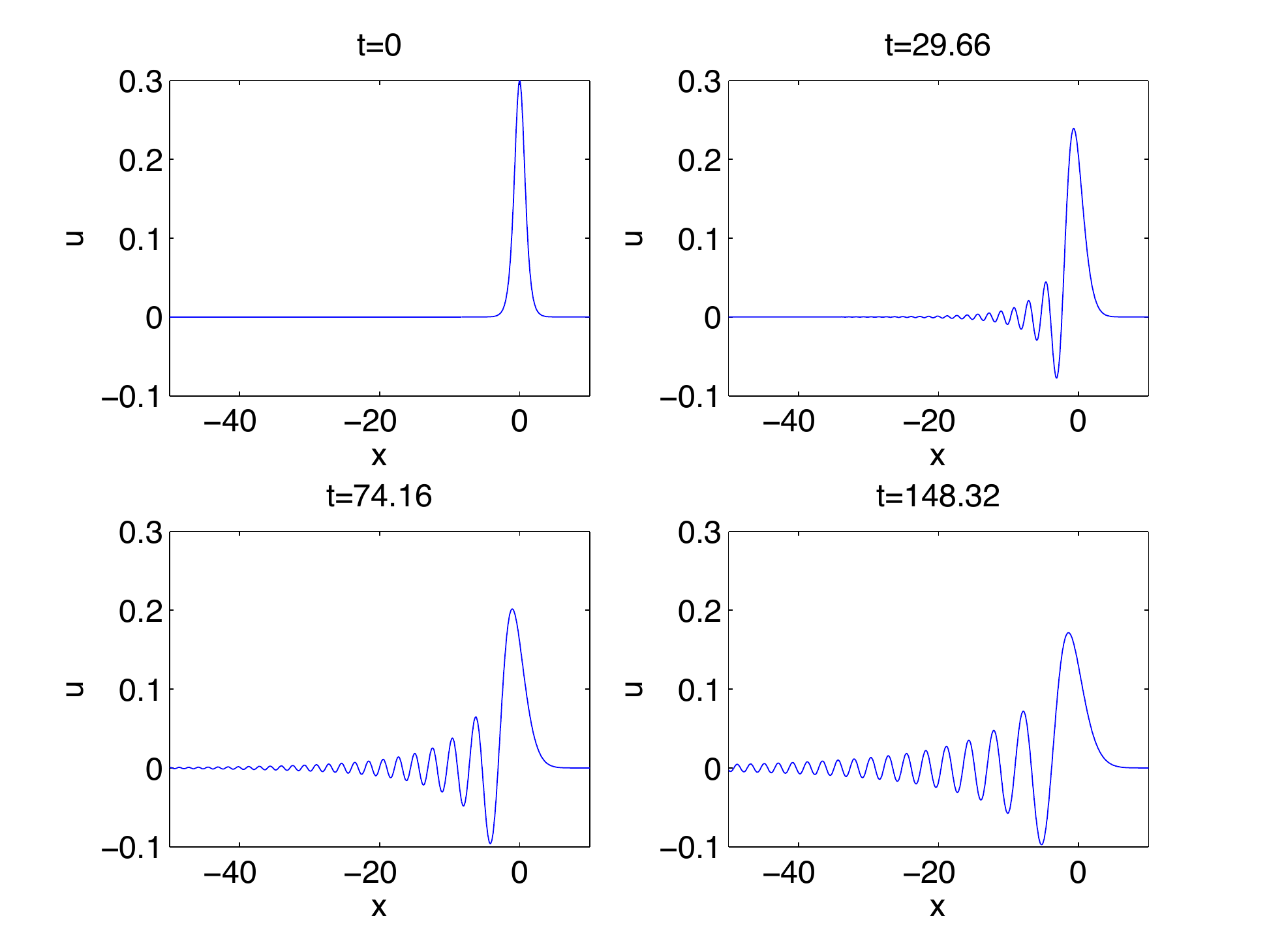}
 \caption{Solution to the gKdV equation (\ref{gKdV}) with 
 $\epsilon=0.1$ for $n=4$ 
 and the initial data $u_{0}=0.3\,\mbox{sech}^{2}x$  
 for several values of $t$.}
 \label{gKdVn4e01b034t}
\end{figure}

In this subsection we study initial data of the form 
$u_{0}=\beta\mbox{ sech}^{2}x$ for $\epsilon=0.1$.
For $u_{0}=0.3\mbox{ sech}^{2}x$ the energy  is positive,  the mass is again smaller 
than the soliton mass, and the critical time (\ref{tch}) is  
$t_{c}\approx74.1577$. No blow-up is expected in this case. 
To directly integrate gKdV numerically for these initial data we use $N_{t}=10^{3}$ 
time steps for $t\leq 2t_{c}$ and $N=2^{12}$ Fourier modes for 
$x\in40[-\pi,\pi]$. As can be 
seen in Fig.~\ref{gKdVn4e01b034t}, the solution develops a tail of 
dispersive oscillations towards $-\infty$, the initial data appear to 
be simply radiated away. The $L_{\infty}$ norm of the solution decreases 
monotonically. Note that the solution  is resolved up to machine 
precision in 
Fourier space, and that the numerically computed energy is conserved 
to the order of $10^{-12}$. 

\begin{figure}[htb!]
  \includegraphics[width=\textwidth]{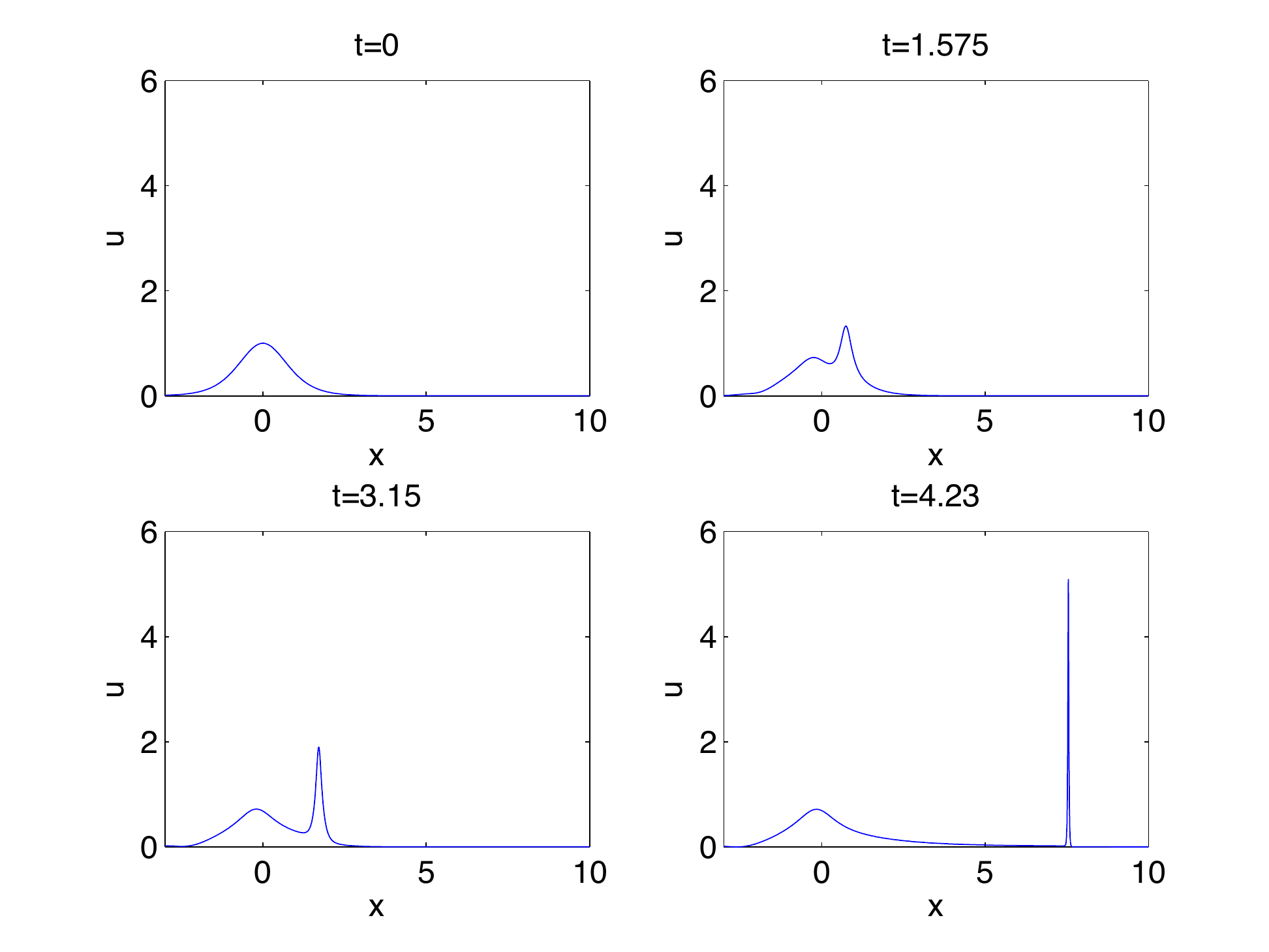}
 \caption{Solution to the gKdV equation (\ref{gKdV}) with 
 $\epsilon=0.1$ for $n=4$ 
 and the initial data $u_{0}=\mbox{sech}^{2}x$  
 for several values of $t$. The time of gradient catastrophe for the 
 solution to the generalized Hopf equation is $t_{c}\approx 0.6007$.}
 \label{gKdVn4e014t}
\end{figure}

The situation is completely different for  $u_{0}=\mbox{sech}^{2}x$ for which 
the energy is negative and for which the mass of the initial data is larger than the soliton mass $M[Q]$. The computation is carried out in this case
with $N_{t}=2*10^{5}$ 
time steps for $t<4.5$ and $N=2^{14}$ Fourier modes for 
$x\in10[-\pi,\pi]$. The code is stopped at $t=4.23$ when the energy conservation 
drops below $10^{-3}$ and the results are no longer reliable. 
In contrast to \cite{DGK2011}, where numerical instabilities stopped 
the code well before a clear indication of blow-up, we reach with the 
IRK4 code unambiguously the blow-up regime. The 
solution is shown for several times in 
Fig.~\ref{gKdVn4e014t}. For small $t$ it is close to
the solution of the generalized Hopf equation for the same initial 
data. The dispersive effects of the third derivative in the gKdV 
equation become important near the critical time of the generalized 
Hopf solution. A first oscillation forms at $t\sim t_{c}$ which then 
develops into a blow-up as for the perturbed soliton. This is to be 
compared to the subcritical initial data in 
Fig.~\ref{gKdVn4e01b034t}. There the dispersive oscillations which 
also appear in Fig.~\ref{gKdVn4e014t} (where they are hardly visible) 
are dominant from the 
beginning. Note that they are numerically fully resolved in both 
cases in contrast to \cite{DixMcKinney}. 

\begin{figure}[htb!]
  \includegraphics[width=0.49\textwidth]{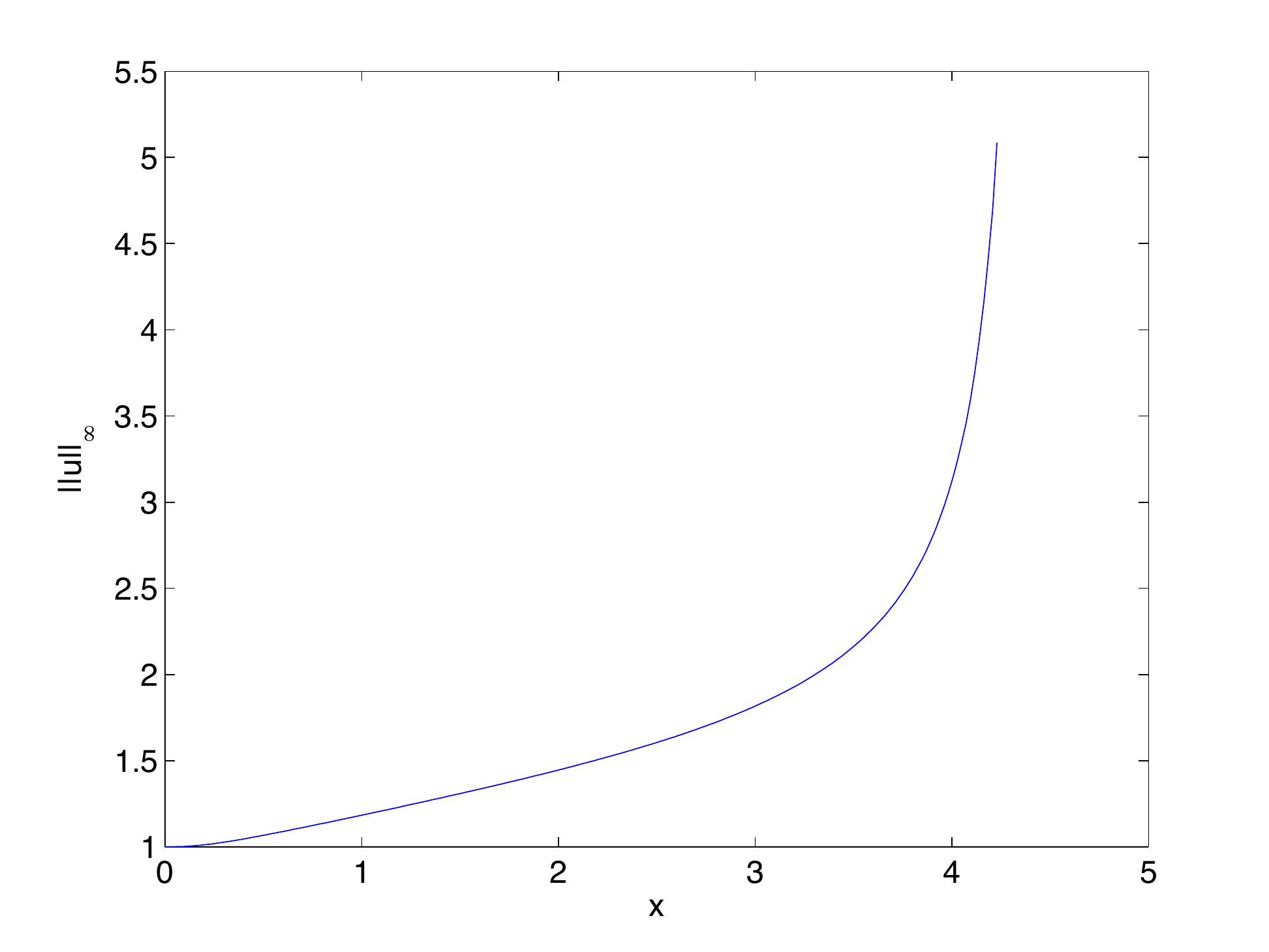}
  \includegraphics[width=0.49\textwidth]{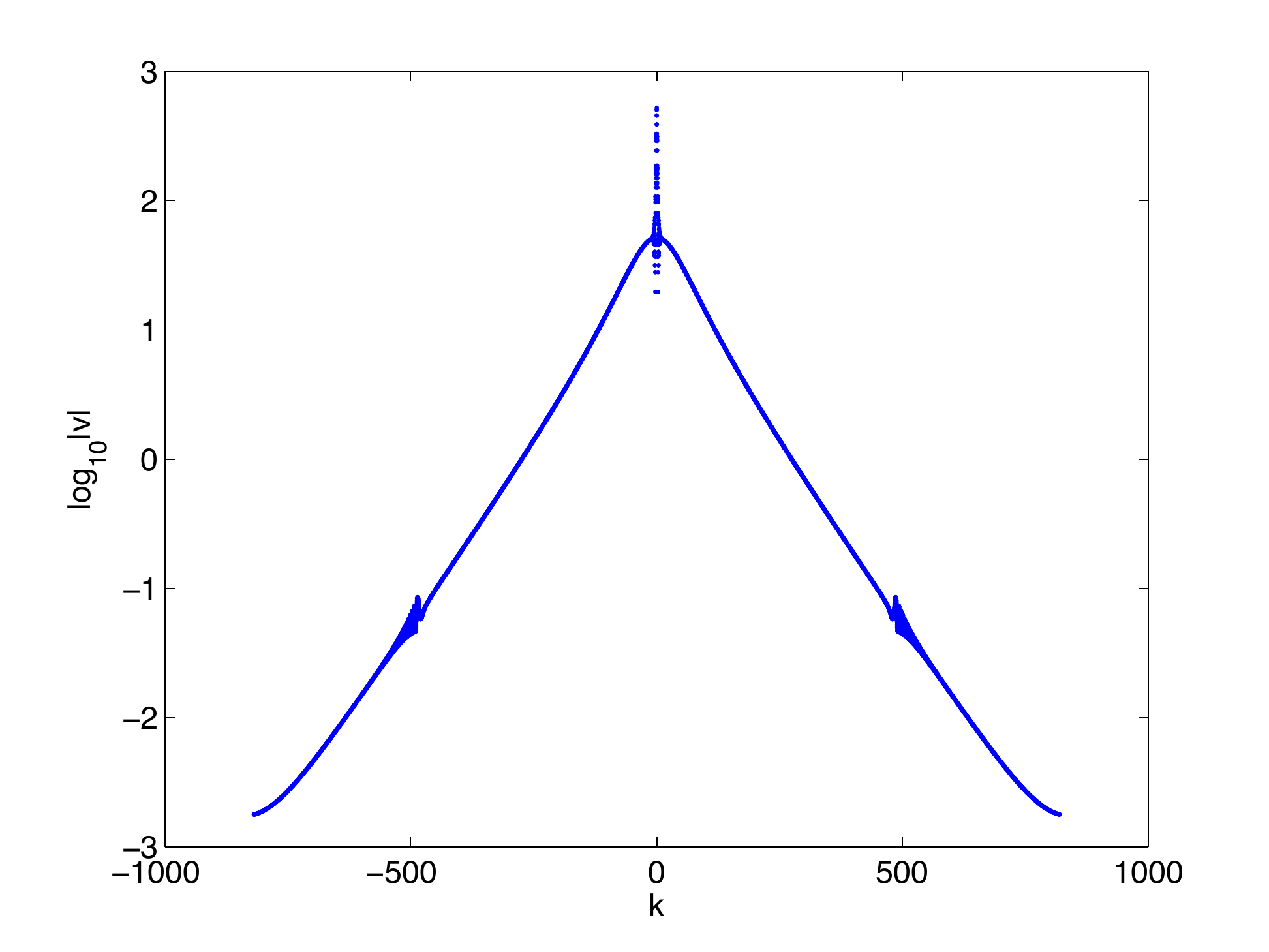}
 \caption{$L_{\infty}$ norm of the solution to the gKdV equation (\ref{gKdV}) 
 with 
 $\epsilon=0.1$ for $n=4$ 
 and the initial data $u_{0}=\mbox{sech}^{2}x$ 
 in dependence of time on the left, and the modulus of the Fourier 
 coefficients of the solution for $t=4.23$ on the right.}
 \label{gKdVn4e01}
\end{figure}

The increase of the $L_{\infty}$ norm in Fig.~\ref{gKdVn4e014t} 
can be clearly seen  
in Fig.~\ref{gKdVn4e01}. As for the perturbed soliton in the previous 
subsection, the code with the used resolution 
is not able to get close enough to the blow-up time. Again there is a 
lack of resolution in Fourier space as can be seen from the Fourier 
coefficients at the final time in Fig.~\ref{gKdVn4e01}. 

\begin{figure}[htb!]
  \includegraphics[width=0.49\textwidth]{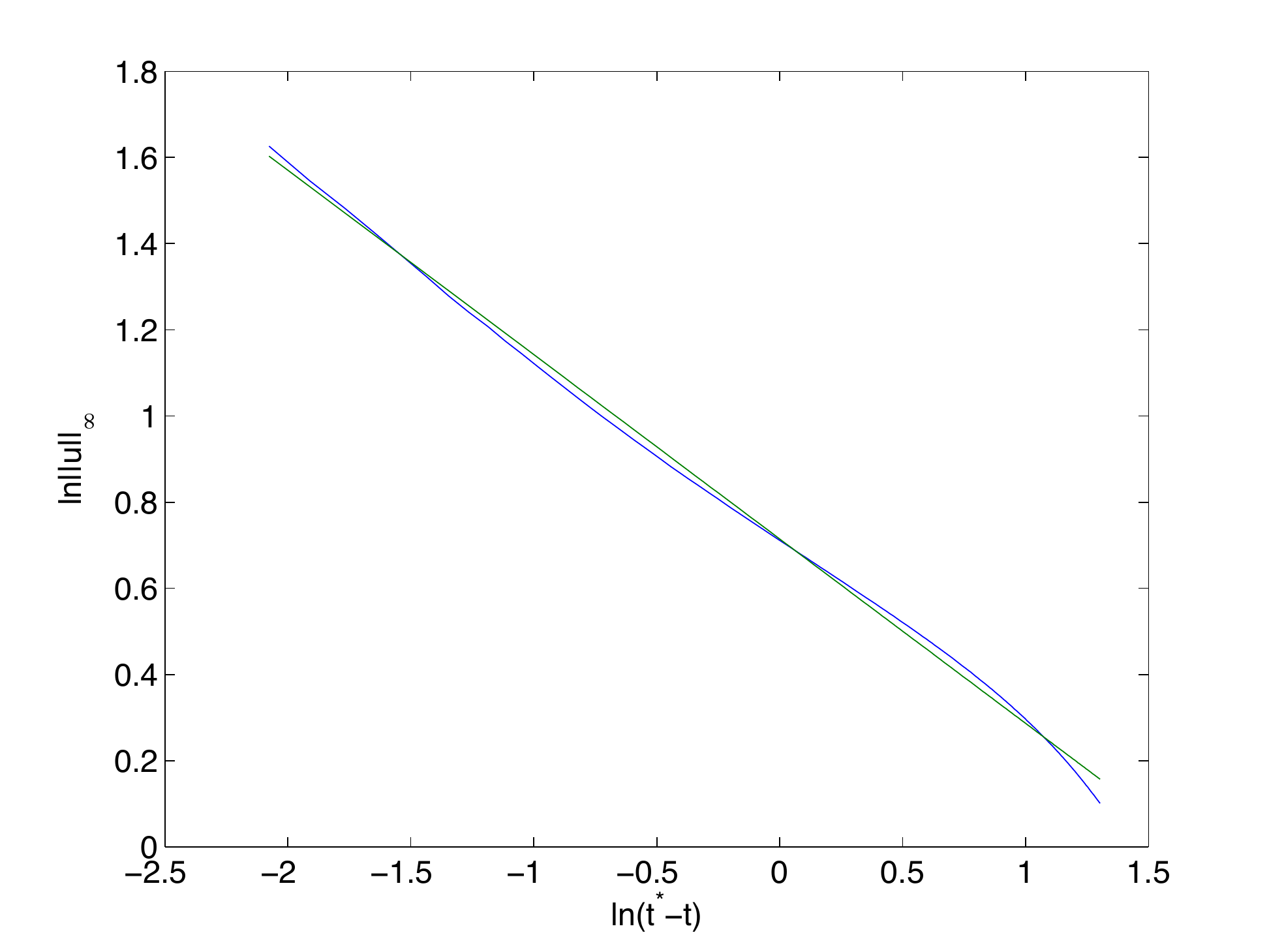}
  \includegraphics[width=0.49\textwidth]{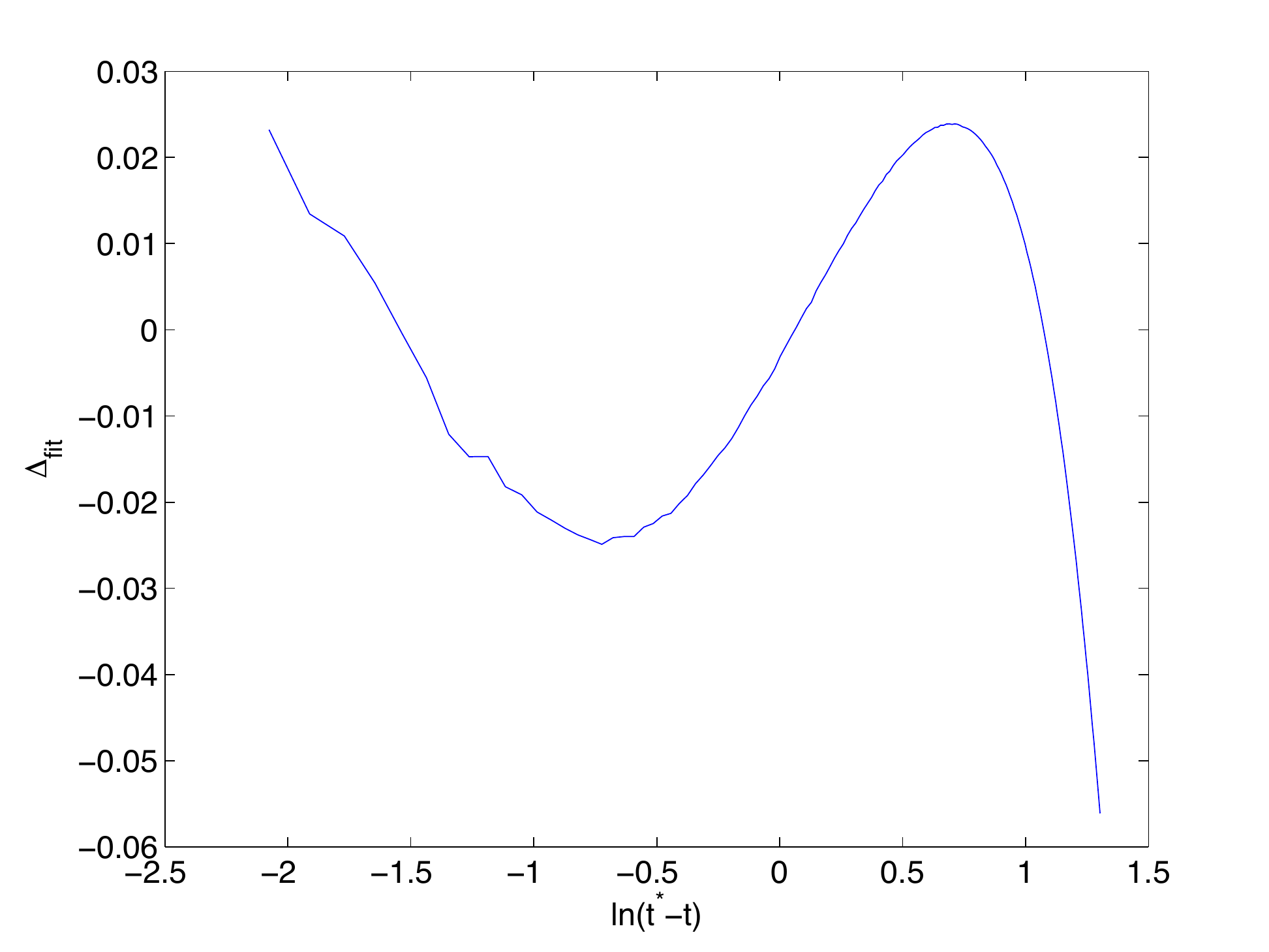}
 \caption{The logarithm of the $L_{\infty}$ norm of the solution in 
 Fig.~\ref{gKdVn4e014t} and the fittd curve $\alpha\ln 
 (t_{c}-t)+\kappa$ on 
 the left, and the difference $\Delta_{fit}$ between both curves on 
 the right.}
 \label{gKdVn4e01fit}
\end{figure}

It does not seem possible to get much closer to the blow-up even with 
considerably higher resolutions in both time and space. Thus we fit 
once more  the  
$L_{\infty}$ norm of the solution  close to blow-up to 
$\ln ||u||_{\infty}\sim \alpha\ln(t^*-t)+\kappa$. Doing this for $t>0.675$ 
(thus greater than the critical time $t_{c}\approx 0.6007$ 
(\ref{tch}) of the generalized Hopf 
solution), we find         
$\alpha=-0.4279$,    $\kappa=0.7147$ and   $t^*=4.3554$. The fitting 
is less good than in the case of the perturbed soliton in the previous 
subsection (of the order of a few percent) as can be seen in 
Fig.~\ref{gKdVn4e01fit}. It indicates a blow-up roughly for $t=4.35$. 
The blow-up rate is again compatible with $||u||_{\infty}\sim 
(t^*-t)^{-1/2}$ as in theorem \ref{MMR} for perturbations of the 
soliton, indicating that the blow-up mechanism given there is valid 
for more general localized initial data than the class $\mathcal{A}$ 
(\ref{A}).

\begin{figure}[ht]
   \centering
   \includegraphics[scale=0.5]{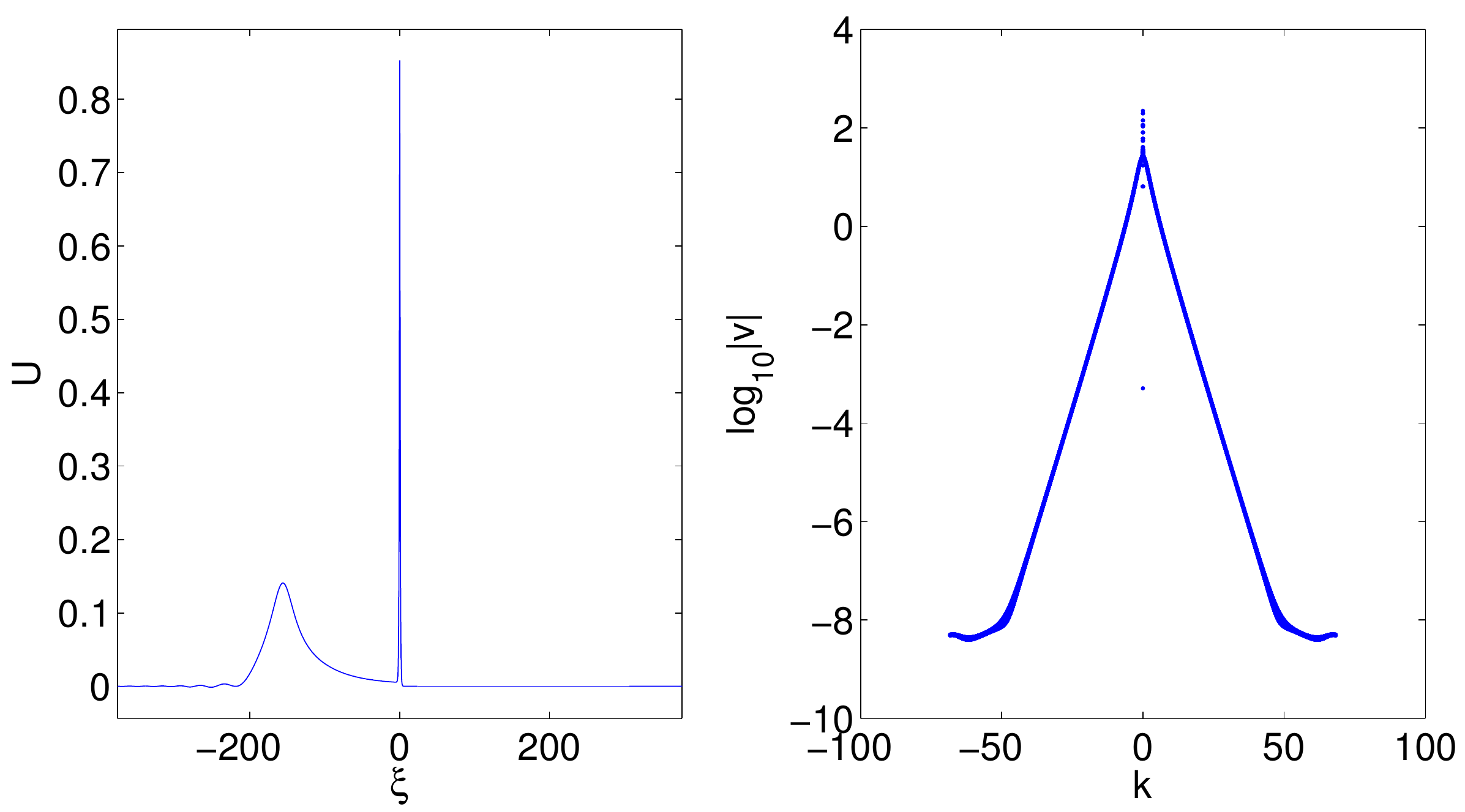}
   \caption{The solution $U(\xi,\tau)$ of the equation (\ref{gKP5}) 
   for the initial data $U(\xi,0) = \mbox{sech}^{2}\xi$  for $n=4$,
    $\tau=1700$ (physical time: $t=4.1787$) and $\epsilon=0.1$ on the left and the corresponding Fourier 
    coefficients on the right.}\label{fig:SechSquare_n4_U_new}
\end{figure}

To analyze the blow-up in more detail, we solve again the dynamically 
rescaled gKdV equation (\ref{gKP5}). With    $N = 2^{14}$ Fourier 
modes for $\xi\in120[-\pi,\pi]$ and  $N_t = 6* 10^{6}$ time steps for 
$\tau \leq 1700$, we get the 
solution shown in Fig.~\ref{fig:SechSquare_n4_U_new}. It can be seen 
that the initial data decompose into the rescaled soliton and a 
remainder which  appears to be ultimately radiated away. The 
Fourier coefficients in  Fig.~\ref{fig:SechSquare_n4_U_new} show the 
resolution up to machine precision of the solution. The computed 
relative energy is conserved to the order of $10^{-4}$. 

\begin{figure}[ht]
   \centering
   \includegraphics[scale=0.5]{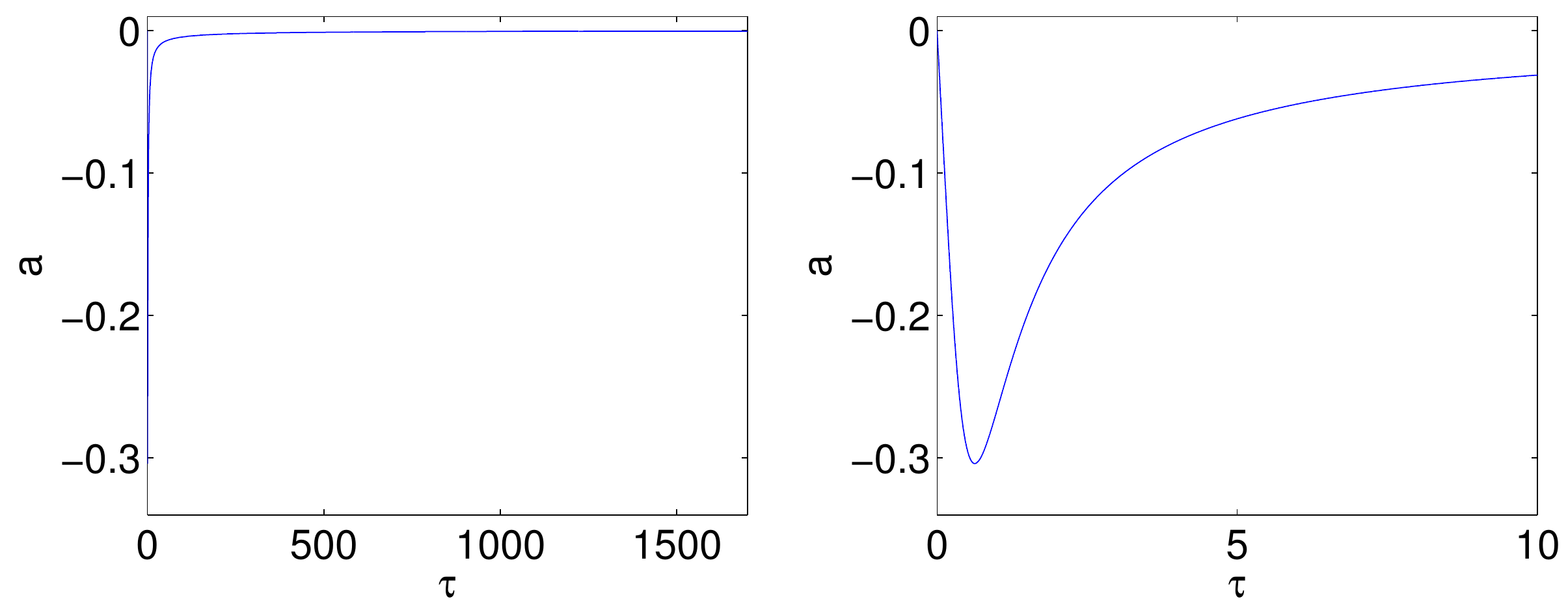}
   \caption{The function $a(\tau)$ on the whole computed $\tau$-range 
   (left) and a close-up in the interval $\tau\in[0,10]$ (right) for the solution of 
    Fig.~\ref{fig:SechSquare_n4_U_new}.}\label{fig:SechSquare_n4_atau}
\end{figure}

\begin{figure}[ht]
   \centering
   \includegraphics[scale=0.5]{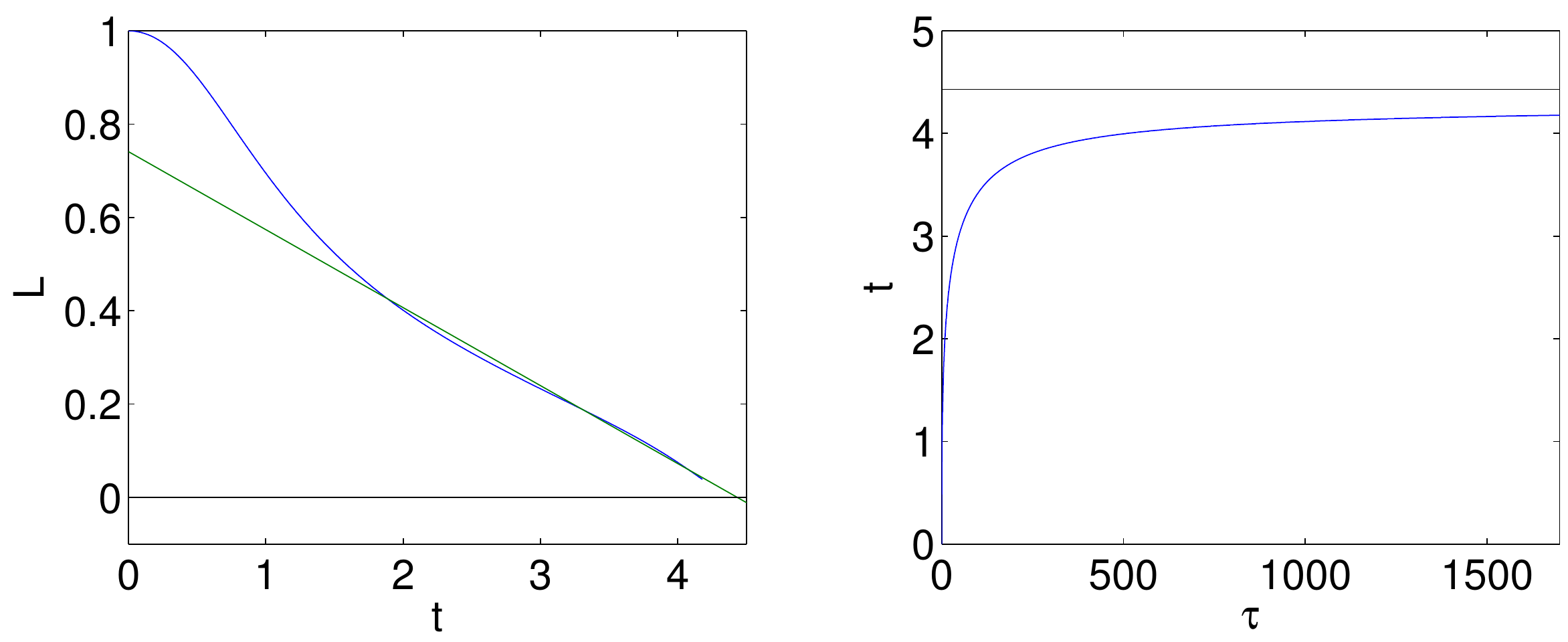}
   \caption{The scaling factor $L$ as a function of the physical time $t$ and its fit for $t>2$ (left). The right figure shows the physical time as a function of the rescaled time $\tau$. The black horizontal line in the right figure is 
    the blow up time $t^*$ determined from the fitting of $L$. Both figures correspond to the solution in 
    Fig.~\ref{fig:SechSquare_n4_U_new}.}\label{fig:SechSquare_n4_t_L_and_tau_t}     
\end{figure}

The function $a(\tau)$ for the solution in 
Fig.~\ref{fig:SechSquare_n4_U_new} is shown in 
Fig.~\ref{fig:SechSquare_n4_atau}. Note that there are no spurious 
oscillations here in contrast to the perturbed soliton in 
Fig.~\ref{fig:SolitonPerturb_n4_atau}. This  is due to the more rapid 
decrease of the initial data with $x$ which implies the same behavior 
for the radiation which caused the oscillations in 
Fig.~\ref{fig:SolitonPerturb_n4_atau} due to the imposed 
periodicity in the computation. We remark that the difference between 
$a(1700)$ and $a(1530)$ is of the order of $10^{-5}$. 
From $a(\tau)$ we get $L(\tau)$ as shown in 
Fig.~\ref{fig:SechSquare_n4_t_L_and_tau_t} where also a least square fit is 
given for $L = mt +L_{0}$. We find  $m = -0.1673$ and $L_0 
= 0.7412$ and thus the blow-up time $t^* = 4.4316$. The 
difference of the fitted line and $L$ is of the order of $10^{-2}$ for $t>2$.


The physical time $t$ in dependence of $\tau$ can be seen in 
Fig.~\ref{fig:SechSquare_n4_t_L_and_tau_t} on the right.

\begin{figure}[ht]
   \centering
   \includegraphics[scale=0.5]{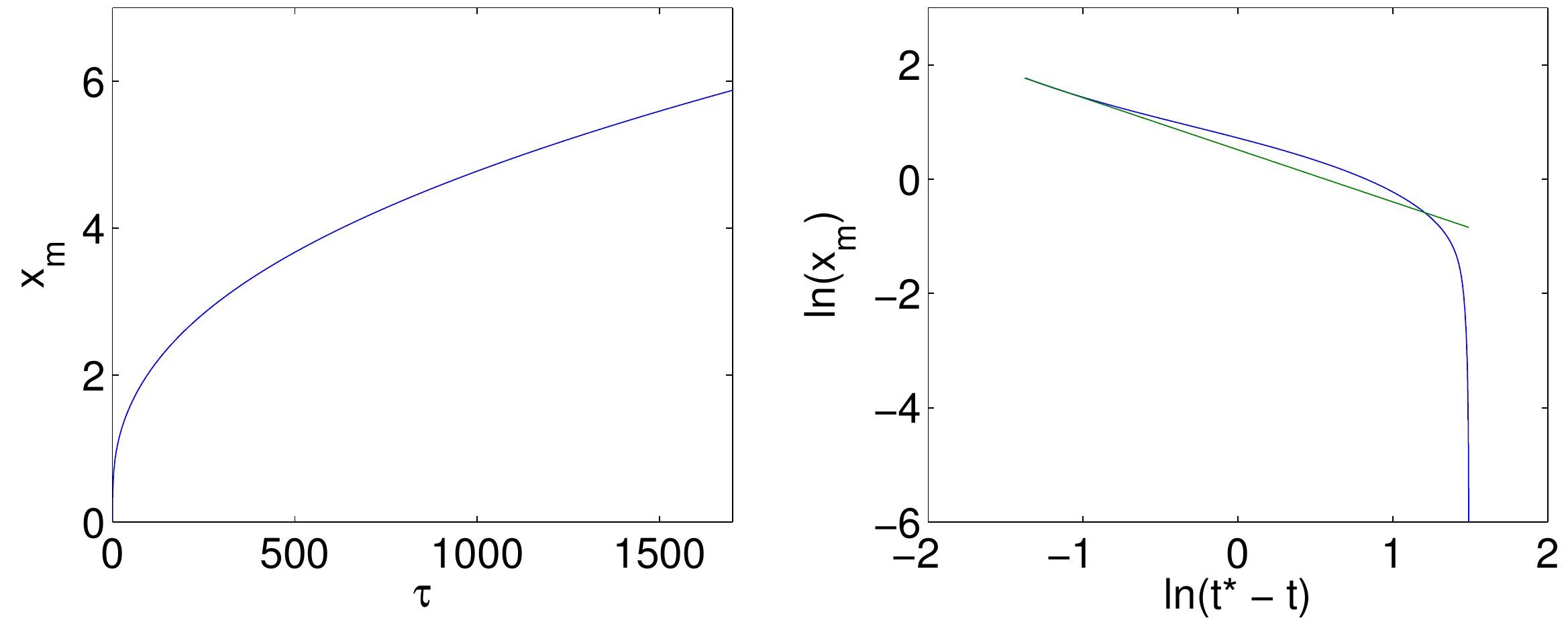}
   \caption{The position $x_m$ of the maximum  for the solution of 
   Fig.~\ref{fig:SechSquare_n4_U_new} as a function of $\tau$ on the 
   left and in dependence of $t$
   and its fit for small $t^* - t$ to a straight 
   line on the right.}\label{fig:SechSquare_n4_xm}
\end{figure}

The position $x_{m}$ of the maximum of the solution in dependence of 
$t$ is shown in Fig.~\ref{fig:SechSquare_n4_xm}. For small $t^{*}-t$ 
it can be fitted to a straight line 
$\ln x_{m}=\gamma_{2}\ln(t^{*}-t)+\ln(C_2)$. We find  $\gamma_2 = 
-0.9117$ and $C_{2} = 
   1.6683$. Thus  
   the maximum shows the same scaling as the $L_{2}$ norm of $u_{x}$, 
   i.e., the same behavior as for the soliton perturbations in 
   theorem \ref{MMR}. This gives a strong indication that the type of 
   blow-up of this theorem might be generic for much larger classes 
   of localized initial data than treated in the theorem.  
   The results of the fitting for $||u_x||_2$ are $\gamma_3 = 
   -1.1921$ and $C_3 = 1.0921$ and therefore also match with the 
   blow-up mechanism of theorem \ref{MMR} which would imply 
   $\gamma_3 = -1$. The conservation of the 
   numerically computed energy and the $L_{2}$ norm of $U_{\xi}$ 
   is of the same order as for the soliton perturbation.

\section{The supercritical case $n=5$}
In this section we study as in the previous section perturbations of 
the soliton and the semiclassical limit for masses in the vicinity of 
the soliton mass (up to three times
this mass). The results  can be  summarized as follows  (see conjecture \ref{conj1}):
\begin{itemize}
    \item  Localized positive smooth initial data with a single 
    maximum and small mass are radiated away.

    \item  Initial data of the same class with sufficiently large 
    mass show self similar blow-up according to (\ref{selfssc}).
\end{itemize}

\subsection{Perturbations of the soliton}
\begin{figure}[htb!]
  \includegraphics[width=\textwidth]{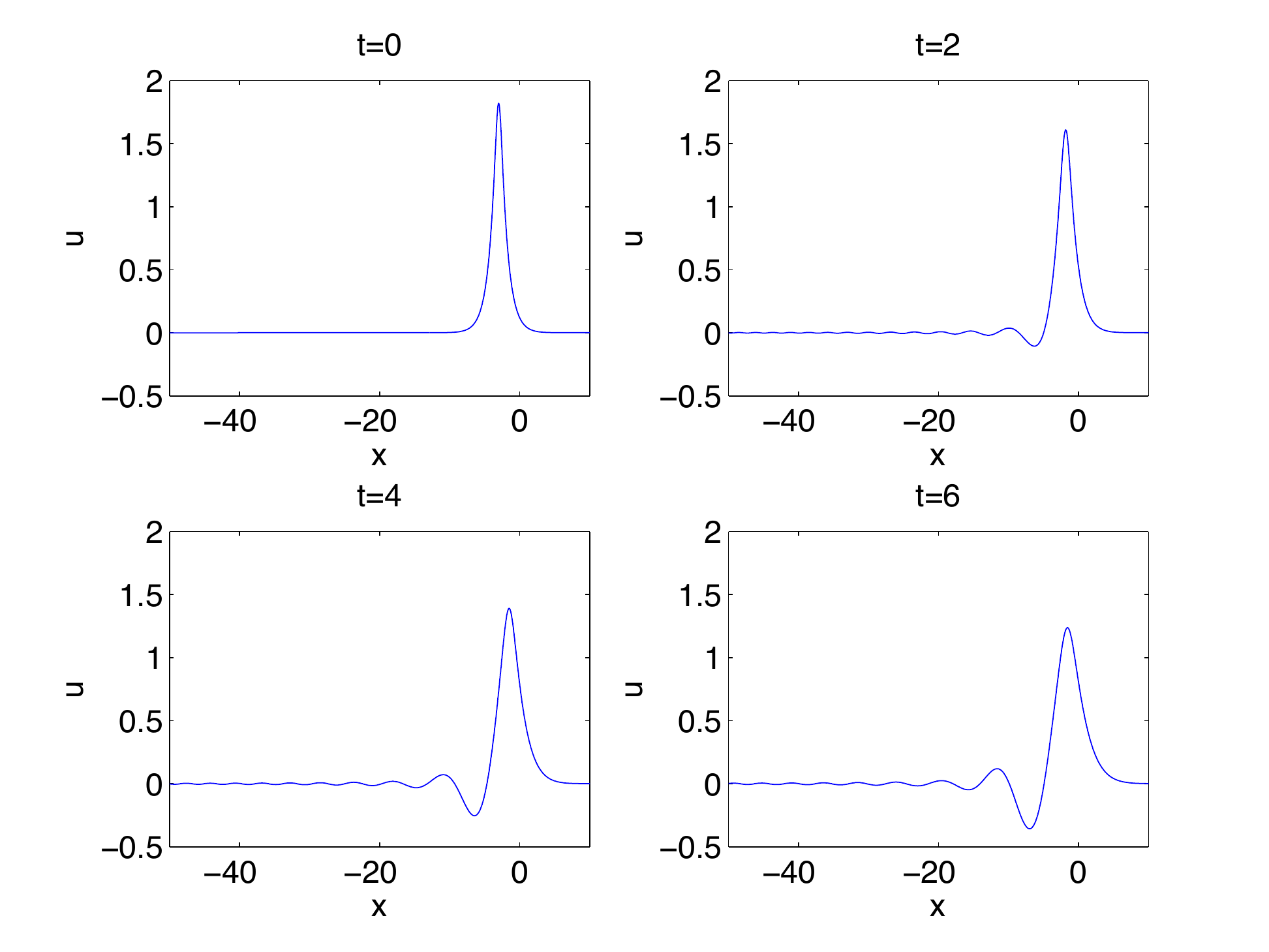}
 \caption{Solution to the gKdV equation (\ref{gKdV}) with 
 $\epsilon=1$ for $n=5$ 
 and the perturbed soliton initial data $0.99\,Q(x+3)$ (\ref{soliton}) 
 for several values of $t$.}
 \label{gKdVn5sol0994t}
\end{figure}

For $n=5$ and $\epsilon=1$, the energy of the 
soliton is $E[Q]=0.5526\ldots$ and the mass is $M[Q] = 4.9738\ldots$. Perturbing the soliton as for $n=4$, i.e., 
considering initial data of the form $u(x,0)=\sigma\, Q(x+3)$ on a large domain, 
$x\in 100[-\pi,\pi]$ with $N=2^{14}$ Fourier modes and $N_{t}=10^{4}$ 
time steps, we find for $\sigma=0.99$ again that the soliton is 
radiated away. The energy of the perturbed initial data is larger 
than the one of the soliton, the mass smaller. 
The numerically computed energy is conserved to better 
than $10^{-12}$.  It can be seen in 
Fig.~\ref{gKdVn5sol0994t} that dispersive oscillations propagating to 
the left form immediately and eventually appear to radiate the 
initial data away. The amplitude of these oscillations 
decreases very slowly which makes the use of a large computational 
domain necessary.  It can be seen  that the $L_{\infty}$ norm of the solution 
for the perturbed soliton decreases  monotonically. 

\begin{figure}[htb!]
  \includegraphics[width=\textwidth]{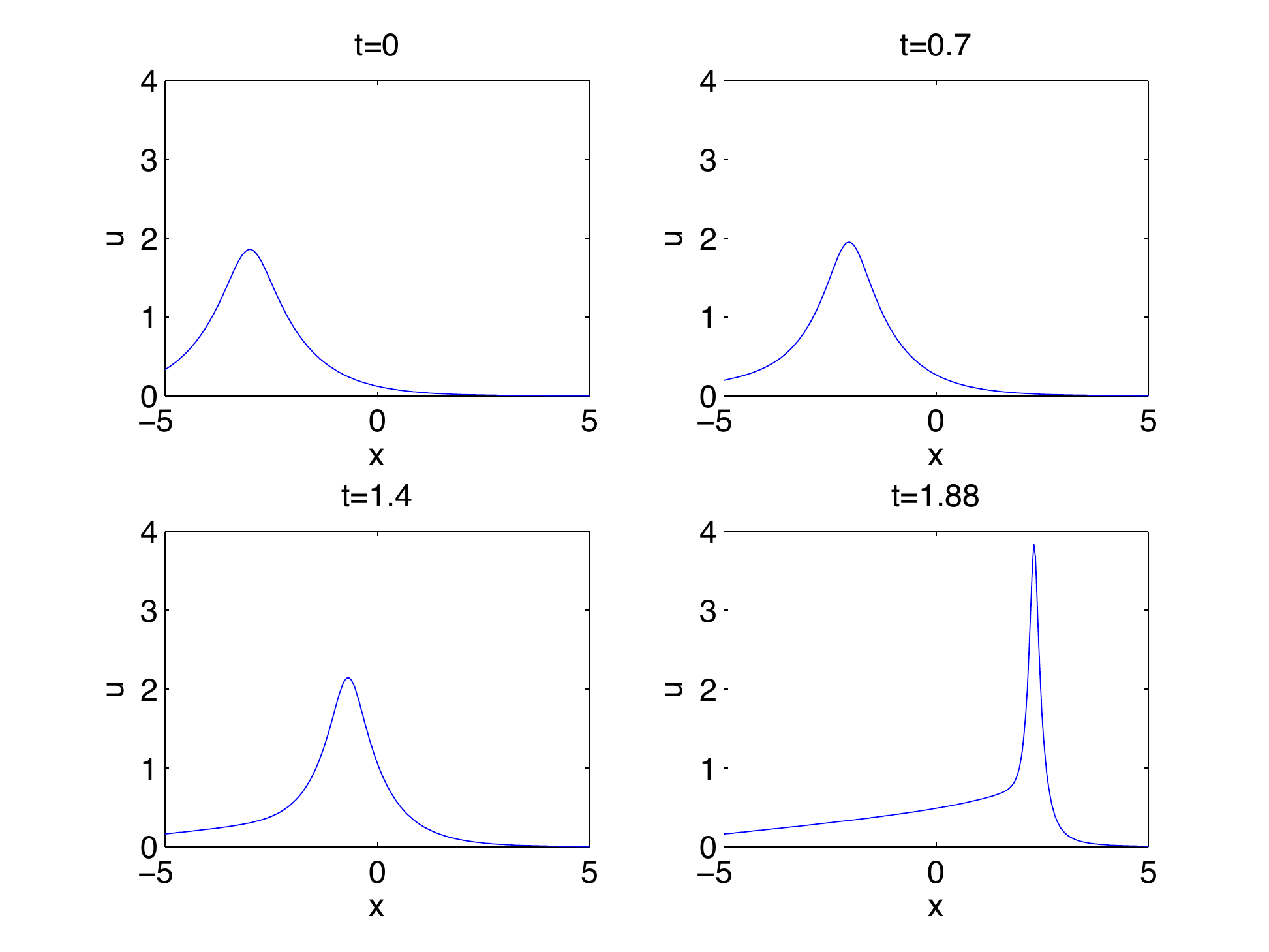}
 \caption{Solution to the gKdV equation (\ref{gKdV}) with 
 $\epsilon=1$ for $n=5$ 
 and the perturbed soliton initial data $1.01\,Q(x+3)$ (\ref{soliton}) 
 for several values of $t$.}
 \label{gKdVn5sol1014t}
\end{figure}

For initial data with $\sigma=1.01$, the soliton is as in the 
critical case unstable against blow-up though the energy in the 
considered example is positive, but smaller than the soliton energy, and the mass is larger than the soliton mass. But this time the blow-up is 
approached much more rapidly as can be inferred from 
Fig.~\ref{gKdVn5sol1014t}.  We compute for 
$x\in 100[-\pi,\pi]$ with $N=2^{14}$ Fourier modes and $N_{t}=2*10^{5}$ 
time steps. The code breaks at $t\approx 1.885$ since the iteration is 
not converging. The numerically computed energy is still conserved to better 
than $10^{-8}$ at $t=1.88$.  Note that the precise blow-up time will 
be determined below.  

\begin{figure}[htb!]
  \includegraphics[width=0.49\textwidth]{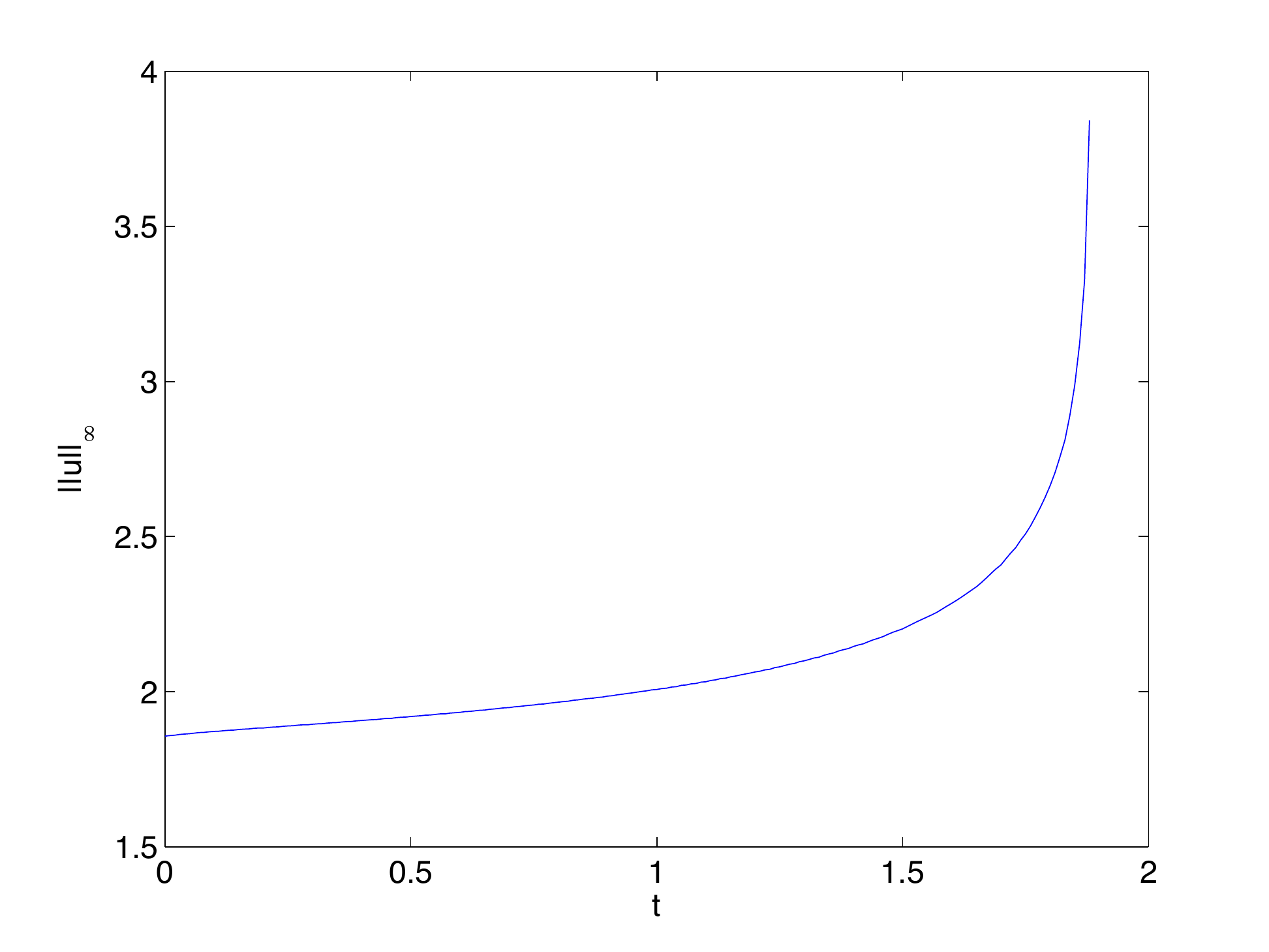}
  \includegraphics[width=0.49\textwidth]{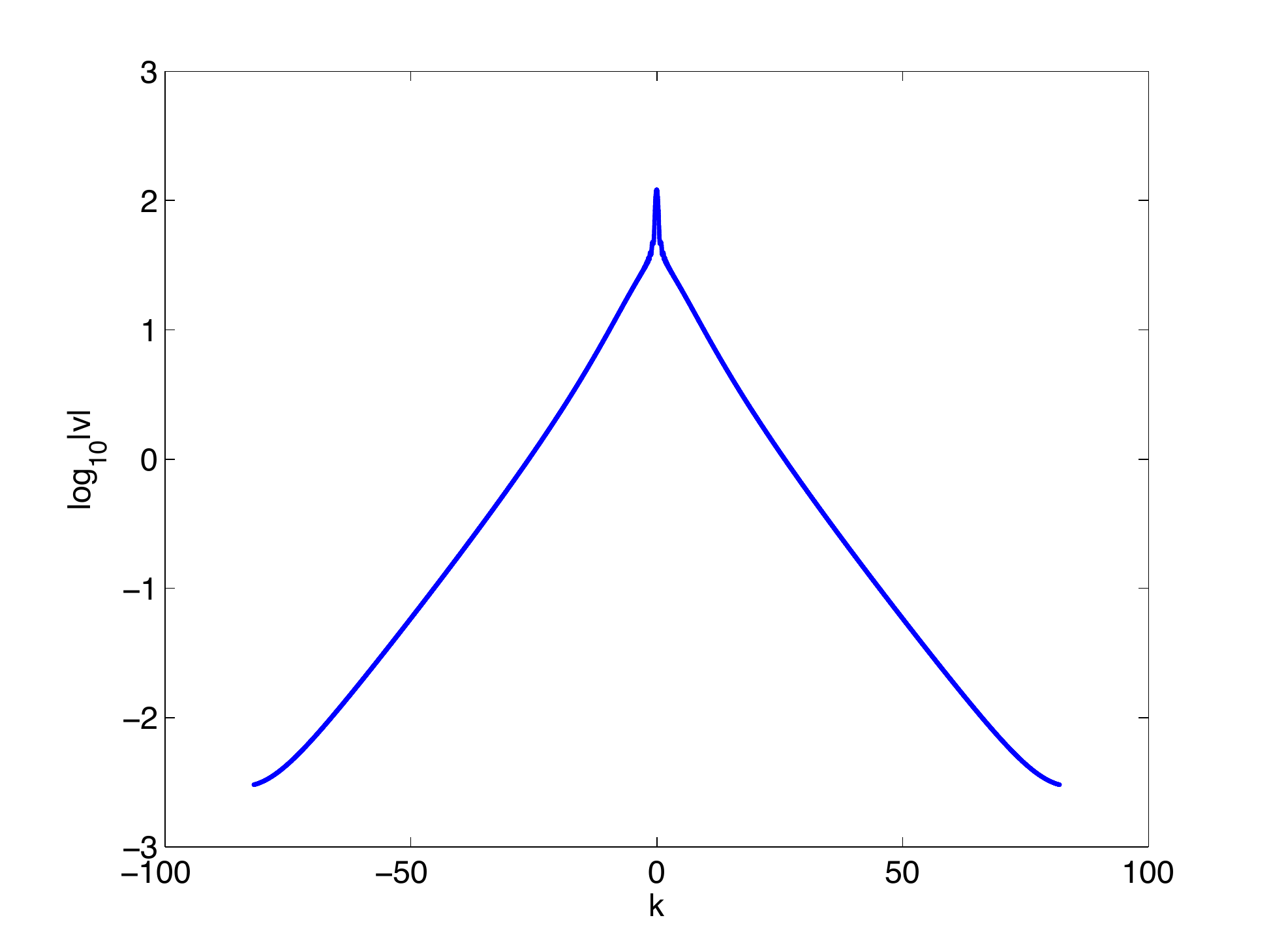}
 \caption{$L_{\infty}$ norm of the solution to the gKdV equation (\ref{gKdV}) 
 with 
 $\epsilon=1$ for $n=5$ 
 and the perturbed soliton initial data $1.01Q(x+3)$ (\ref{soliton}) 
 in dependence of time on the left, and the modulus of the Fourier 
 coefficients of the solution for $t=1.88$ on the right.}
 \label{gKdVn5sol101}
\end{figure}

It is clear from Fig.~\ref{gKdVn5sol101} that sufficient
resolution in Fourier space is no longer given near blow-up. Overall 
the way the blow-up is approached is  different from the case 
$n=4$, the loss of resolution in both space and time happens on very 
small $t^{*}-t$ scales. The $L_{\infty}$ norm of the solution 
for the perturbed soliton is monotonically increasing which also 
indicates a blow-up, see Fig.~\ref{gKdVn5sol101}. 

\begin{figure}[ht]
   \centering
    \includegraphics[scale=0.5]{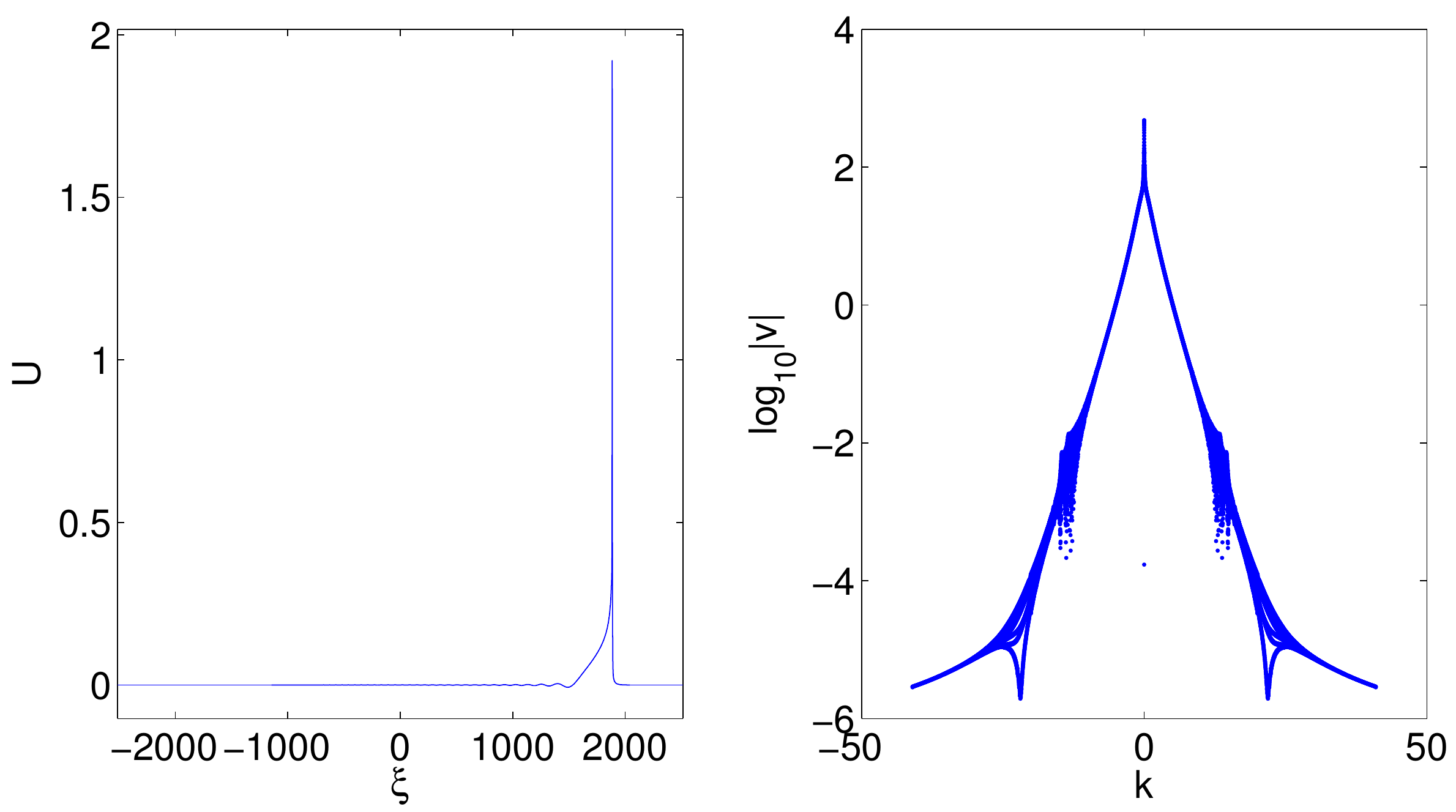}
   \caption{Solution $U(\xi,\tau)$ of the equation (\ref{gKP5}) with 
   $n=5$
   for the initial data $U(\xi,0) = 1.01\,Q(\xi)$ (\ref{soliton}) for 
   $\tau=12$ (physical time: $t = 1.8854$) (left) and the corresponding Fourier coefficients (right).}\label{fig:SolitonPerturb_n5_U_and_FC}
\end{figure}

Note that the type 
of blow-up is very different from the one in the $L_{2}$ critical case 
for the perturbed soliton in Fig.~\ref{gKdVn4sol1014t}. There the 
$L_{2}$ norm is invariant under the rescaling (\ref{gKP4}), and the 
blow-up profile has essentially the mass of the initial data. In the case $n=5$ 
the $L_{2}$ norm of the part of the solution blowing up vanishes in 
the limit $t\to t^{*}$ as follows from (\ref{L2}). This can be 
recognized already in Fig.~\ref{gKdVn5sol1014t} where the mass 
escapes to the left of the peak. 
To study the blow-up in more detail, we again use 
dynamic rescaling and solve gKdV in the form (\ref{gKP5}). With $N = 
2^{16}$ Fourier modes for $\xi\in800[-\pi,\pi]$ and $N_t = 2* 10^6$ 
time steps for $\tau\leq 12$, 
we obtain the solution shown in 
Fig.~\ref{fig:SolitonPerturb_n5_U_and_FC}. Since the 
mass of the solution spreads out to the left of the peak, we shift 
the initial data to the right to allow longer simulation times 
(the code typically breaks if the modulus of the solution at the 
boundaries of the computational domain is of the order of $10^{-3}$). 
We fix the location of the maximum at $\xi_0 = 1884.88$. The Fourier coefficients 
which can be seen in the same figure indicate that the solution is 
well resolved. The relative computed mass is 
conserved to better than $10^{-4}$ (as before, the relative computed 
energy is conserved only to the order 
of $4\%$).   Our results here are in accordance with the ones by 
Bona et al.~\cite{BDKMcK1995} as well as Dix and McKinney 
\cite{DixMcKinney} in what concerns the blow-up. But in contrast to 
these works, also the dispersive radiation is fully resolved for all 
recorded values of $t$. 

\begin{figure}[ht]
   \centering
   \includegraphics[scale=0.5]{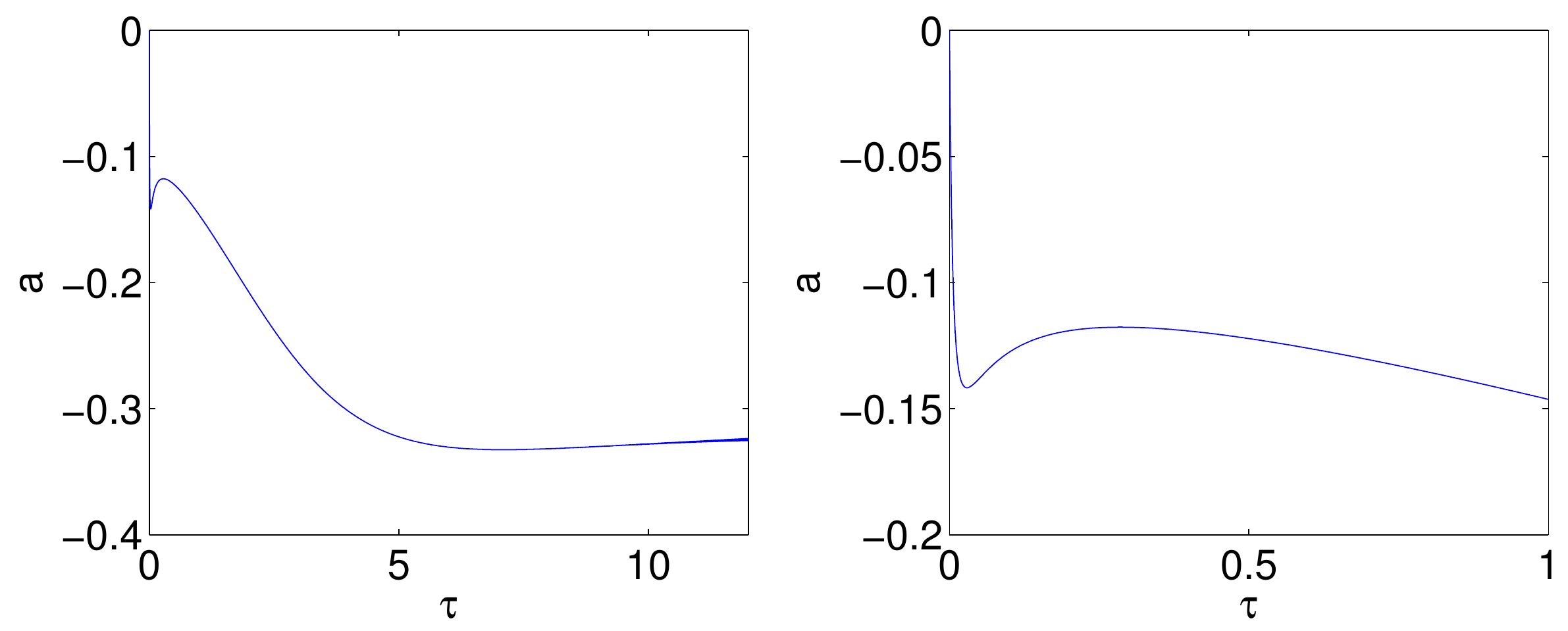}
      \caption{The function $a(\tau)$ on the whole $\tau$-range (left) and a detailed view of $a(\tau)$ in the interval $\tau\in[0,1]$ (right) for the solution shown in Fig.~\ref{fig:SolitonPerturb_n5_U_and_FC}.}\label{fig:SolitonPerturb_n5_atau}
\end{figure}

\begin{figure}[ht]
   \centering
   \includegraphics[scale=0.5]{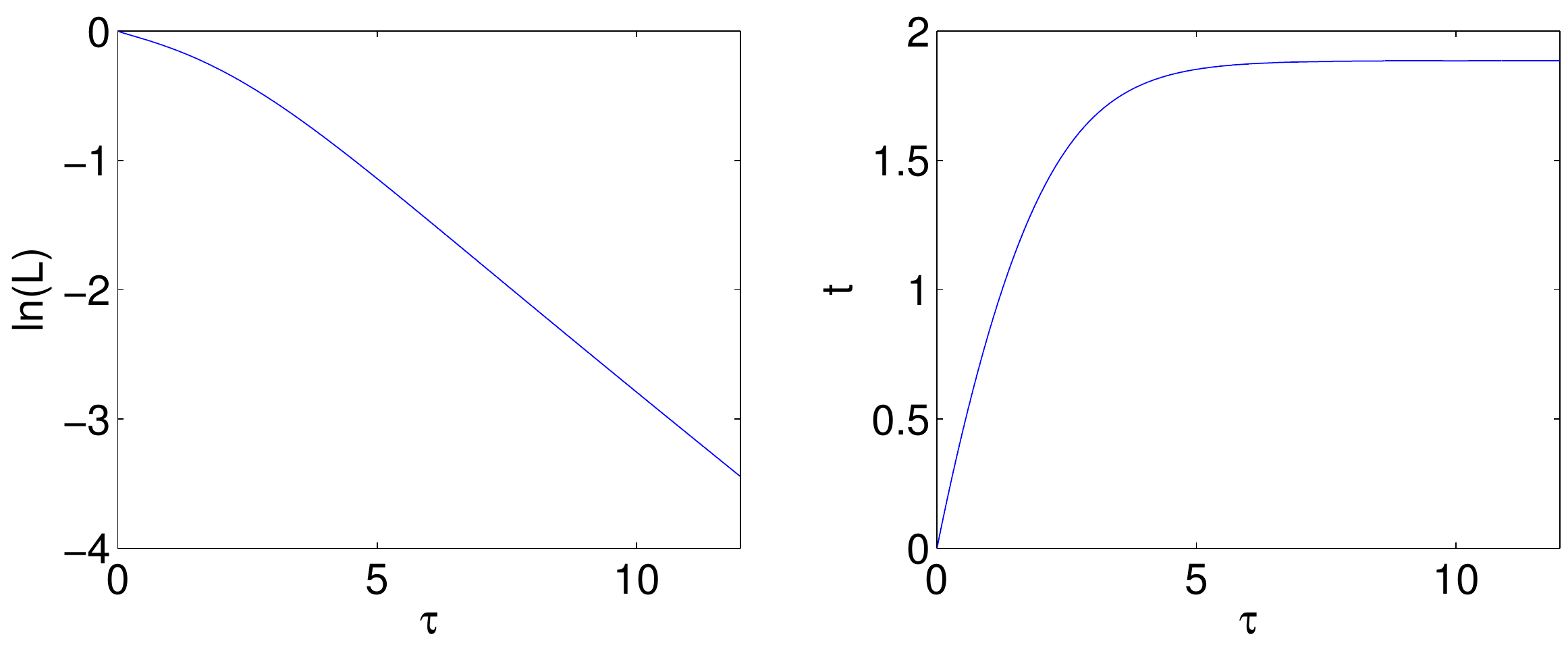}
   \caption{The scaling factor $L$ (left) and the physical time $t$ (right) as functions of the rescaled time $\tau$ for the solution of Fig.~\ref{fig:SolitonPerturb_n5_U_and_FC}.} 
\label{fig:SolitonPerturb_n5_tau_lnL_and_tau_t}
\end{figure}

The corresponding function $a(\tau)$ can be seen in 
Fig.~\ref{fig:SolitonPerturb_n5_atau}. As in the case $n=4$ in 
Fig.~\ref{fig:SolitonPerturb_n4_atau}, there are small oscillations 
in $a$ due to the periodic boundary conditions for $a$ which make it 
hard to read off a precise asymptotic value for $a(\tau)$. 
But since this value is definitely not zero,  the function $L(\tau)$ given in 
Fig.~\ref{fig:SolitonPerturb_n5_tau_lnL_and_tau_t} goes exponentially to zero as 
can be seen in the logarithmic plot.

\begin{figure}[ht]
   \centering
   \includegraphics[scale=0.5]{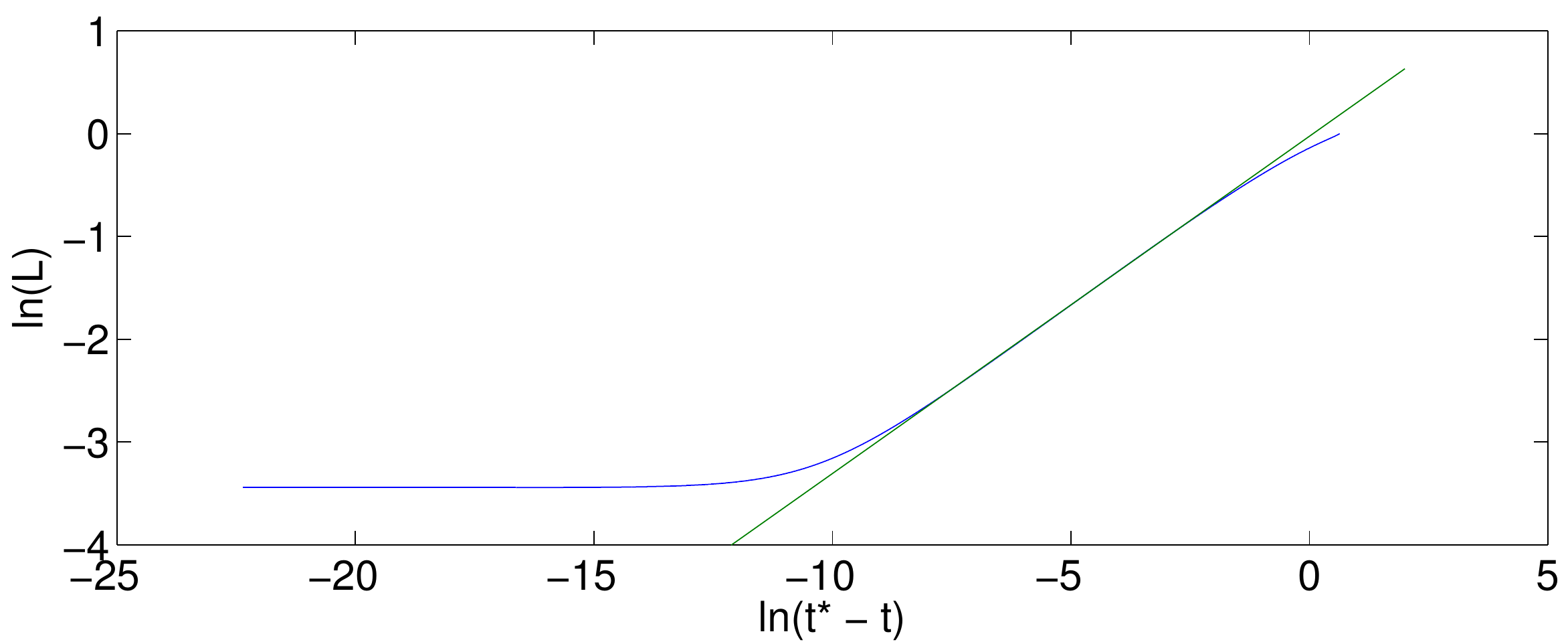}
   \caption{The scaling factor $L$ 
   in Fig.~\ref{fig:SolitonPerturb_n5_U_and_FC} as a function of the physical time $t$ and its corresponding fit.}\label{fig:SolitonPerturb_n5_lnt_lnL}
\end{figure}

If we consider $L$ as a function of the physical time $t$, the 
coordinate transform (\ref{gKP4}) implies a power law for $L(t)$ (\ref{exp}). In 
Fig.~\ref{fig:SolitonPerturb_n5_lnt_lnL} we can see that in a doubly 
logarithmic plot $L(t)$ approaches a linear regime. For values 
$\ln(t^* - t)< -8$ the systems runs into saturation and the 
corresponding values have to be neglected. We plot the linear part of 
$\ln L(t)$, 
$\ln L = \gamma\ln(t^* - t) + C$ and obtain: $\gamma 
= 0.3281$ and $C = -0.0256$ in accordance with the expectation 
$\gamma = 1/3$ (\ref{exp}).

In contrast to the case $n=4$, the blow-up time can directly be read off from the physical time in dependence of $\tau$ as can also be seen in Fig.~\ref{fig:SolitonPerturb_n5_tau_lnL_and_tau_t}. We get the final value  $t^* = 1.8854$. The  relative change compared to $\tau = 11$ 
is of the order $10^{-5}$. This also shows that the direct integration 
in this case reaches values of $t$ very close to $t^{*}$. 

\begin{figure}[ht]
   \centering
   \includegraphics[scale=0.5]{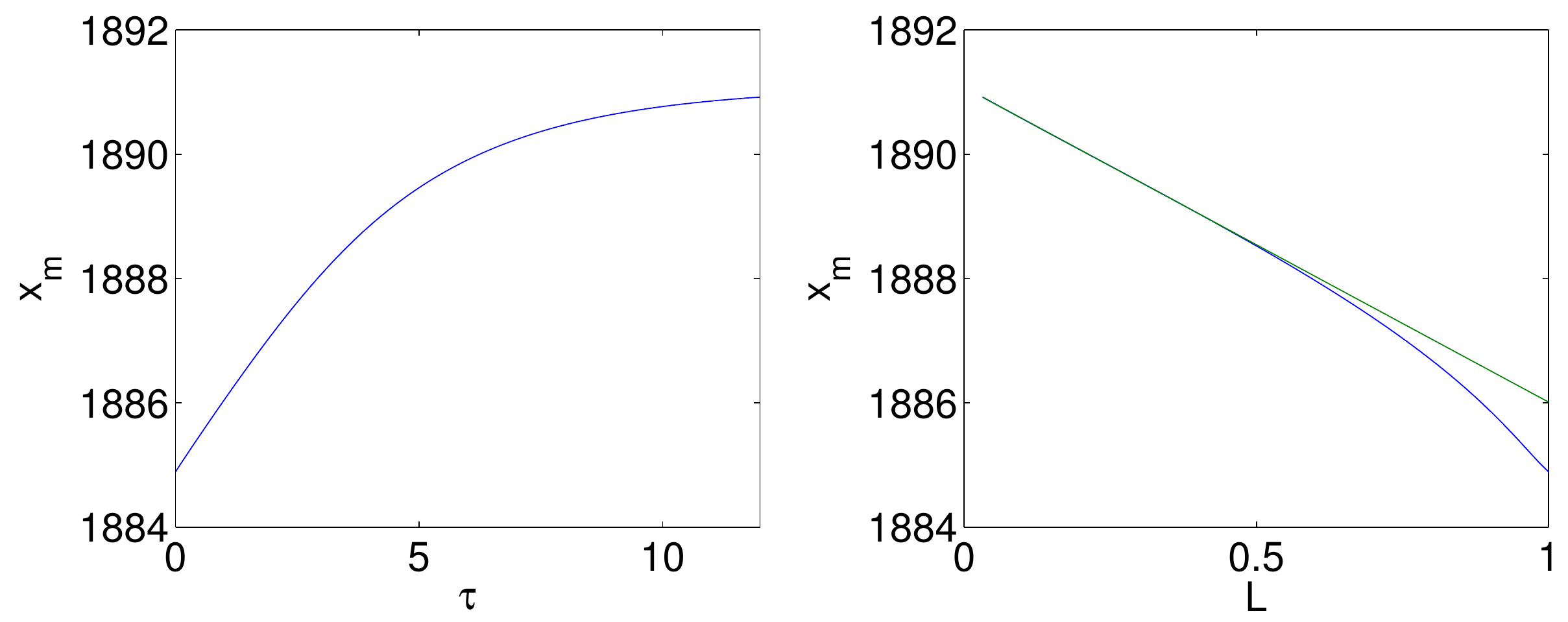}
   \caption{The position $x_m$ of the maximum of the solution shown 
   in Fig.~\ref{fig:SolitonPerturb_n5_U_and_FC} as a function of the 
   rescaled time  $\tau$ on the left and in dependence of $L$ on the 
   right, with the corresponding fit. }\label{fig:SolitonPerturb_n5_lnL_xm}
\end{figure}

A further difference to the $L_{2}$ critical case $n=4$ is the 
location of the blow-up. Whereas it was infinite in the former case, 
it is clearly finite for $n=5$ as can be seen in 
Fig.~\ref{fig:SolitonPerturb_n5_lnL_xm}.
Plotting $x_{m}$ in dependence of the scaling factor $L$ in 
Fig.~\ref{fig:SolitonPerturb_n5_lnL_xm}, one can recognize an
essentially linear dependence for small $L$ ($L<0.4$), i.e., close to 
blow-up. We get $x_m = \gamma L + x_m^0$ with $\gamma = -5.0665$ and 
$x_m^0 = 1891.08$, i.e., \ $\Delta x_m = 6.1983$ which can be 
interpreted as the blow-up position. The 
conservation of the numerically computed quantities is here of order 
$10^{-5}$ for both the $L_2$ norm of $U_{\xi}$ and for the derivative 
$U_{\xi}$ at $\xi_0$. Since $a$ tends asymptotically to a more 
negative value than in the other considered examples (for $n=4$ it 
tends to zero), comparatively small times $\tau$ are sufficient to 
come close to blow-up. This leads to better numerical results for the 
computed conserved quantities, but the effect is partially offset by 
the oscillations in $a$ due to the dispersive radiation and the 
periodic boundary conditions.


\subsection{Small dispersion limit}
\begin{figure}[htb!]
  \includegraphics[width=\textwidth]{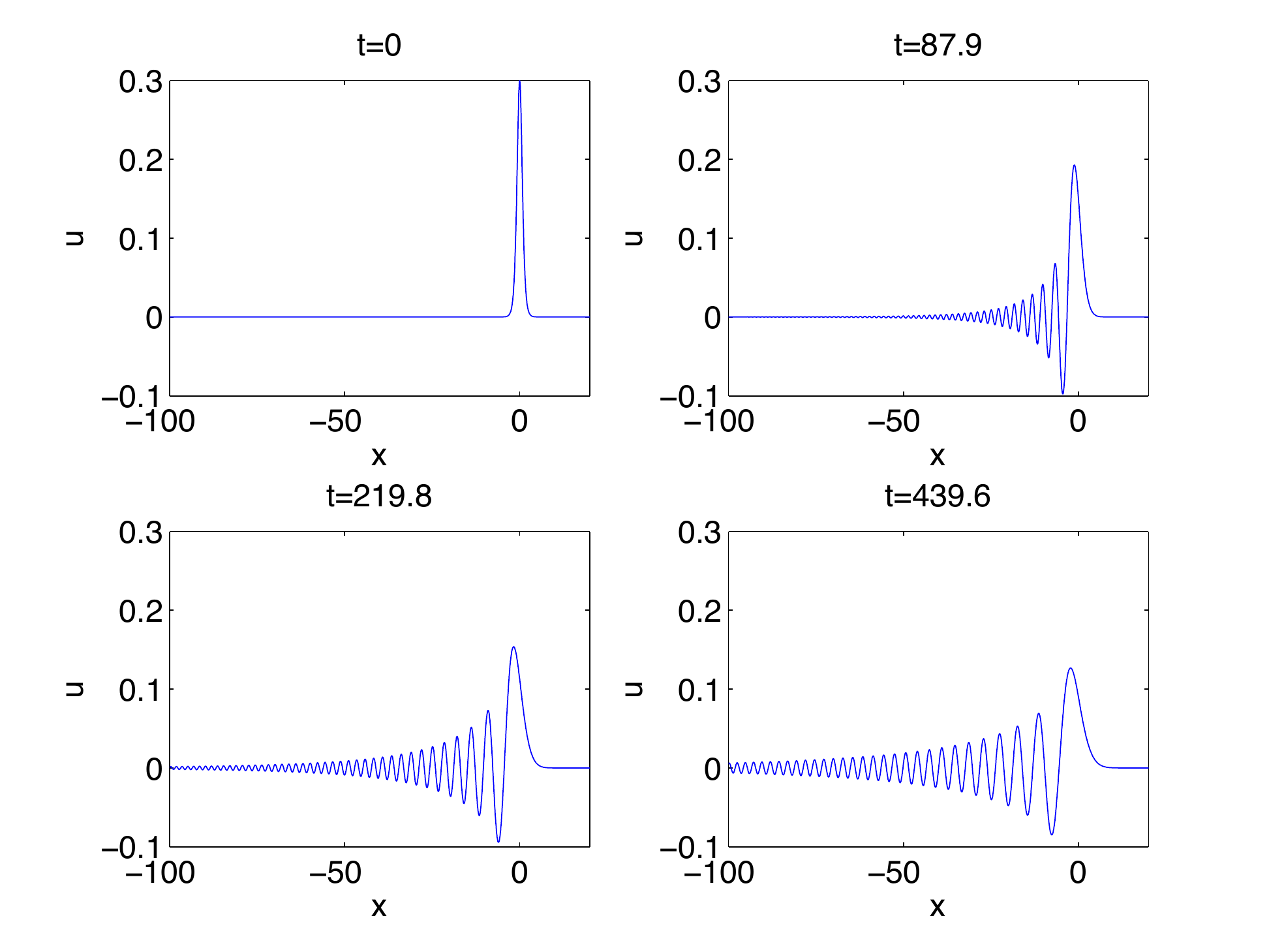}
 \caption{Solution to the gKdV equation (\ref{gKdV}) with 
 $\epsilon=0.1$ for $n=5$ 
 and the initial data $u_{0}=0.3\,\mbox{sech}^{2}x$  
 for several values of $t$.}
 \label{gKdVn5e01b034t}
\end{figure}

As for the $L_{2}$ critical case $n=4$ we discuss the initial data 
$u_{0}=\beta\, \mbox{sech}^{2}x$ in the semiclassical limit 
$\epsilon\ll 1$. For $\beta=0.3$ the energy  is larger than the  soliton energy, 
but positive, the mass is smaller than the mass of the soliton (since 
the mass has to be rescaled with a factor
$1/\epsilon$ to allow  comparison with the soliton, it is here equal to 4) 
and the critical time (\ref{tch}) is  
$t_{c}\approx219.8131$.
We use 
$N_{t}=10^{4}$ 
time steps for $t\leq 2t_{c}$ and $N=2^{12}$ Fourier modes for 
$x\in100[-\pi,\pi]$ to directly integrate the gKdV equation for these 
initial data. As can be 
seen in Fig.~\ref{gKdVn5e01b034t}, the solution develops again a tail of 
dispersive oscillations towards $-\infty$. The $L_{\infty}$ norm of 
the solution decreases monotonically.

%

\begin{figure}[htb!]
  \includegraphics[width=\textwidth]{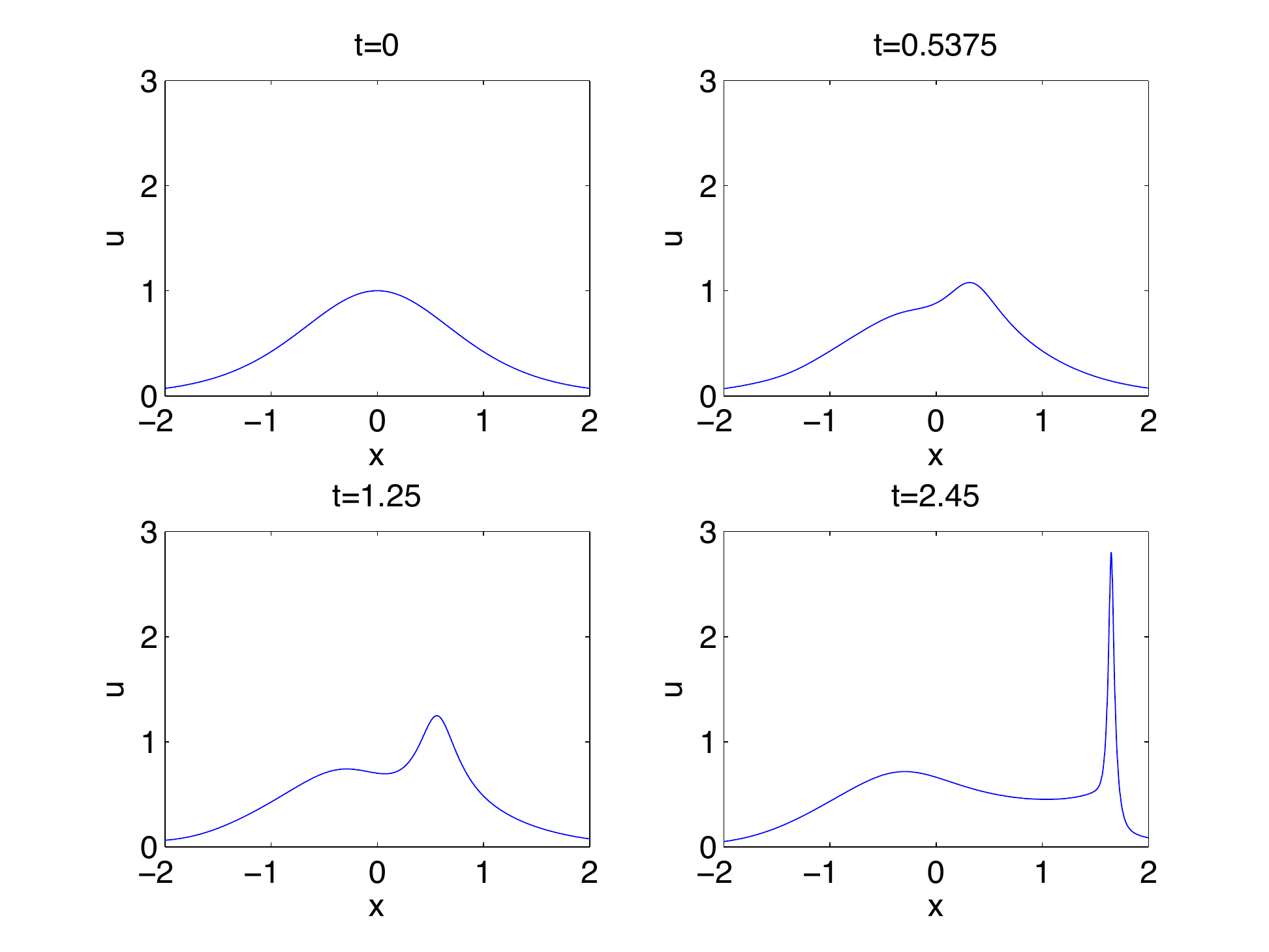}
 \caption{Solution to the gKdV equation (\ref{gKdV}) with 
 $\epsilon=0.1$ for $n=5$ 
 and the initial data $u_{0}=\mbox{sech}^{2}x$  
 for several values of $t$.}
 \label{gKdVn5e014t}
\end{figure}

For $\beta=1$  
the energy is again negative, and instead of dispersive radiation 
dominating as for $\beta=0.3$, we observe blow-up. The mass of the initial data is again larger than the soliton mass. The computation is carried out in this case
with $N_{t}=10^{5}$ 
time steps for $t<2.5$ and $N=2^{14}$ Fourier modes for 
$x\in5[-\pi,\pi]$. The code breaks at $t=2.45$ since the iteration  
no longer converges. The 
solution is shown for several times in 
Fig.~\ref{gKdVn5e014t}. For $t\ll t_{c} \approx 0.5341$ (\ref{tch}) it is very 
close to 
the solution of the generalized Hopf equation for the same initial 
data. The dispersive effects of the third derivative in the gKdV 
equation become important near the critical time $t_{c}$.  
A first oscillation forms at this time which then 
develops into a blow-up as for the perturbed soliton. This is to be 
compared to the subcritical initial data in 
Fig.~\ref{gKdVn5e01b034t} where the initial data is just radiated 
away. 

\begin{figure}[htb!]
  \includegraphics[width=0.49\textwidth]{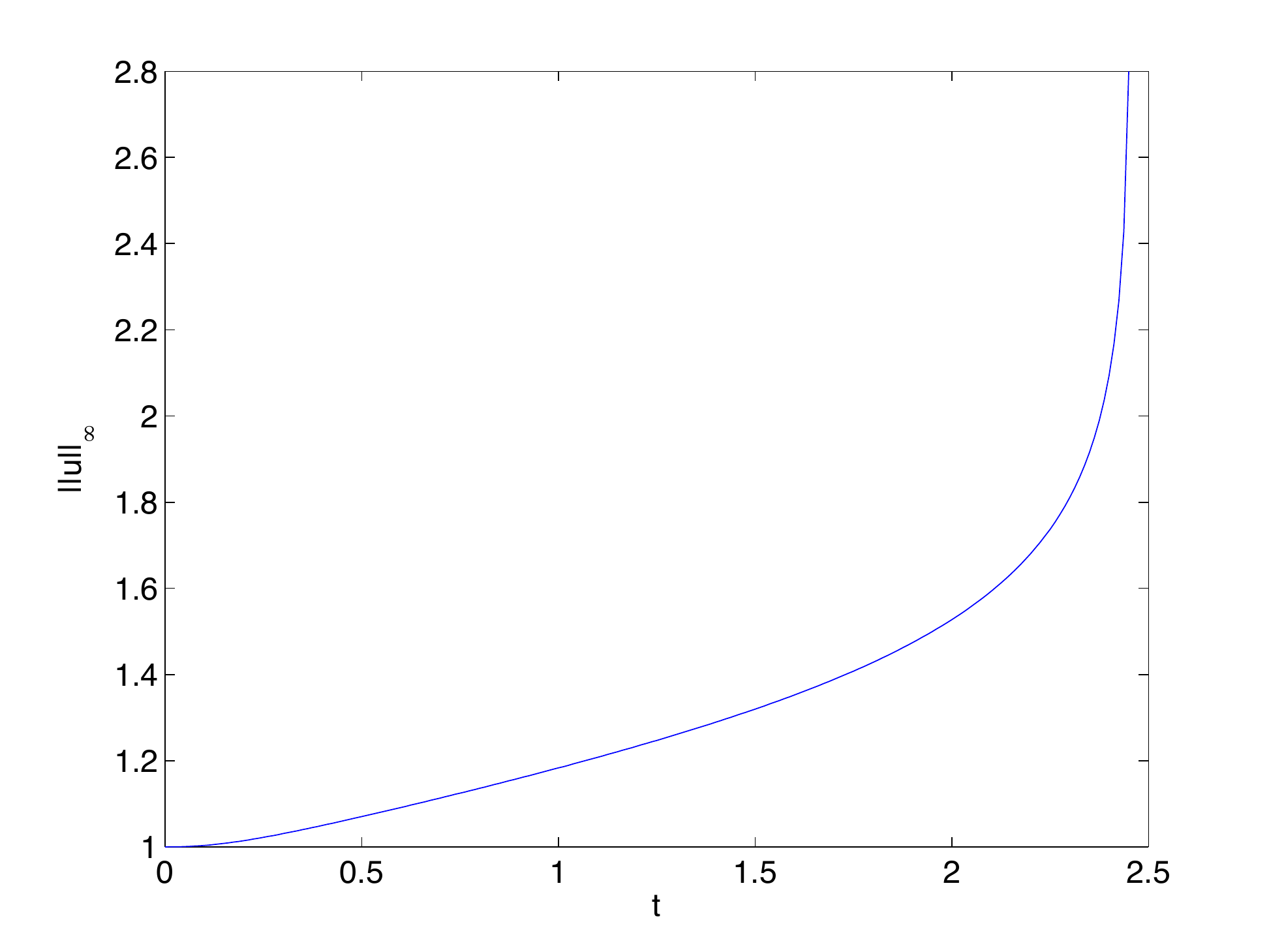}
  \includegraphics[width=0.49\textwidth]{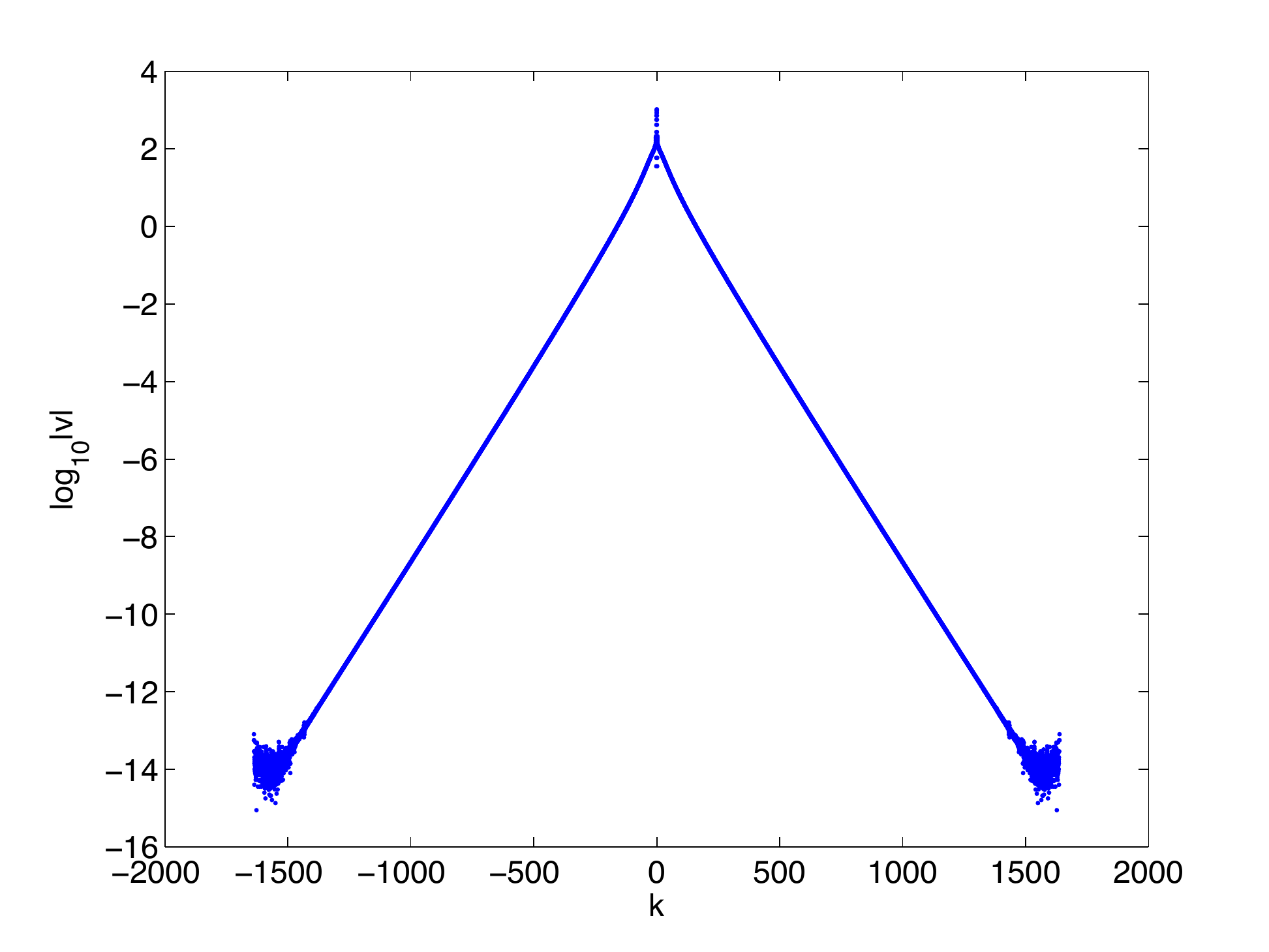}
 \caption{$L_{\infty}$ norm of the solution to the gKdV equation (\ref{gKdV}) 
 with 
 $\epsilon=0.1$ for $n=5$ 
 and the initial data $u_{0}=\mbox{sech}^{2}x$ 
 in dependence of time on the left, and the modulus of the Fourier 
 coefficients of the solution for $t=2.45$ on the right.}
 \label{gKdVn5e01}
\end{figure}

The monotonous increase of the $L_{\infty}$ norm in Fig.~\ref{gKdVn5e014t} 
is even more obvious from Fig.~\ref{gKdVn5e01}. In contrast to the 
$L_{2}$ critical case $n=4$ we can get very close to the blow-up time. Even 
for $t\sim t^{*}$ there is still considerable resolution in Fourier space 
as can be seen from the Fourier 
coefficients at the final recorded time in Fig.~\ref{gKdVn5e01}. 

\begin{figure}[ht]
   \centering
   \includegraphics[scale=0.5]{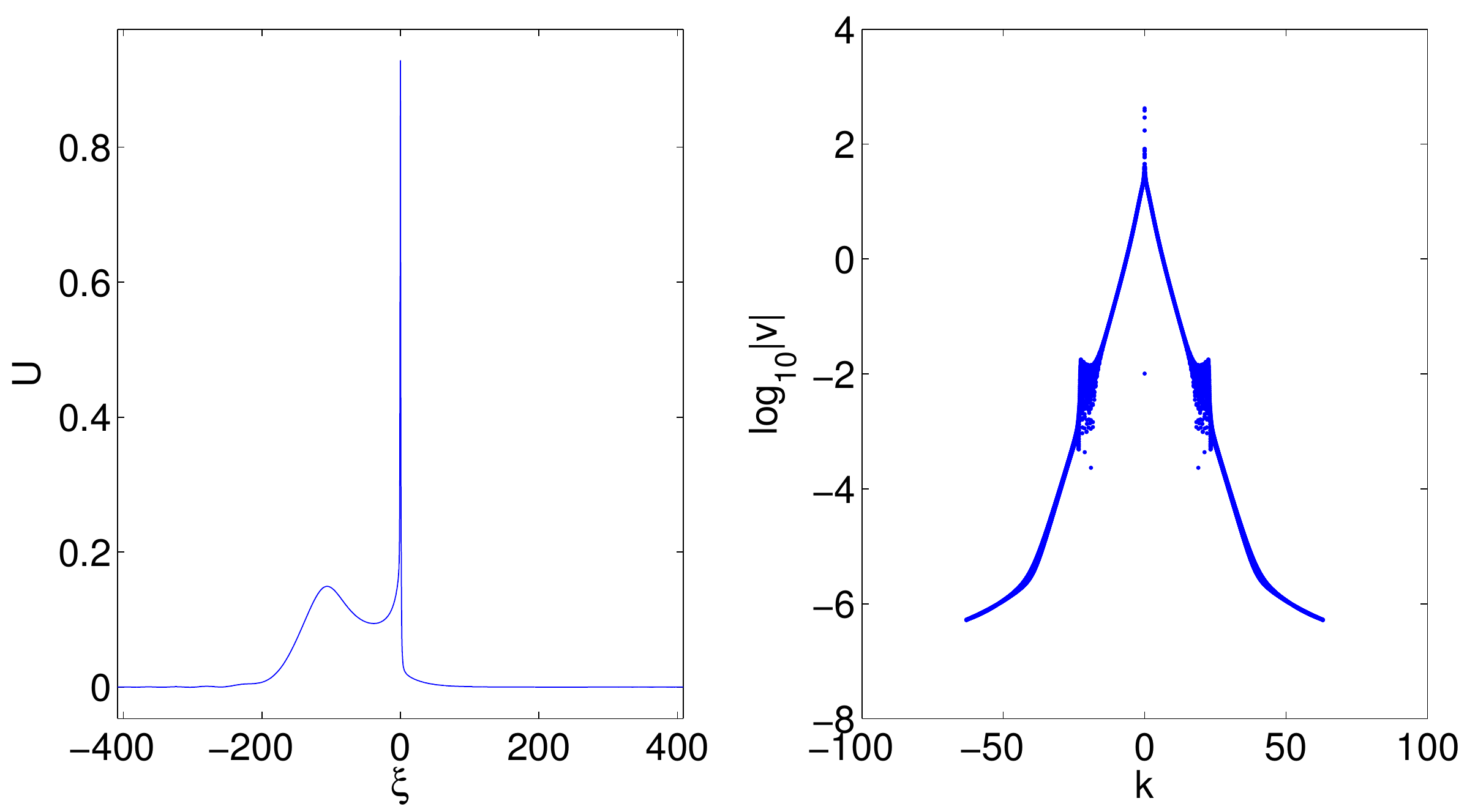}
   \caption{The solution $U(\xi,\tau)$ for the equation (\ref{gKP5}) 
   for the initial data $U(\xi,0) = \mbox{sech}^2\xi$ at $\tau = 200$ (physical time: $t = 2.4564$) for $n=5$ and $\epsilon=0.1$ (left) and the corresponding Fourier coefficients (right).}\label{fig:SechSquare_n5_U}
\end{figure}

To analyze the blow-up in more detail, we study the dynamically 
rescaled equation (\ref{gKP5}). We use $N = 2^{14}$ Fourier modes 
for $\xi\in130[-\pi,\pi]$ and  $N_t =6* 10^6$ time steps for  
$\tau\leq 200$.
The solution at the 
final time can be seen in Fig.~\ref{fig:SechSquare_n5_U} where also 
the Fourier coefficients are given. They indicate that the solution 
is well resolved. The relative computed mass containing no 
derivative is conserved to better 
than $10^{-4}$ (as before, the relative computed energy is conserved 
only to the 
order of $7\%$).  

\begin{figure}[ht]
   \centering
   \includegraphics[scale=0.5]{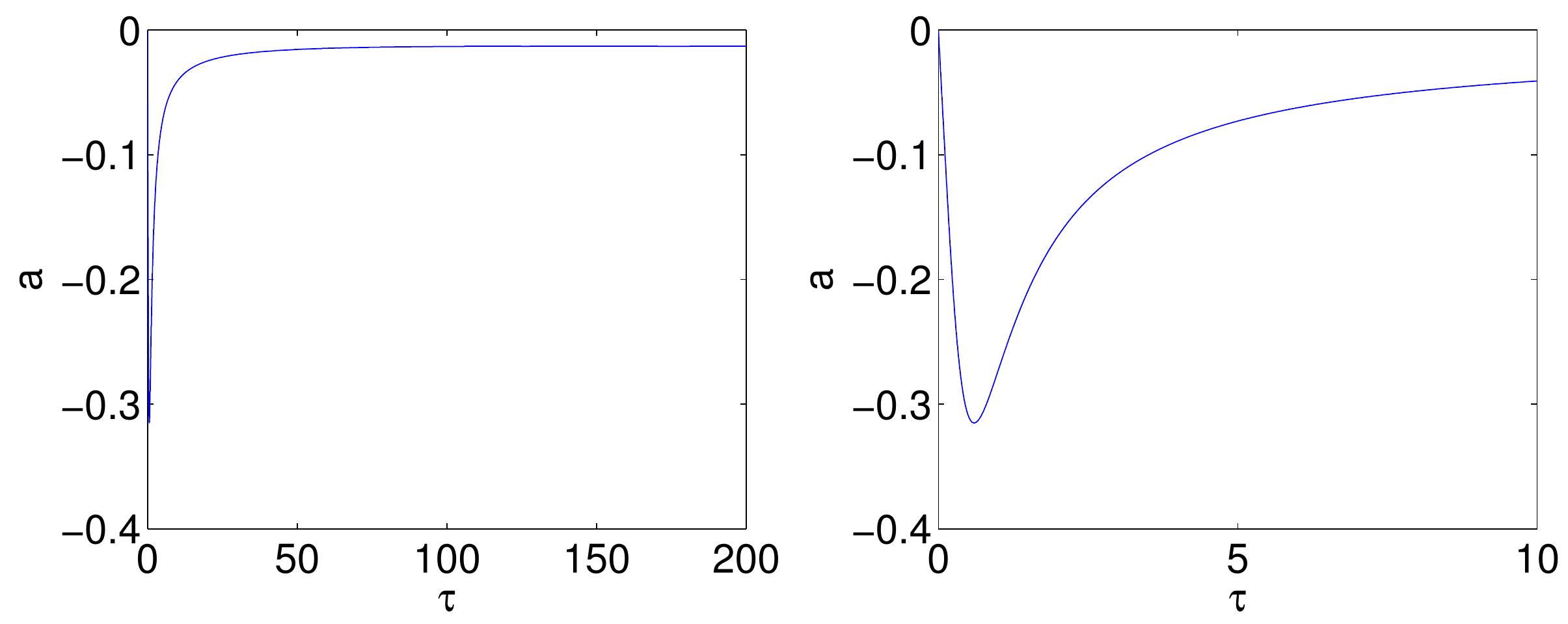}
   \caption{The function $a(\tau)$ on the whole computed $\tau$-range (left) 
   and a close-up view in the interval $\tau\in[0,10]$ (right) for the solution of Fig.~\ref{fig:SechSquare_n5_U}.}\label{fig:SechSquare_n5_atau}
\end{figure}

\begin{figure}[ht]
   \centering
   \includegraphics[scale=0.5]{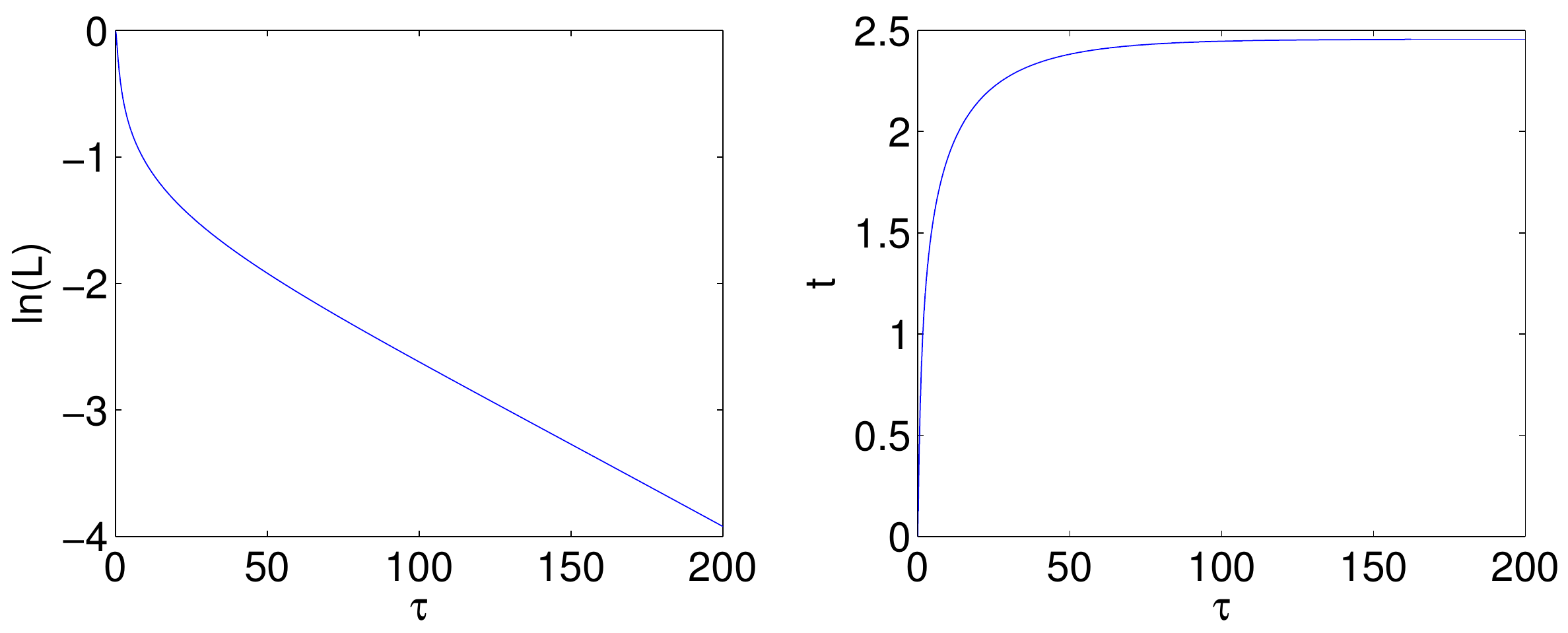}
   \caption{The scaling factor $L$ (left) and the physical time $t$ (right) as functions of the rescaled time $\tau$ for the solution of Fig.~\ref{fig:SechSquare_n5_U}.} 
\label{fig:SechSquare_n5_tau_lnL_and_tau_t}
\end{figure}

The corresponding function $a(\tau)$ tends here clearly to a negative 
constant as can be seen in Fig.~\ref{fig:SechSquare_n5_atau}. The final 
value for $a$ is considerably smaller than for the perturbed 
soliton,  $a(200) = -0.0131$, but the relative change 
compared to $a(180)$ is of the order of $10^{-3}$. This indicates 
that the shown situation is close to blow-up.
This can be also seen from the scaling $L(\tau)$ in 
Fig.~\ref{fig:SechSquare_n5_tau_lnL_and_tau_t} which 
follows for larger $\tau$ an exponential law. 

\begin{figure}[ht]
   \centering
   \includegraphics[scale=0.5]{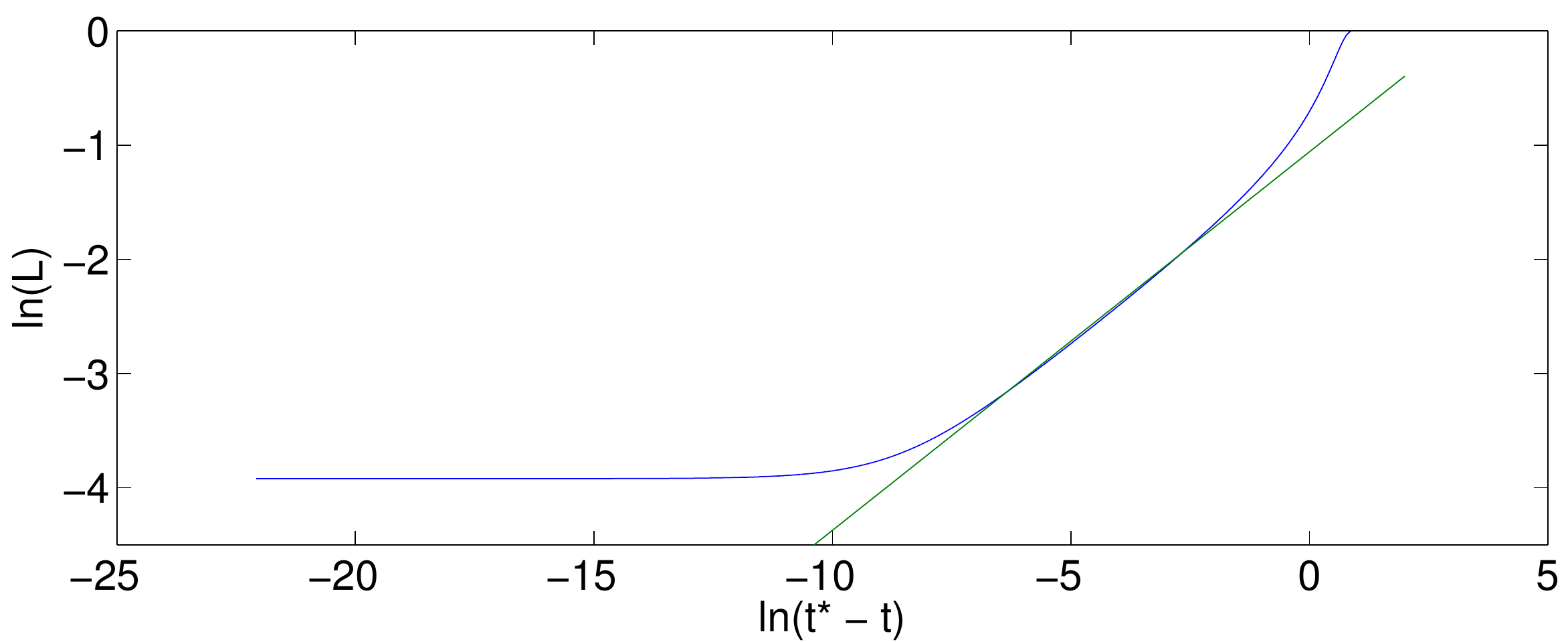}
   \caption{The scaling factor $L$ 
   in Fig.~\ref{fig:SechSquare_n5_U} as a function of the physical time $t$ and its corresponding fit.}\label{fig:SechSquare_n5_lnt_lnL}
\end{figure}

As in the case of the soliton perturbation, we also perform a fitting 
of $L(t)$ to the expected behavior given in formula 
(\ref{exp}). In Fig.~\ref{fig:SechSquare_n5_lnt_lnL} we can 
see that in a doubly logarithmic plot $L(t)$ approaches the linear 
regime. For values $\ln(t^* - t)< -8$ the system runs also here into 
saturation, the corresponding values have to be neglected. We plot the 
linear part of $\ln L(t)$ and 
obtain: $\gamma = 0.3316$ and $C = -1.0585$ in accordance with the 
expectation $\gamma = 1/3$ (\ref{exp}).


The dependence of the physical time on $\tau$ can be seen in 
Fig.~\ref{fig:SechSquare_n5_tau_lnL_and_tau_t} on the right. The final value can be again 
interpreted as the blow-up time $t^* = 2.4564$. The relative 
change compared to $t(180)$ is of the order of $10^{-5}$ which 
confirms that the shown situation is very close to the blow-up.

\begin{figure}[ht]
   \centering
   \includegraphics[scale=0.5]{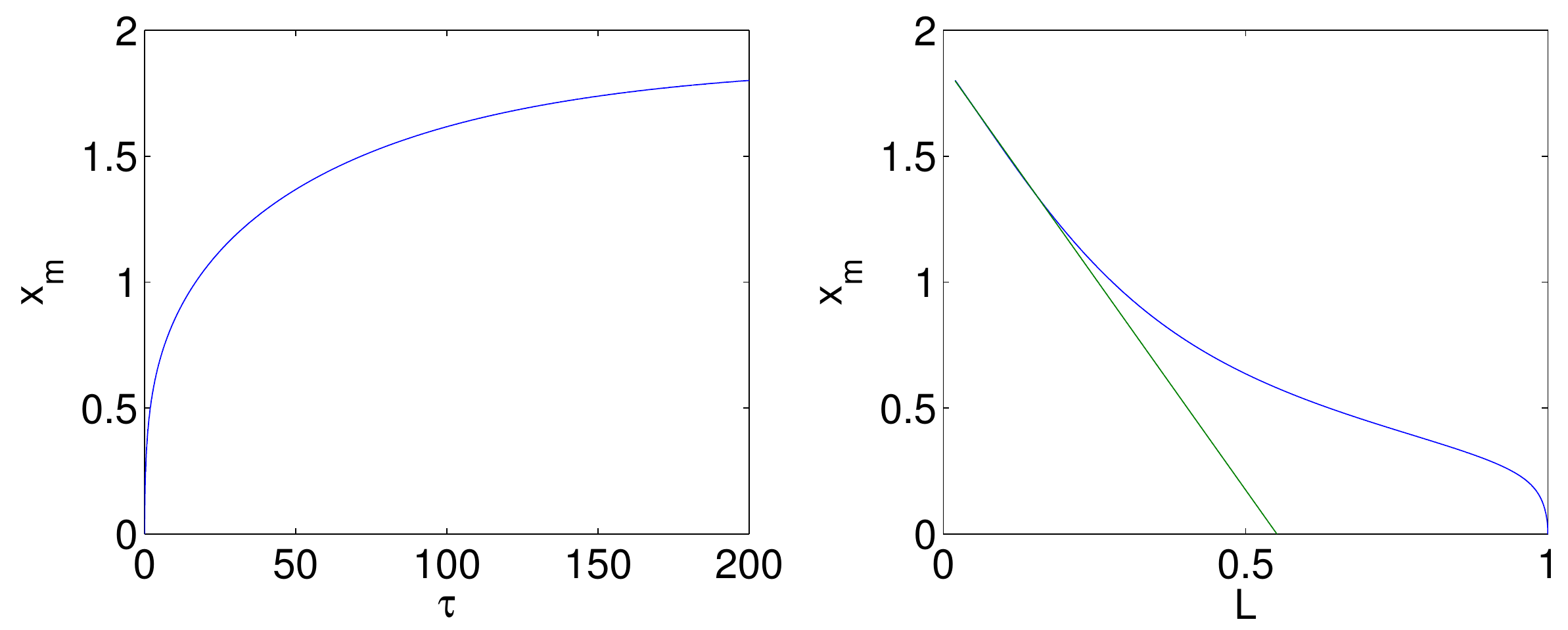}
   \caption{The position $x_m$ of the maximum of the solution shown 
   in Fig.~\ref{fig:SechSquare_n5_U} as a function of the rescaled 
   time $\tau$ (left) and in dependence of $L$ (right).}\label{fig:SechSquare_n5_xm}
\end{figure}

As before this is not the case for the location $x_{m}$ of the 
maximum of the solution as can be seen in 
Fig.~\ref{fig:SechSquare_n5_xm}, where the asymptotic regime is not 
yet reached. 
Plotting $x_{m}$ as a function of $L$ and fitting it for small $L$ 
($L<0.2$) to a straight 
line  $x_m = \gamma L + x_m^0$, we find $\gamma = -3.3824$ and 
$x_m^0 = 1.8656$. At the blow-up time $t^*$, the scaling factor $L$ 
vanishes, and therefore $x_m^0$ can be interpreted as the position of the blow-up $x^*$.
The $L_2$ norm of $U_{\xi}$ and the value of $U_{\xi}(0,\tau)=0$  are both preserved to the order of $10^{-4}$.

\section{Small dispersion limit}\label{smalldisp}
In this section we address the case of large masses (more than twice
the soliton mass). This allows also to study the small 
dispersion limit for $\epsilon<0.1$ for the initial data $u_{0}= 
\mbox{sech}^{2} x$. The results of this study can be summarized as 
follows (see 
conjecture \ref{sdconj}):
\begin{itemize}
    \item  Localized smooth, positive initial data with a single 
    maximum of large mass show self-similar blow-up as 
    detailed in conjectures \ref{conj4} and \ref{conj1}.

    \item  The blow-up times for such initial data depend 
    exponentially on $\epsilon$ and are strictly greater than the 
    critical time $t_{c}$ of the corresponding solution to the 
    generalized Hopf equation. This implies there is always a 
    dispersive shock in this case. 
\end{itemize}

Since the dynamically rescaled code is not applicable for large 
masses for stability reasons as discussed in the previous sections, 
we directly integrate the gKdV equation and trace the $L_{\infty}$ 
norm of $u$ as well as the $L_{2}$ norm of $u_{x}$. As before the code is stopped once 
the conservation of the numerically computed energy drops below 
$10^{-3}$ indicating that the solution is no longer reliable. For 
the resulting data, we fit the last 1000 time steps to the expected 
asymptotic behavior (\ref{exp}) or (\ref{alg}). 
We consider $\ln ||u_{x}||_{2}^{2}$ 
and fit it to $\alpha\ln(t^{*}-t)+\kappa$. The quantity 
$\alpha$ is fixed by the 
theoretical expectation (\ref{exp}) or (\ref{alg}), $\kappa$ and $t^{*}$ 
are determined in a way to minimize $||\ln 
||u_{x}||_{2}^{2}-\alpha\ln(t^{*}-t)+\beta||_{2}$. It would be possible to 
determine also $\alpha$ in this way, but since the result depends very
sensitively  on $t^{*}$ especially for the values of $t$ closest to 
$t^{*}$ where the numerical accuracy is lowest, this is in practice 
not a good idea. Instead we check the consistency with an analogous 
fit for $||u||_{\infty}$. And since we are interested in the 
$\epsilon$-dependence of $t^{*}$, it is mainly important that the 
estimate for $t^{*}$ is determined in a consistent way.

\subsection{The $L_{2}$ critical case $n=4$}
\begin{figure}[htb!]
   \includegraphics[width=\textwidth]{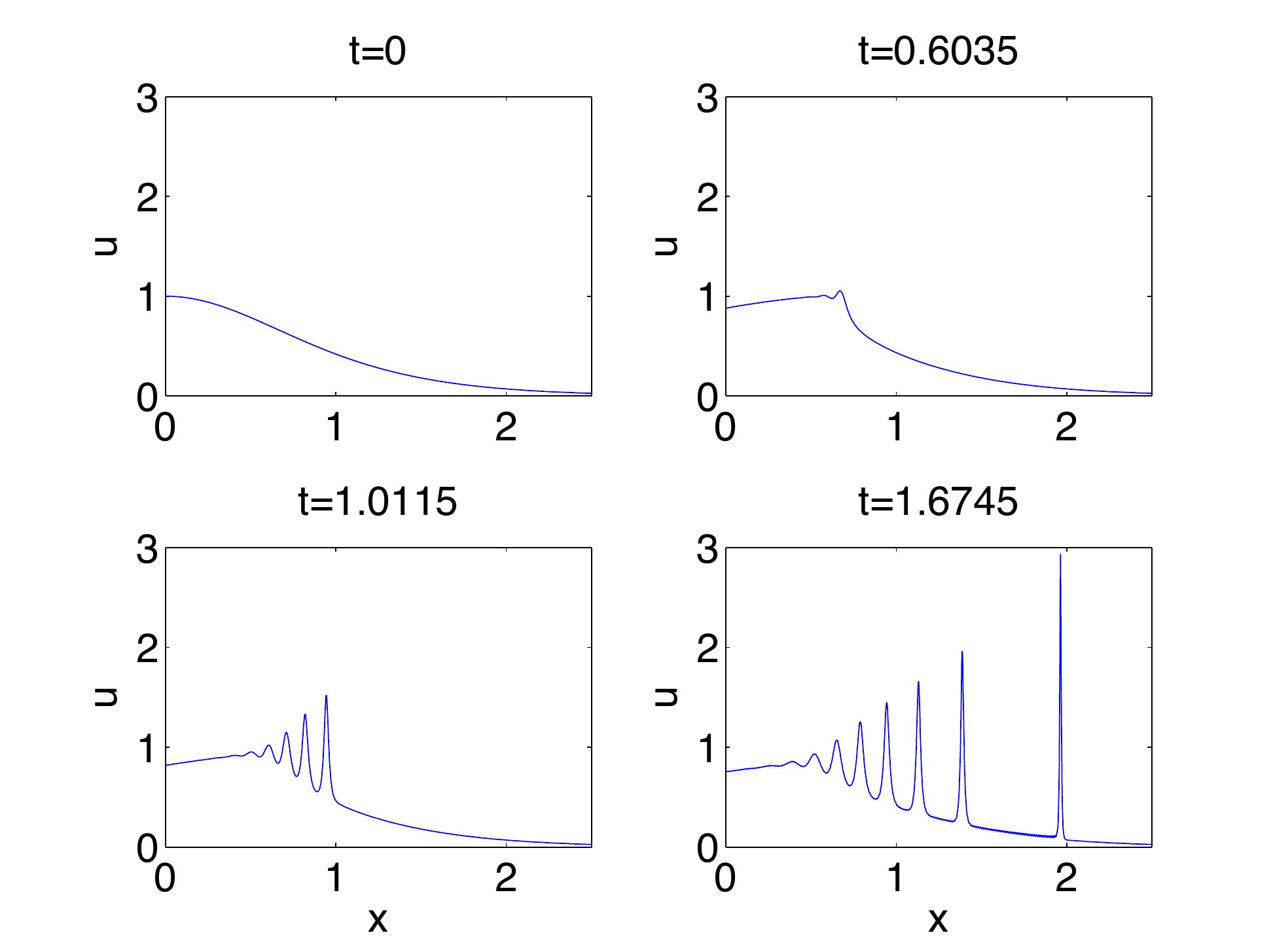}
 \caption{Solution to the gKdV equation (\ref{gKdV}) with $n=4$ for the initial 
 data $u_{0}=\mbox{sech}^{2}x$ and $\epsilon=0.01$ for different times.}
 \label{gKdVn4sech2e0014t}
\end{figure}

In the $L_{2}$ critical case $n=4$,
we solve the gKdV equation for the initial data 
$u_{0}=\mbox{sech}^{2}x$ and $\epsilon=0.1,0.09,\ldots,0.001$. We 
always make first an exploratory run to roughly determine the time, 
when the code breaks for a given $\epsilon$. Then we run it with 
$N=2^{14}$ Fourier modes for $x\in5[-\pi,\pi]$
and $N_{t}=10^{5}$ time steps to the 
estimated blow-up time. Only the data with $\Delta<10^{-3}$ will be 
considered. For $\epsilon=0.1$ the solution was shown for different times 
in Fig.~\ref{gKdVn4e014t}. It can be seen that close to the 
critical time $t_{c}\sim0.6007$ (\ref{tch}) one soliton forms which eventually turns 
into a blow-up. 
The same situation is considered for $\epsilon=0.01$ in 
Fig.~\ref{gKdVn4sech2e0014t}. It can be seen that in this case 
several oscillations form before the one appearing first will blow up.

\begin{figure}[htb!]
   \includegraphics[width=0.49\textwidth]{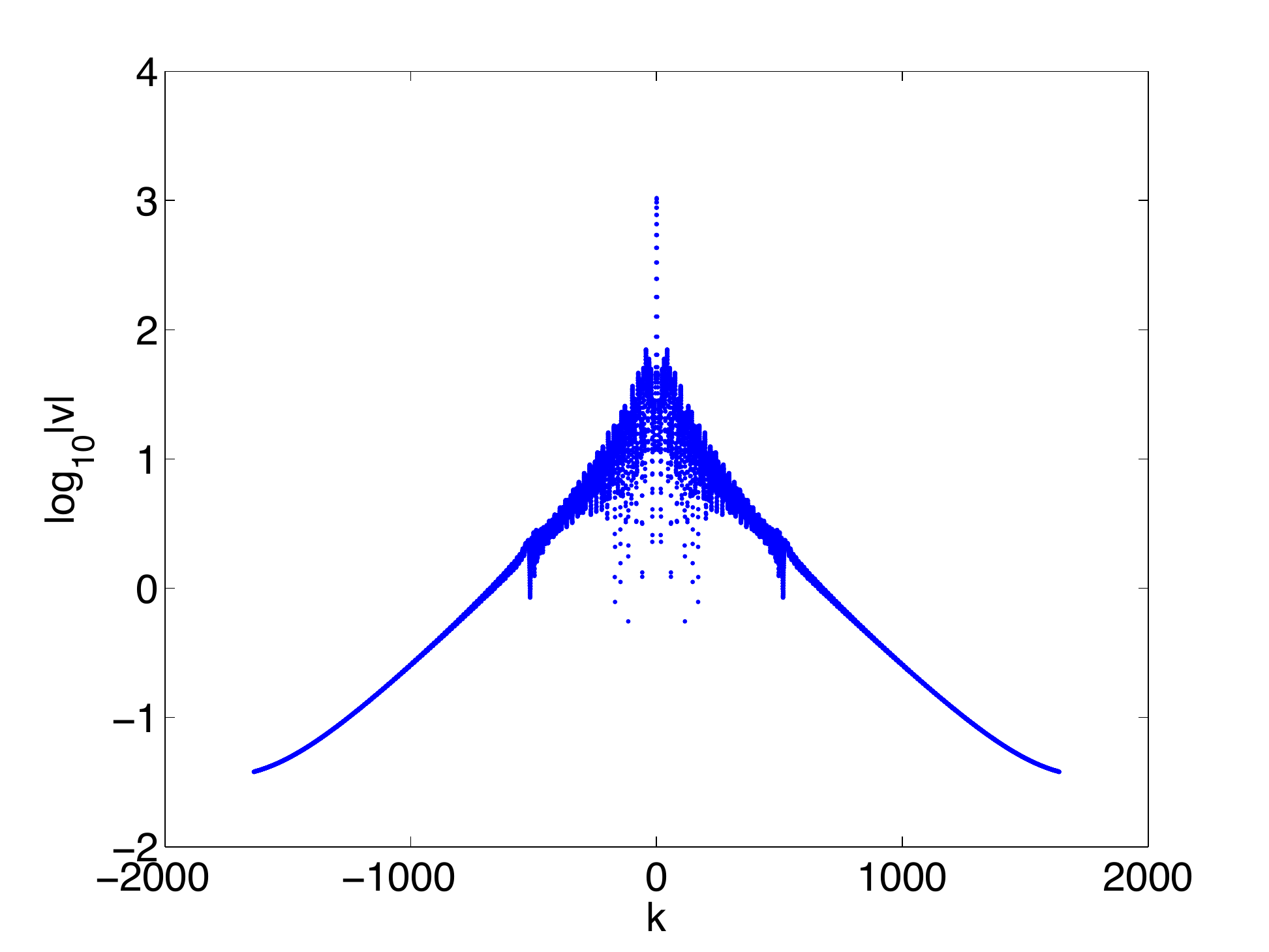}
   \includegraphics[width=0.49\textwidth]{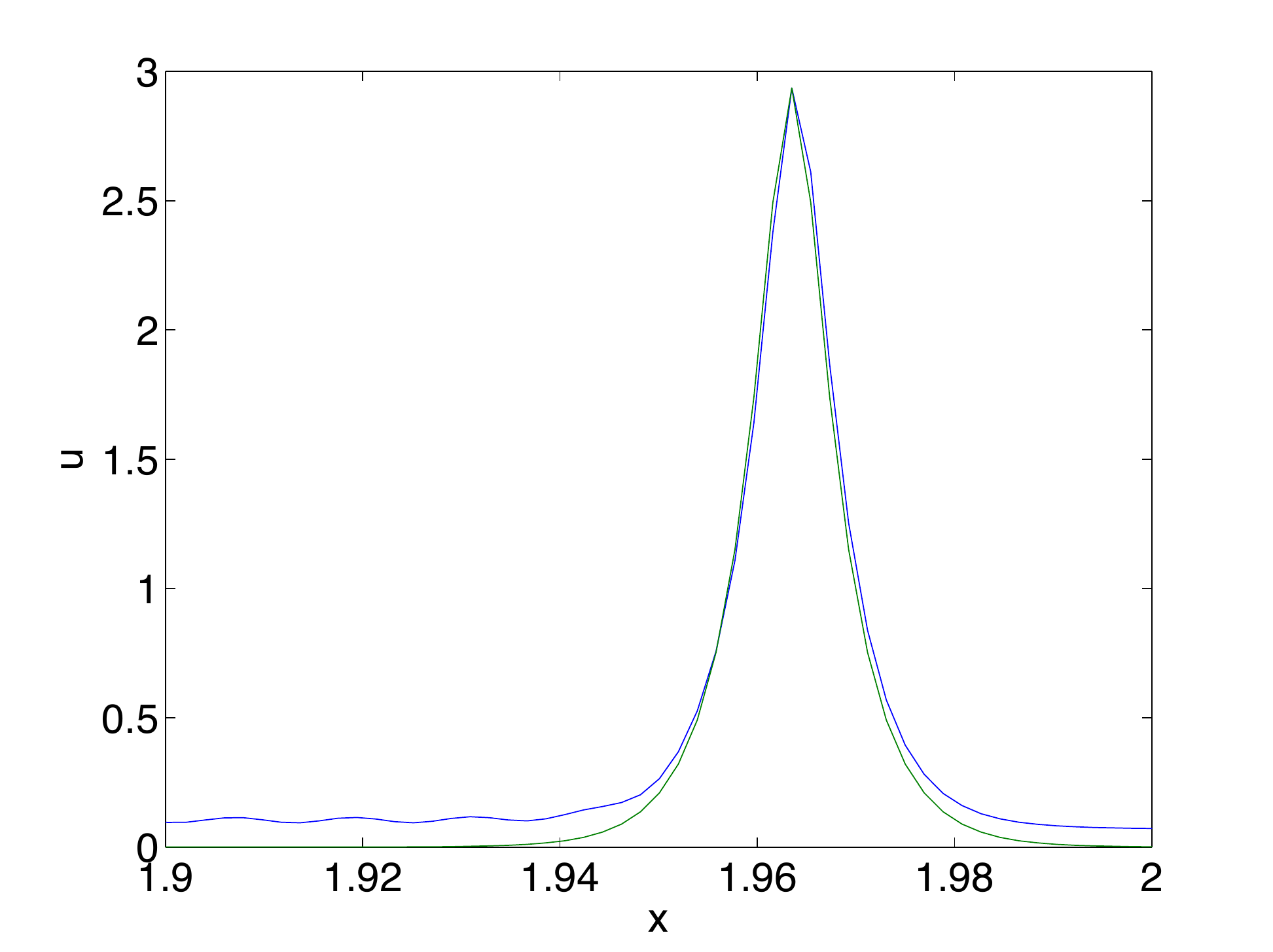}
 \caption{Modulus of the  Fourier coefficients of the solutions 
 to the gKdV equation (\ref{gKdV}) with $n=4$  for $\epsilon=0.01$ 
 for the initial 
 data $u_{0}=\mbox{sech}^{2}x$ at the final times shown in
 Fig.~\ref{gKdVn4sech2e0014t} on the 
 left; on the right a close up to the soliton blowing up at 
 $t=1.6745$ in blue and the rescaled (according to (\ref{selfs}))
 soliton (\ref{soliton}) in green.}
 \label{gKdVn4sech2e01fourier}
\end{figure}

In all cases the codes run out of resolution in Fourier space as can 
be seen in Fig.~\ref{gKdVn4sech2e01fourier} where the modulus of the 
Fourier coefficients is given for the last recorded time with a 
$\Delta<10^{-3}$ for $\epsilon=0.01$.  In the same figure we show a 
close up of the `soliton' (in  as light abuse of notation, we call 
the fist oscillation in the following soliton though the latter is 
not stable and will eventually blow up) blowing up in Fig.~\ref{gKdVn4sech2e0014t} at the last 
recorded time and the rescaled (according to (\ref{selfs}) with an 
appropriate rescaling of $x$ with $\epsilon$) soliton 
(\ref{soliton}). This indicates that the blow-up mechanism of Theorem 
\ref{MMR} applies also to initial data with much larger mass than the 
soliton mass, i.e., to cases not covered by Theorem \ref{MMR}.

\begin{figure}[htb!]
   \includegraphics[width=0.49\textwidth]{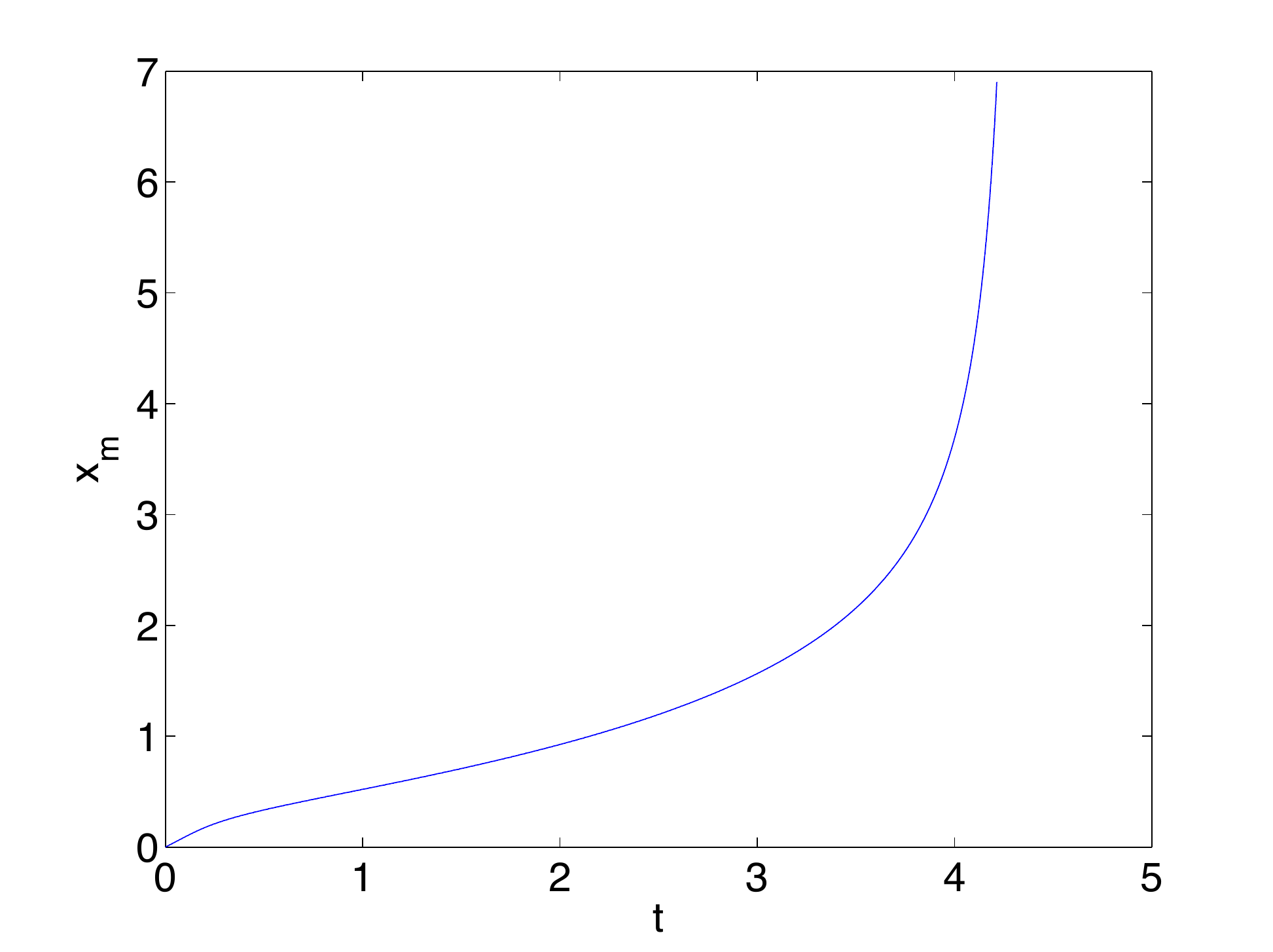}
   \includegraphics[width=0.49\textwidth]{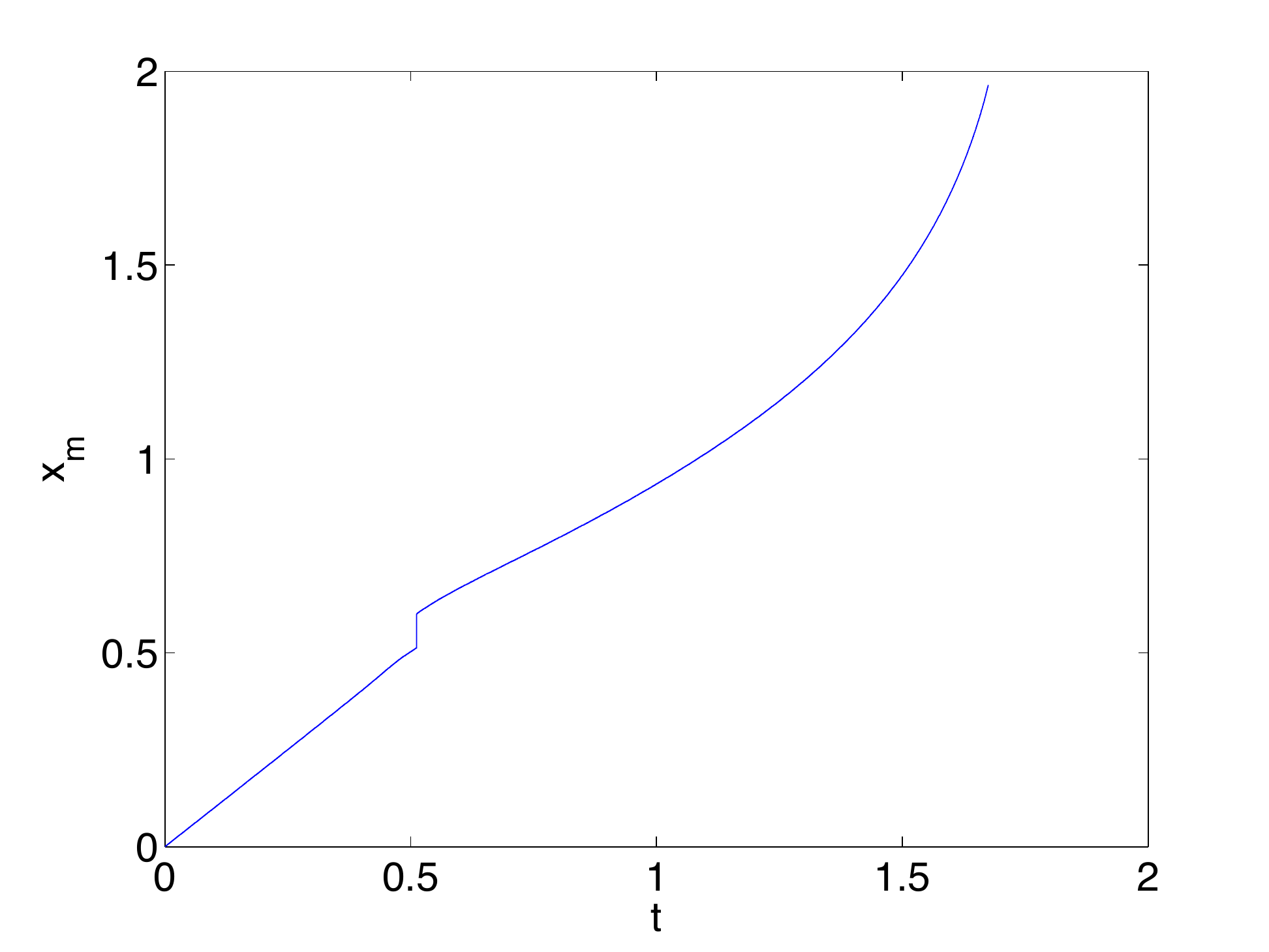}
 \caption{Location of the maximum of the solutions 
 to the gKdV equation (\ref{gKdV}) with $n=4$ for the initial 
 data $u_{0}=\mbox{sech}^{2}x$ at the final times shown in 
 Fig.~\ref{gKdVn4e014t} and Fig.~\ref{gKdVn4sech2e0014t}; on the 
 left for $\epsilon=0.1$, on the right for $\epsilon=0.01$.}
 \label{gKdVn4sech2e01xm}
\end{figure}

The location $x_{m}(t)$ of the maximum for both solutions can be seen 
in Fig.~\ref{gKdVn4sech2e01xm}. It appears that the maximum tends to 
infinity at blow-up as is rigorously established by \cite{MMR2012_I} for 
initial data close to the soliton. Thus this seems to be a rather 
universal feature for localized initial data with a single maximum. 
Note, however, that we cannot compute long enough to check this in 
more detail. It was already seen in section 3 that the predicted 
asymptotics for $x_{m}$ is reached considerably later than the one 
for $||u_{x}||_{2}$. This is not surprising, since the rescaled 
soliton will move faster and faster close to blow-up, a behavior that 
is difficult to catch for any numerical approach.

\begin{figure}[htb!]
   \includegraphics[width=\textwidth]{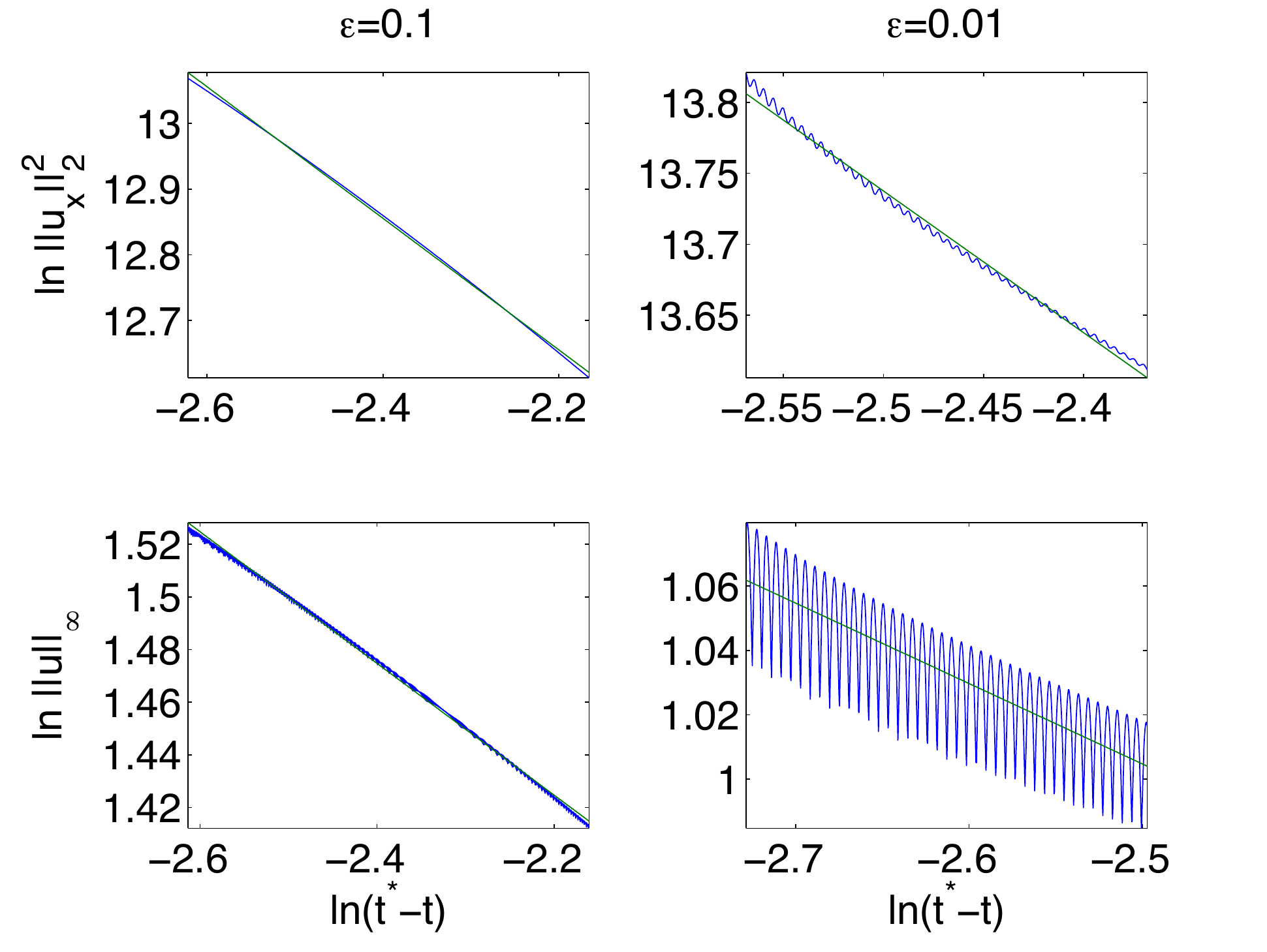}
 \caption{Fitting of $\ln||u_{x}||_{2}^{2}$ and 
 $\ln||u||_{\infty}$ (in blue) to $\alpha \ln(t^{*}-t)+\beta$ (in green) 
 for the situations shown in 
 Fig.~\ref{gKdVn4e014t} and Fig.~\ref{gKdVn4sech2e0014t}.}
 \label{gKdVn4sech2efit}
\end{figure}

In Fig.~\ref{gKdVn4sech2efit} we show the results of the fitting of 
the logarithm of the various norms to  $\alpha\ln (t^{*}-t)+\beta$. For 
$\epsilon=0.1$ we get $t^{*}=4.2727$ and $\beta=10.4555$ for     
$||u_{x}||_{2}^{2}$ and $t^{*}=4.2733$ and $\beta=0.8747$ for     
$||u_{x}||_{\infty}$ which shows a remarkable consistency between the 
two values obtained for $t^{*}$. It can be seen that the fitting is 
worst at early times where the asymptotic regime is not yet reached, and 
also for $t\sim t^{*}$ where the accuracy is getting smaller. 
It can be also recognized that there are small 
oscillations in the $L_{\infty}$ norm of $u$. This is due to the 
dispersive oscillations propagating to the left and reentering at the 
right due to the imposed periodicity of the problem, and to the fact 
that we only determine the maximum on the grid points. This is even 
more visible for $\epsilon=0.01$, where the oscillations in the 
$L_{2}$ norm of $u_{x}$ are considerably smaller than for 
$||u||_{\infty}$. The fitting gives $t^{*}=1.7511$ and 
$\beta=11.2375$ in the former case, and $t^{*}=1.7398$ and 
$\beta=0.3797$ in the latter. Note again the coincidence 
of the values for $t^{*}$ despite the strong oscillations in 
$||u||_{\infty}$.     

\begin{figure}[htb!]
   \includegraphics[width=.49\textwidth]{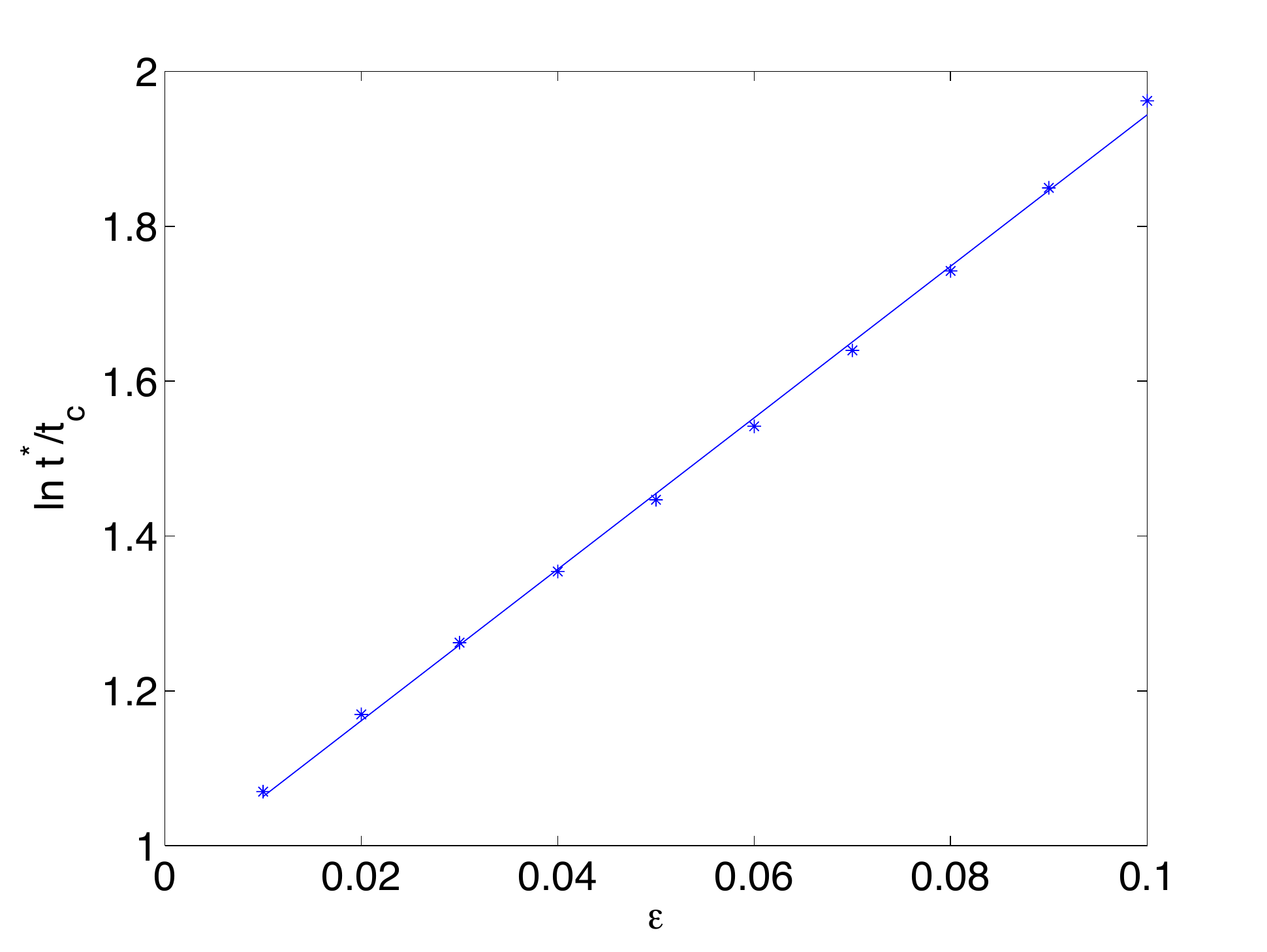}
   \includegraphics[width=0.49\textwidth]{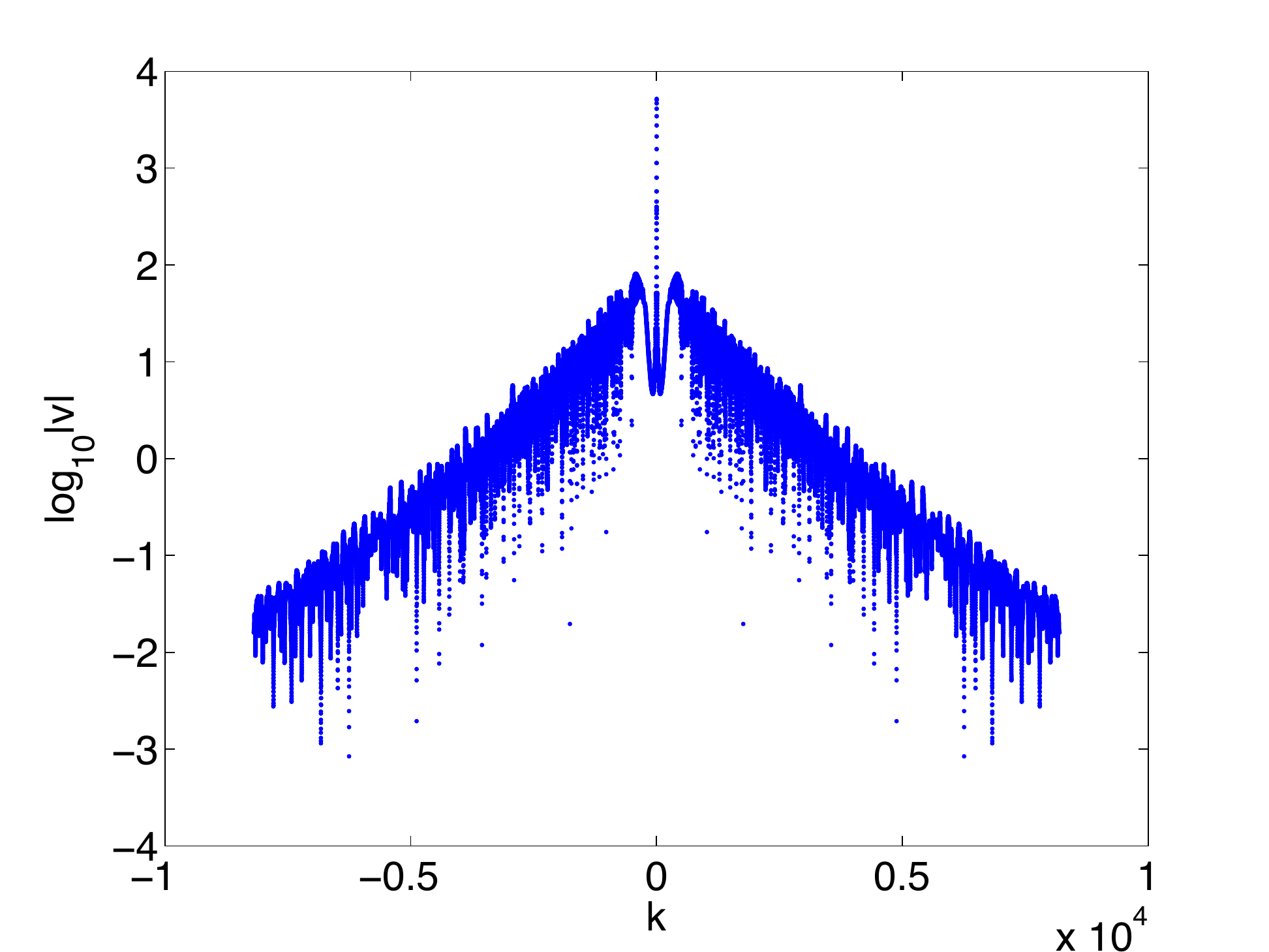}
 \caption{Least square fit of $\ln (t^{*}/t_{c})$ to 
 $\gamma\epsilon+\delta$ for solutions to the gKdV equation 
 (\ref{gKdV}) and $n=4$ on the left; on the right the 
 modulus of the Fourier coefficients for the solution in 
 Fig.~\ref{gKdVn4sech2e0001u}.}
 \label{gKdVn4sech2tfit}
\end{figure}

This procedure is repeated for all values of $\epsilon$ we consider. 
The resulting values of $t^{*}(\epsilon)$ are then further analyzed. 
If the conjectured time dependence in \cite{DGK2011} proportional to 
$\epsilon^{4/7}$ close to the critical time $t_{c}$ 
were relevant up to blow-up, one would expect an algebraic 
dependence of $t^{*}$ on $\epsilon$. But as already mentioned, this 
cannot really be assumed since the solution of the PI2 equation which 
is relevant for the asymptotic description close to $t_{c}$ is 
regular on the whole real line. In fact we find that a 
fitting of $\ln(t^{*}(\epsilon)-t_{c})$ to $\gamma\ln \epsilon+\delta$ does not 
lead to a good correlation. Instead we get as shown in 
Fig.~\ref{gKdVn4sech2tfit} that there is an exponential dependence of 
$t^{*}$ on $\epsilon$. This means that a least square analysis for 
$\ln (t^{*}/t_{c})$ to $\gamma\epsilon +\delta$ gives $\gamma=9.7807$ 
    and $\delta=0.9658$ with standard deviation $\sigma_{\gamma}=0.0068$ 
and a correlation coefficient $0.9995$. 

\begin{figure}[htb!]
   \includegraphics[width=\textwidth]{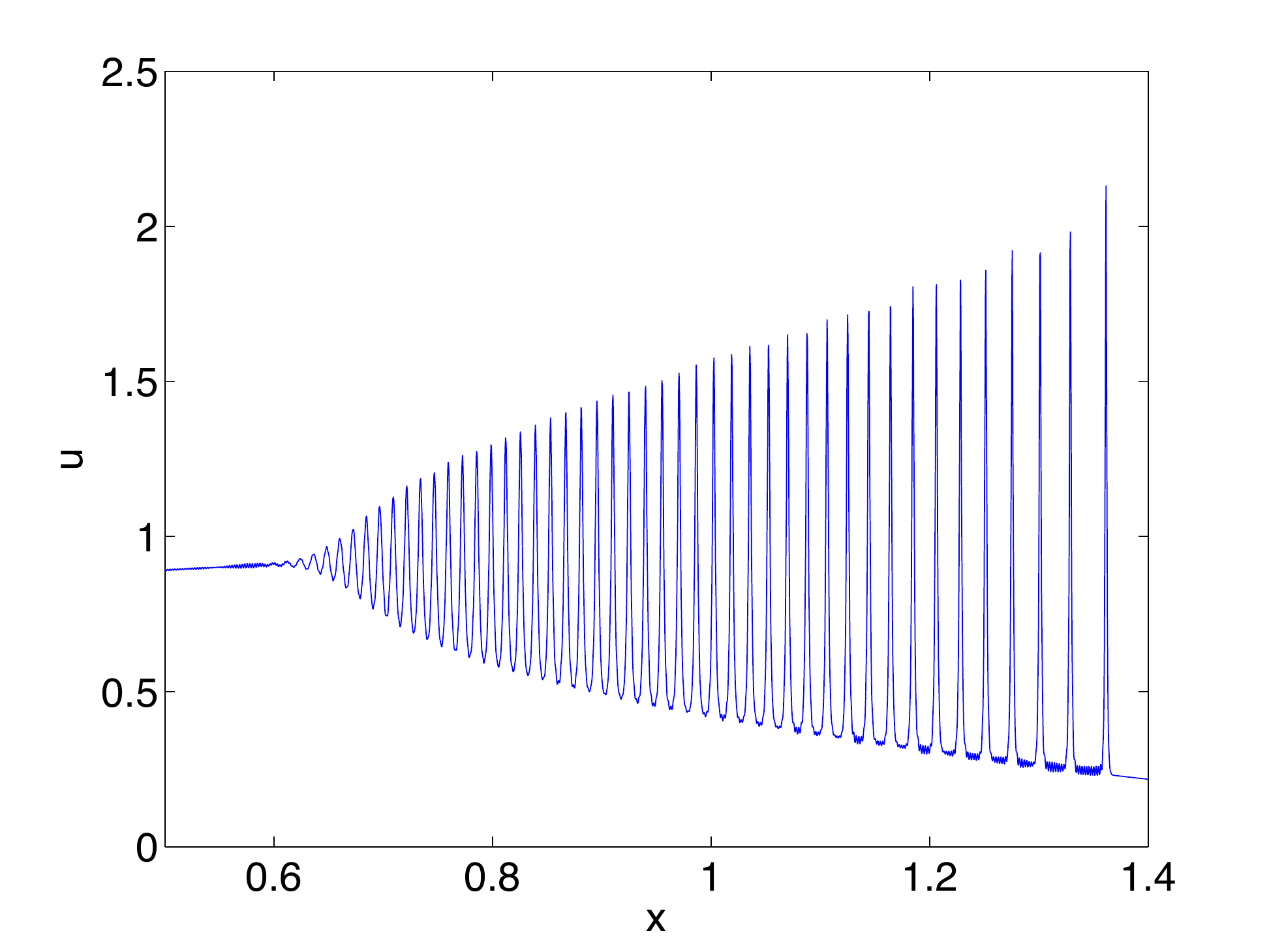}
 \caption{Solution to the gKdV equation (\ref{gKdV}) with $n=4$ for the initial 
 data $u_{0}=\mbox{sech}^{2}x$ and $\epsilon=0.001$ at the final time 
 ($t= 1.33$)
 for which $\Delta<10^{-3}$.}
 \label{gKdVn4sech2e0001u}
\end{figure}

This implies not only that $t^{*}>t_{c}$, but that $\lim_{\epsilon\to0}t^{*}
=t^{*}_{0}\sim 1.5780>t_{c}\sim0.6007$. In other words there will be always a gap between the 
critical time of the generalized Hopf solution and the blow-up time 
of the corresponding gKdV solution. This is intuitively clear if one 
looks at the gKdV solutions for different values of $\epsilon$. The 
smaller $\epsilon$, the more oscillations form before the rightmost 
soliton blows up. This becomes even more obvious when one looks at 
even smaller values of $\epsilon$ as in Fig.~\ref{gKdVn4sech2e0001u}. 
It can be seen that a zone of rapid modulated oscillations as in KdV 
forms, see for instance \cite{GK07}, for which the rightmost peak 
eventually blows up. The smaller $\epsilon$, the more oscillations 
visibly form before this happens, but it appears that the $t^{*}$ 
tends to a constant for small $\epsilon$. The calculation is carried 
out with $N=2^{16}$ Fourier modes on $x\in4[-\pi,\pi]$ with 
$N_{t}=10^{5}$ time steps. The above fitting procedure for the last 
5000 time steps gives for this 
case a blow-up time $t^{*}=1.5427$ which is roughly in accordance 
with expectations given the numerical inaccuracies (we do not get 
close enough to blow-up in this case). 

The small oscillations which can be seen between the peaks 
appear to be dispersive oscillations shed from the 
onset of blow-up. 
These oscillations are related to the fact that several 
singularities in the complex plane approach the real axis in this 
case. As is well known, this leads to oscillations in the Fourier 
spectrum, see for instance \cite{KR13} for the small dispersion limit 
of KdV. The strong oscillations in the Fourier spectrum are clearly 
visible in Fig.~\ref{gKdVn4sech2tfit} on the right.

\subsection{The supercritical case $n=5$}
\begin{figure}[htb!]
   \includegraphics[width=\textwidth]{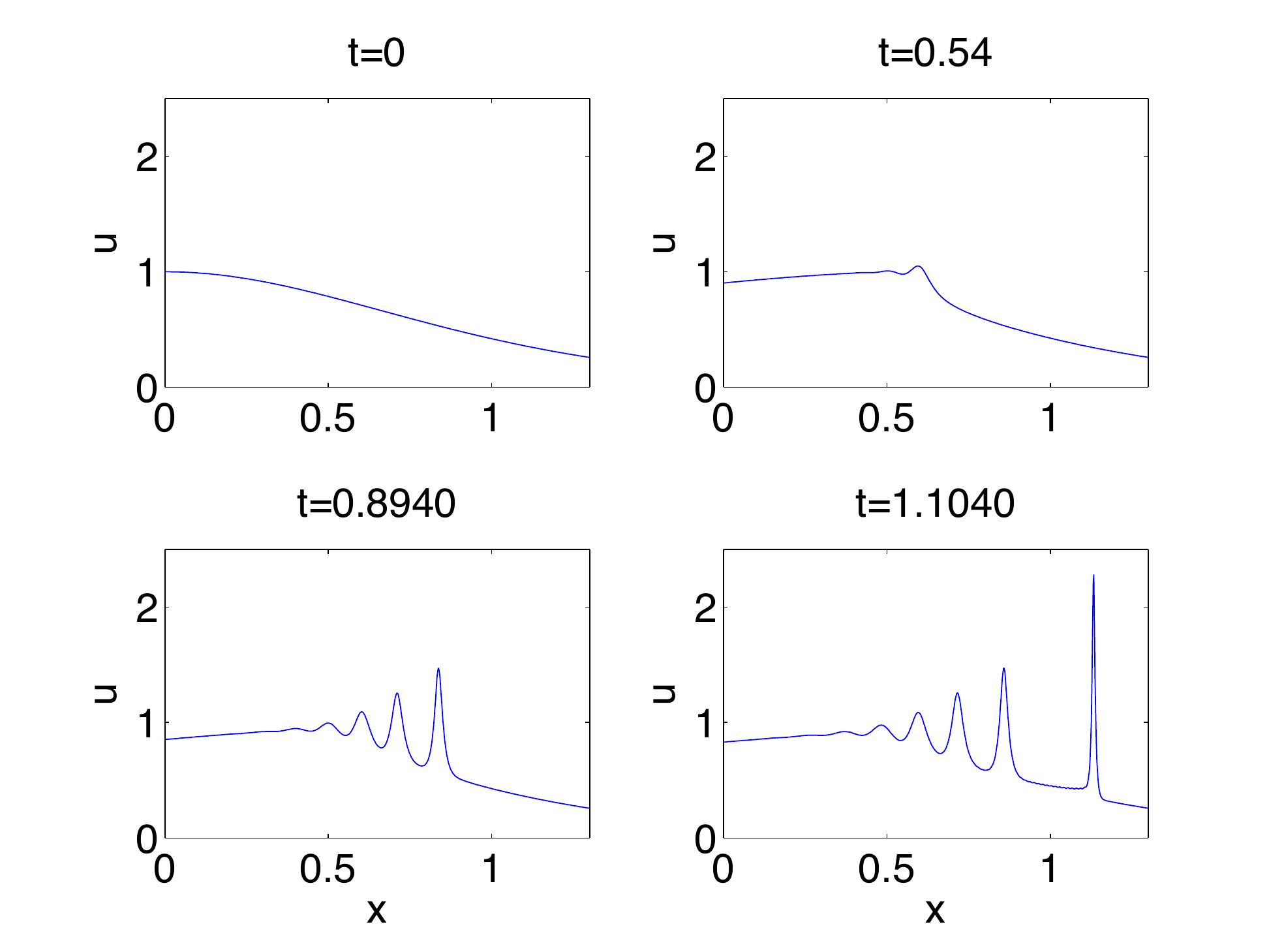}
 \caption{Solution to the gKdV equation (\ref{gKdV}) with $n=5$ for the initial 
 data $u_{0}=\mbox{sech}^{2}x$ and $\epsilon=0.01$ for different times.}
 \label{gKdVn5sech2e0014t}
\end{figure}

In this subsection we repeat the considerations for gKdV with $n=4$ 
for $n=5$. The same approaches and parameters are used for the numerical experiments 
as before.
In Fig.~\ref{gKdVn5e014t} the solution for gKdV for the 
initial data $u_{0}=\mbox{sech}^{2}x$ for several values of $t$ was 
shown. 
As for $n=4$ a first oscillation forms close to the 
critical time $t_{c}\sim0.5341$ (\ref{tch}). This soliton will eventually 
blow up, and the rest of the hump appears to be radiated away. 
The same situation is shown for $\epsilon=0.01$ in 
Fig.~\ref{gKdVn5sech2e0014t}. Whereas the type of blow-up is the same 
as in Fig.~\ref{gKdVn5e014t},
several oscillations form before.

\begin{figure}[htb!]
   \includegraphics[width=0.49\textwidth]{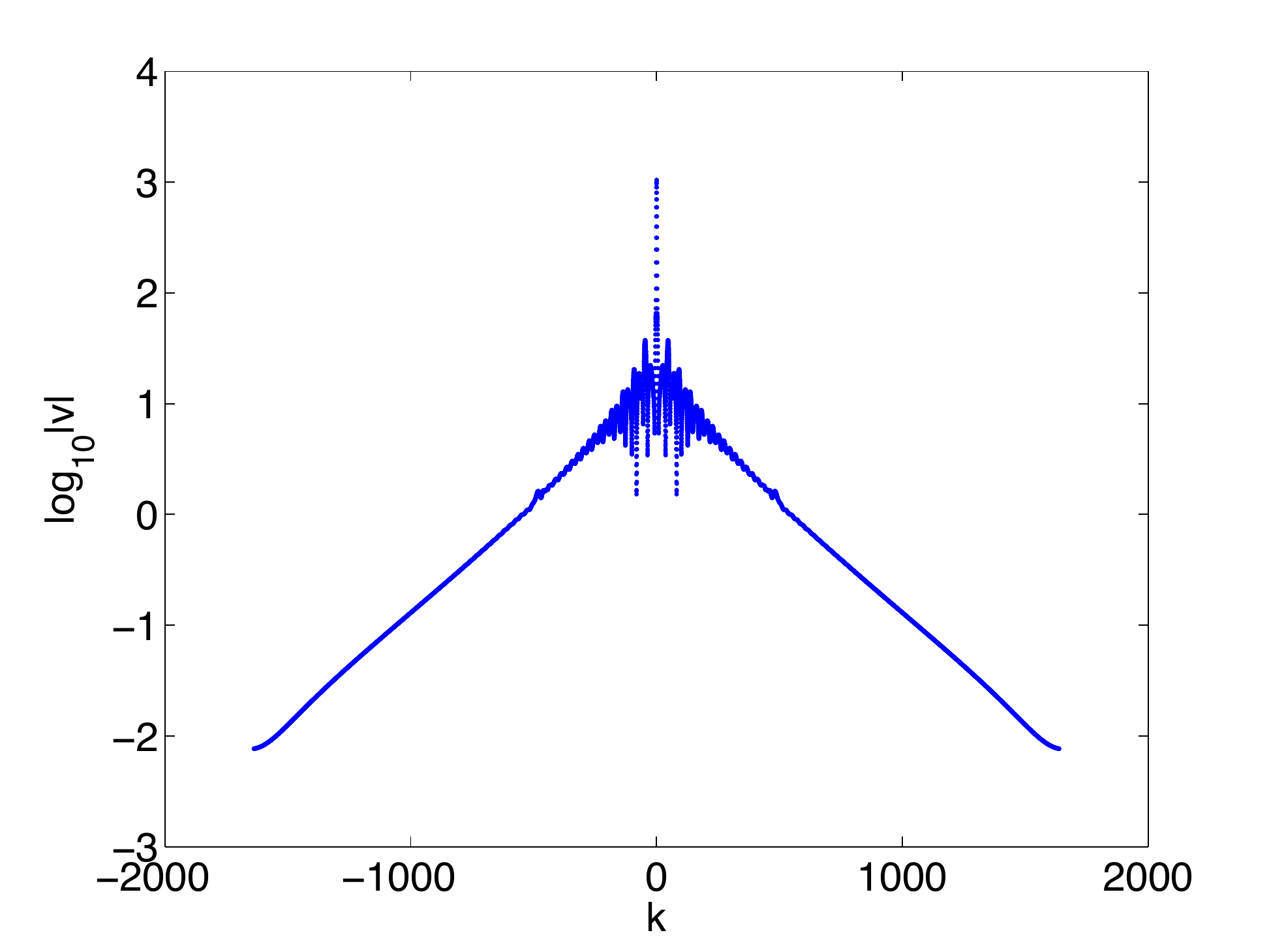}
   \includegraphics[width=0.49\textwidth]{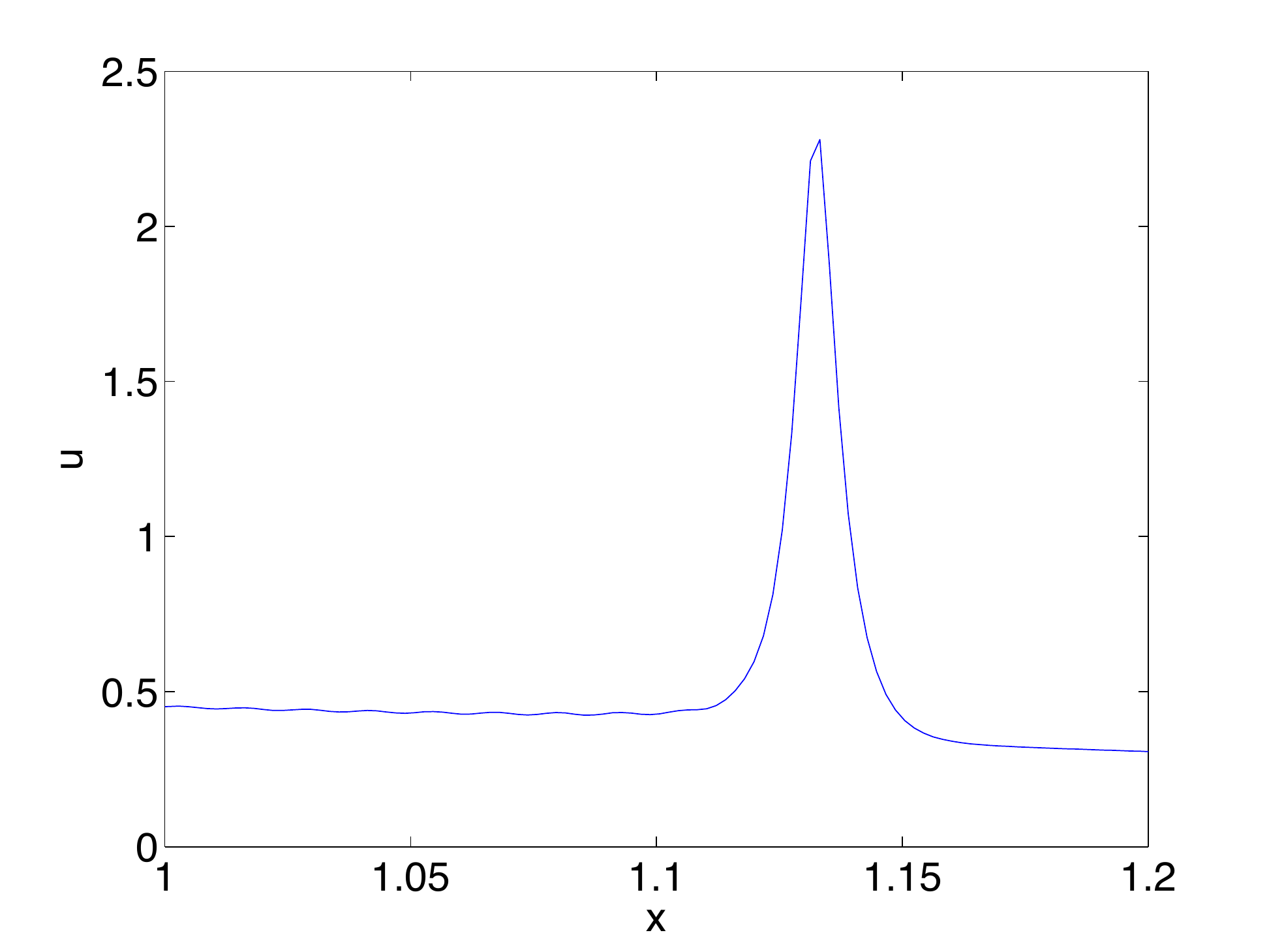}
 \caption{Modulus of the  Fourier coefficients of the solutions 
 to the gKdV equation (\ref{gKdV}) with $n=5$ for the initial 
 data $u_{0}=\mbox{sech}^{2}x$ for $\epsilon=0.01$ at the final times shown in 
 Fig.~\ref{gKdVn5sech2e0014t} on the 
 left; close up of the soliton blowing up in Fig.~\ref{gKdVn5sech2e0014t}  on the right.}
 \label{gKdVn5sech2e01fourier}
\end{figure}

Again the codes run out of resolution in Fourier space as can 
be seen on the left in Fig.~\ref{gKdVn5sech2e01fourier} where the modulus of the 
Fourier coefficients is given for  $\epsilon=0.01$ 
for the last recorded time with a 
$\Delta<10^{-3}$. The blow-up is reached here more rapidly than in 
the case 
$n=4$ due to the exponential dependence of the scaling factor $L$ on 
the rescaled time $\tau$. A close up on the right in 
Fig.~\ref{gKdVn5sech2e01fourier} of the rightmost soliton in 
Fig.~\ref{gKdVn5sech2e0014t} 
illustrates the blow-up profile which is 
different from the case $n=4$ in Fig.~\ref{gKdVn4sech2e01fourier}. 

\begin{figure}[htb!]
   \includegraphics[width=0.49\textwidth]{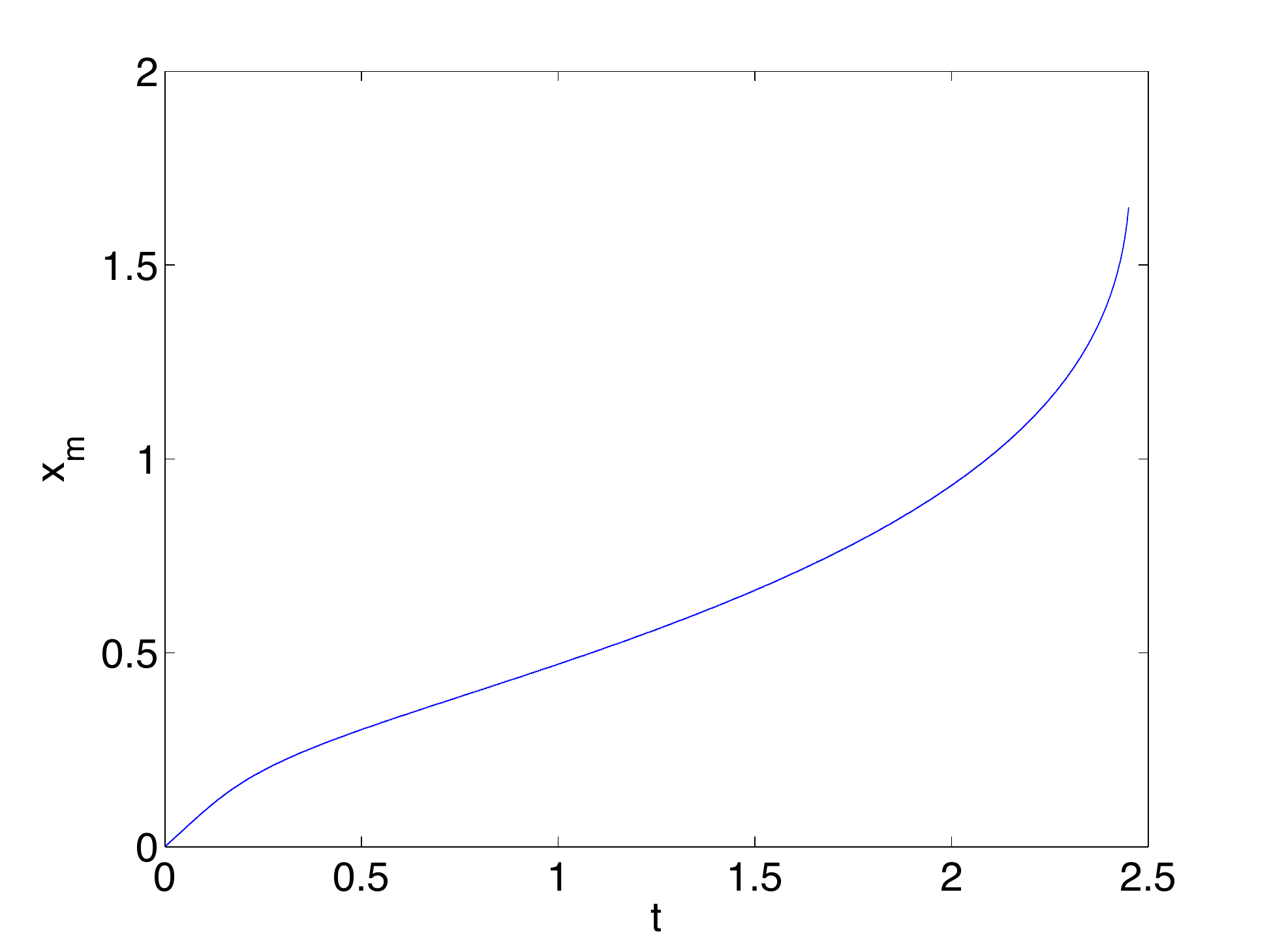}
   \includegraphics[width=0.49\textwidth]{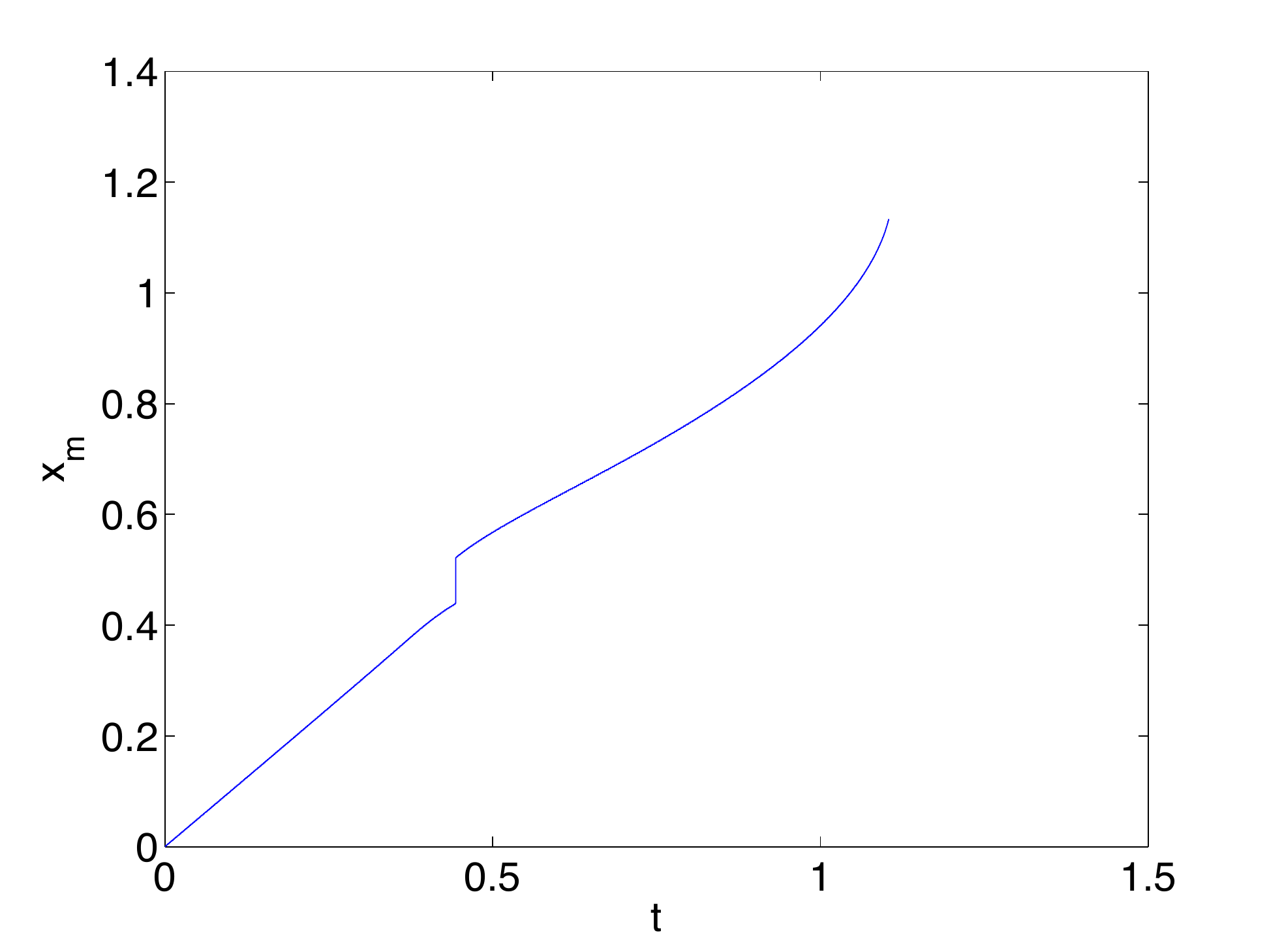}
 \caption{Location of the maximum of the solutions 
 to the gKdV equation (\ref{gKdV}) with $n=5$ for the initial 
 data $u_{0}=\mbox{sech}^{2}x$ at the final times shown in 
 Fig.~\ref{gKdVn5e014t} and Fig.~\ref{gKdVn5sech2e0014t}; on the 
 left for $\epsilon=0.1$, on the right for $\epsilon=0.01$.}
 \label{gKdVn5sech2e01xm}
\end{figure}

The location $x_{m}(t)$ of the maximum for both solutions can be seen 
in Fig.~\ref{gKdVn5sech2e01xm}. Here it is expected from numerical 
results in \cite{DixMcKinney} (see also section 4) that the blow-up 
will occur at finite values of $x$. But again we are not able to 
compute long enough to obtain conclusive results for $x^{*}$.

\begin{figure}[htb!]
   \includegraphics[width=\textwidth]{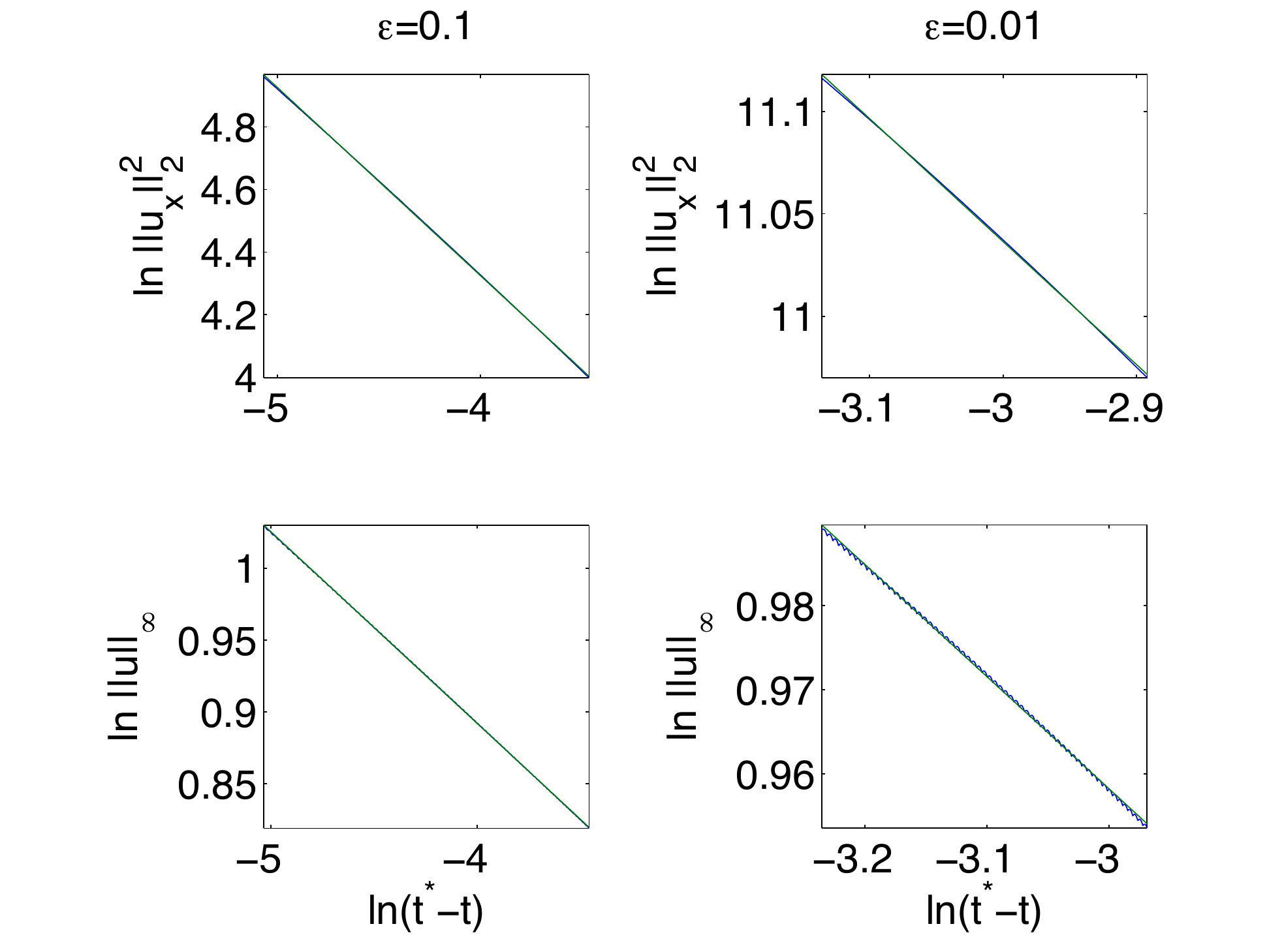}
 \caption{Fitting of $\ln||u_{x}||_{2}^{2}$ and 
 $\ln||u||_{\infty}$ (in blue) to $\alpha \ln(t^{*}-t)+\beta$ (in green) 
 for the situations shown in 
 Fig.~\ref{gKdVn5e014t} and Fig.~\ref{gKdVn5sech2e0014t}.}
 \label{gKdVn5sech2efit}
\end{figure}

In Fig.~\ref{gKdVn5sech2efit} we show the results of the fitting of 
the logarithm of the various norms to  $\alpha\ln (t^{*}-t)+\beta$. For 
$\epsilon=0.1$ we get $t^{*}=2.4563$ and $\beta=1.9253$ for 
$||u_{x}||_{2}^{2}$ and $t^{*}=2.4565$ and $\beta=0.3587$ for 
$||u_{x}||_{\infty}$. Again the values for $t^{*}$ are consistent.
Once more the fitting is 
worst at early times where the asymptotic regime is not yet reached, and 
also for $t\sim t^{*}$ where the accuracy is getting smaller. But 
overall the agreement is much better than for $n=4$ which is due to 
the fact that the blow-up is reached in the rescaled time $\tau$ 
exponentially, whereas this happens algebraically for $n=4$. Also the 
oscillations in the norms are much smaller than in the latter case. 
For $\epsilon=0.01$, the fitting gives $t^{*}=1.1475$ and 
$\beta=9.2364$ for $\ln||u_{x}||_{2}^{2}$, and $t^{*}=1.1434$ and 
$\beta=0.55821$ for $\ln||u||_{\infty}$. Note again the coincidence 
of the values for $t^{*}$.

The resulting values of $t^{*}(\epsilon)$ for the remaining values of 
$\epsilon$ are considered in a least square fit. As for $n=4$ 
we do not obtain good correlation for an algebraic dependence of 
$t^{*}$ on $\epsilon$. But we find also here that a 
fitting of
$\ln (t^{*}/t_{c})$ to $\gamma\epsilon +\delta$ gives $\gamma=8.4596$ 
and $\delta=     0.6794$ with standard deviation $\sigma_{\gamma}=0.0031$ 
and a correlation coefficient $r=0.9999$. 
\begin{figure}[htb!]
   \includegraphics[width=.49\textwidth]{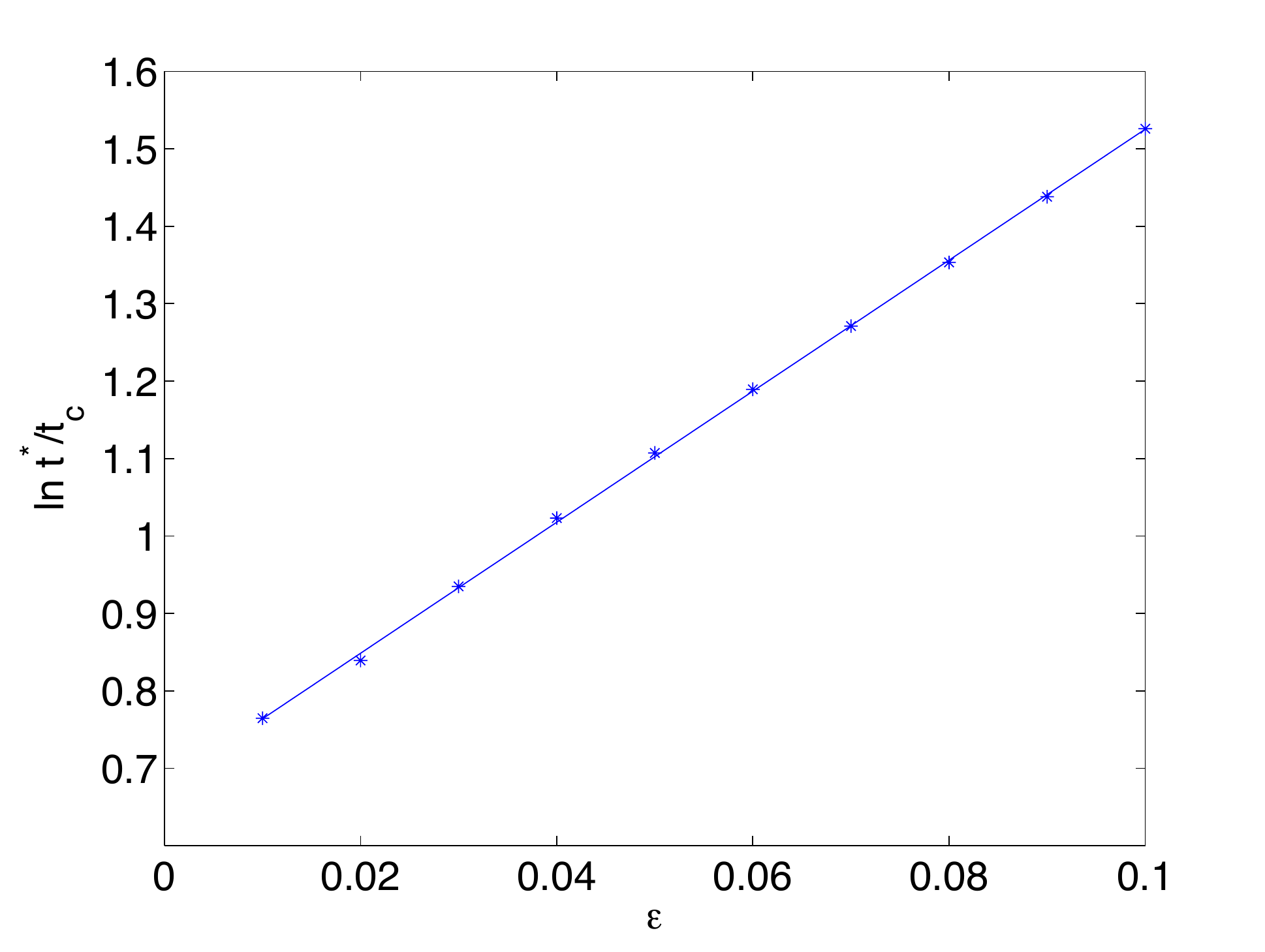}
   \includegraphics[width=0.49\textwidth]{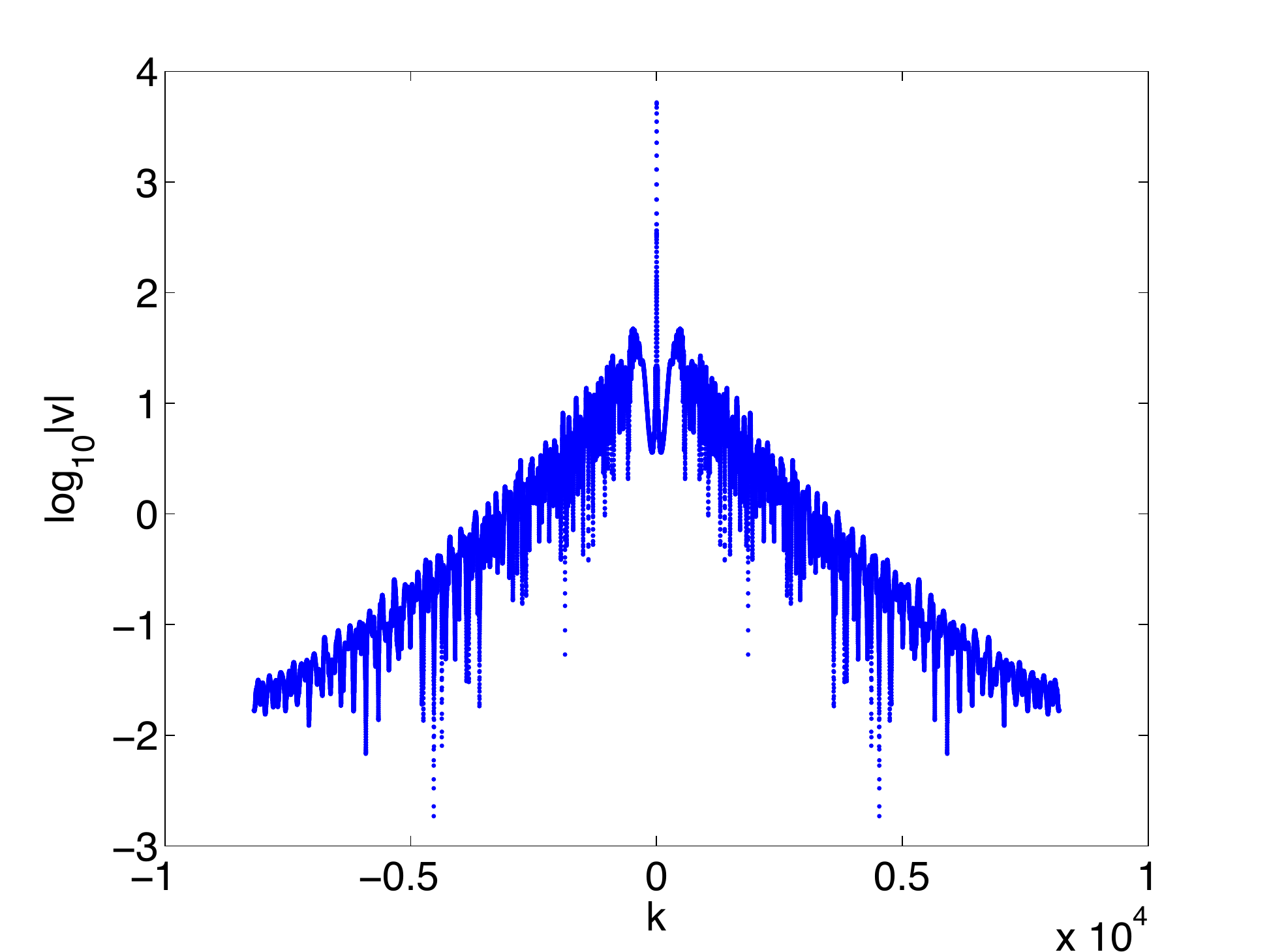}
 \caption{Least square fit of $\ln (t^{*}/t_{c})$ to 
 $\gamma\epsilon+\delta$ for solutions to the gKdV equation 
 (\ref{gKdV}) and $n=5$ on the left; the modulus of the Fourier coefficients for the solution in 
 Fig.~\ref{gKdVn5sech2e0001u} on the right.}
 \label{gKdVn5sech2tfit}
\end{figure}

\begin{figure}[htb!]
   \includegraphics[width=\textwidth]{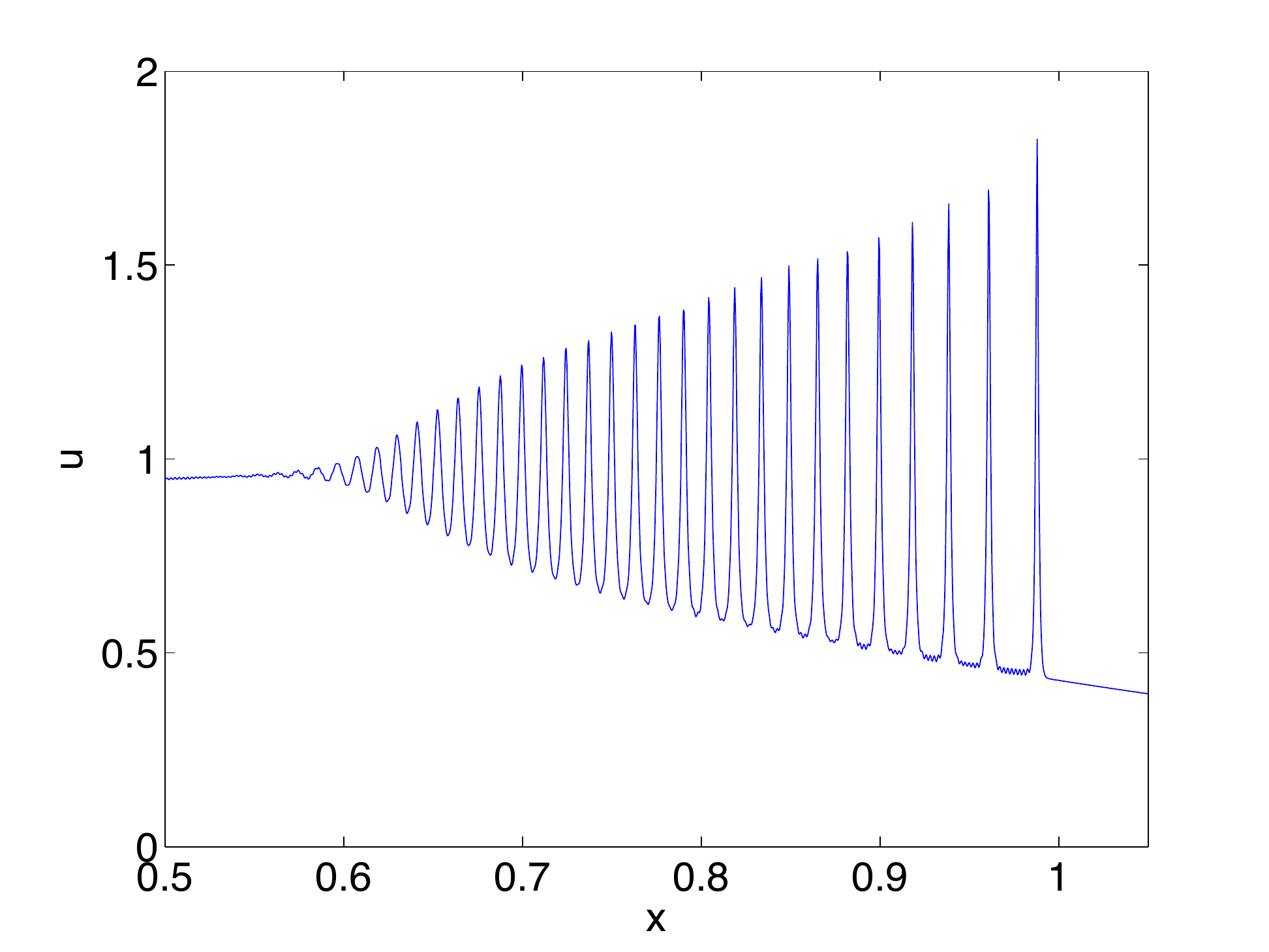}
 \caption{Solution to the gKdV equation (\ref{gKdV}) with $n=5$ for the initial 
 data $u_{0}=\mbox{sech}^{2}x$ and $\epsilon=0.001$ at the final time 
 for which $\Delta<10^{-3}$.}
 \label{gKdVn5sech2e0001u}
\end{figure}

This implies  that $\lim_{\epsilon\to0}t^{*}
=t^{*}_{0}\sim 1.0536>t_{c}\sim0.5341$. Thus there will be also here a gap 
between the 
critical time of the generalized Hopf solution and the blow-up time 
of the corresponding gKdV solution.  In Fig.~\ref{gKdVn5sech2e0001u} 
we show the solution for gKdV with $n=5$ for the initial data 
$u_{0}=\mbox{sech}^{2}x$ at the final recorded time before 
blow-up for $\epsilon=0.001$. Again a zone of rapid modulated 
oscillations forms before the rightmost oscillation blows up.
There are once more small oscillations between the peaks 
marking the onset of the self 
similar blow-up. They are related again to 
oscillations in the Fourier spectrum as can be seen 
on the right in Fig.~\ref{gKdVn5sech2tfit}.

\bibliographystyle{elsart-num-sort}
\bibliography{Num_gKdV_bib}

\begin{thebibliography}{10}
\expandafter\ifx\csname url\endcsname\relax
  \def\url#1{\texttt{#1}}\fi
\expandafter\ifx\csname urlprefix\endcsname\relax\def\urlprefix{URL }\fi

\bibitem{BDK1986}
J.~L. Bona, V.~A. Dougalis, O.~A. Karakashian, Fully discrete {G}alerkin
  methods for the {K}orteweg-de~{V}ries equation, Comp. \& Maths. with Appls.
  12A~(7) (1986) 859--884.

\bibitem{BDKMcK1992}
J.~L. Bona, V.~A. Dougalis, O.~A. Karakashian, W.~R. Mc{K}inney, Computations
  of blow-up and decay for periodic solutions of the generalized
  {K}orteweg-de~{V}ries-{B}urgers equation, Appl. Num. Maths. 10~(3--4) (1992)
  335--355.

\bibitem{BDKMcK1995}
J.~L. Bona, V.~A. Dougalis, O.~A. Karakashian, W.~R. Mc{K}inney, Conservative,
  high-order numerical schemes for the generalized {K}orteweg-de~{V}ries
  equation, Phil. Trans. R. Soc. Lond. A 351~(1695) (1995) 107--164.

\bibitem{BSW1987}
J.~L. Bona, P.~E. Souganidis, W.~A. Strauss, Stability and instability of
  solitary waves of {K}orteweg-de~{V}ries type, Proc. R. Soc. Lond. A 411
  (1987) 395--412.

\bibitem{BonaWeissler1999}
J.~L. Bona, F.~B. Weissler, Similarity solutions of the generalized
  {K}orteweg-de~{V}ries equation, Math. Proc. Camb. Phil. Soc. 127~(2) (1999)
  323--351.

\bibitem{CoxMatthews2002}
S.~M. Cox, P.~C. Matthews, Exponential time differencing for stiff systems, J.
  Comp. Phys. 176 (2002) 430--455.

\bibitem{DixMcKinney}
D.~B. Dix, W.~R. Mc{K}inney, Numerical computations of self-similar blow-up
  solutions of the generalized {K}orteweg-de~{V}ries equations, Differ.
  Integral Equ. 11~(5) (1998) 679--723.

\bibitem{Dub06}
B.~Dubrovin, On {H}amiltonian perturbations of hyperbolic systems of
  conservation laws, {II}: universality of critical behaviour, Comm. Math.
  Phys. 267 (2006) 117 --139.

\bibitem{DGK2011}
B.~Dubrovin, T.~Grava, C.~Klein, Numerical study of breakup in generalized
  {K}orteweg-de {V}ries and {K}awahara equations, SIAM J. Appl. Math. 71~(4)
  (2011) 983--1008.

\bibitem{DGK2013}
B.~Dubrovin, T.~Grava, C.~Klein, A.~Moro, On critical behaviour in systems of
  {H}amiltonian partial differential equations, Preprint available at:
  \href{http://arxiv.org/abs/1204.4625/}{\texttt{arXiv:1204.4625}}.

\bibitem{GK07}
T.~Grava, C.~Klein, Numerical solution of the small dispersion limit of
  {K}orteweg de {V}ries and {W}hitham equations, Comm. Pure Appl. Math. 60~(11)
  (2007) 1623--1664.

\bibitem{HO}
M.~Hochbruck, A.~Ostermann, Exponential {R}unge-{K}utta methods for semilinear
  parabolic problems, SIAM J. Numer. Anal. 43 (2005) 1069--1090.

\bibitem{HO09}
M.~Hochbruck, A.~Ostermann, Exponential integrators, Acta Numerica 19 (2010)
  209--286.

\bibitem{KassamTrefethen2005}
A.-K. Kassam, L.~N. Trefethen, Fourth order time-stepping for stiff pdes, SIAM
  J. Sci. Comput. 26~(4) (2005) 1214--1233.

\bibitem{Kato83}
T.~Kato, On the {C}auchy problem for the (generalized) {K}orteweg-de~{V}ries
  equation, in: I.~E. Segal, V.~Guillemin (eds.), Studies in applied
  mathematics: A volume dedicated to {I}rving {S}egal, vol.~8 of Advances in
  Mathematics: Supplementary Studies, Academic Press, 1983.

\bibitem{KPV93}
C.~E. Kenig, G.~Ponce, L.~Vega, Well-posedness and scattering results for the
  generalized {K}orteweg-de~{V}ries equation via the contraction principle,
  Comm. Pure Appl. Math. 46 (1993) 527--620.

\bibitem{Klein2008}
C.~Klein, Fourth order time-stepping for low dispersion {K}orteweg-de {V}ries
  and nonlinear {S}chr{\"o}dinger equations, ETNA 29 (2008) 116--135.

\bibitem{KP14}
C.~Klein, R.~Peter, Numerical study of blow-up in solutions to generalized
  {K}adomtsev-{P}etviashvili equations, Discr. Cont. Dyn. Syst. B 19(6) (2014)
  doi:10.3934/dcdsb.2014.19.1689.

\bibitem{KleinRoidot2011}
C.~Klein, K.~Roidot, Fourth order time-stepping for {K}adomtsev-{P}etviashvili
  and {D}avey-{S}tewardson equations, SIAM J. Sci. Comput. 33~(6) (2011)
  3333--3356.

\bibitem{KR13}
C.~Klein, K.~Roidot, Numerical study of shock formation in the dispersionless
  {K}adomtsev-{P}etviashvili equation and dispersive regularizations, Physica D
  265 (2013) 1--25.

\bibitem{KS2014}
C.~Klein, J.-C. Saut, A numerical approach to blow-up issues for dispersive
  perturbations of {B}urgers' equation, Preprint available at:
  \href{http://arxiv.org/abs/1401.1390/}{\texttt{arXiv:1401.1390}}.

\bibitem{KS2014b}
C.~Klein, J.-C. Saut, A numerical approach to blow-up issues for
  {D}avey-{S}tewartson {II} type systems, Preprint available at:
  \href{http://arxiv.org/abs/1406.1146/}{\texttt{arXiv:1406.1146}}.

\bibitem{KSM14}
C.~Klein, C.~Sparber, P.~Markowich, Numerical study of fractional nonlinear
  {S}chr{\"o}dinger equations, Proc. R. Soc. A 470 (2014) 20140364.

\bibitem{Koch}
H.~Koch, Self similar solution to super-critical g{K}d{V}, Preprint available
  at: \href{http://arxiv.org/abs/1404.4591/}{\texttt{arXiv:1404.4591}}.

\bibitem{KPSZ1988}
N.~E. Kosmatov, I.~V. Petrov, V.~F. Shvets, V.~E. Zakharov, Large amplitude
  simulation of wave collapses in nonlinear {S}chr{\"o}dinger equations,
  preprint from the Space Research Institute Moscow.

\bibitem{KSZ1991}
N.~E. Kosmatov, V.~F. Shvets, V.~E. Zakharov, Computer simulation of wave
  collapses in the nonlinear {S}chr{\"o}dinger equation, Physica D 52~(1)
  (1991) 16--35.

\bibitem{fminsearch}
J.~C. Lagarias, J.~A. Reeds, M.~H. Wright, P.~E. Wright, Convergence properties
  of the {N}elder-{M}ead simplex method in low dimensions, SIAM Journal of
  Optimization 9 (1998) 112--147.

\bibitem{LPSS1988}
M.~J. Landman, G.~C. Papanicolaou, C.~Sulem, P.~L. Sulem, Rate of blowup for
  solutions of the nonlinear {S}chr{\"o}dinger equation at critical dimension,
  Phys. Rev. A 38~(8) (1988) 3837--3843.

\bibitem{LeMPSS1987}
B.~Le{M}esurier, G.~Papanicolaou, C.~Sulem, P.-L. Sulem, The focusing
  singularity of the nonlinear {S}chr{\"o}dinger equation, in: M.~G. Crandall,
  P.~H. Rabinowitz, R.~E.~L. Turner (eds.), Directions in partial differential
  equations, vol.~54 of Mathematics Research Center Symposium, Academic Press,
  1987.

\bibitem{MarMer2000}
Y.~Martel, F.~Merle, A {L}iouville theorem for the critical generalized
  {K}orteweg-de {V}ries equation, J. Maths. Pures Appl. 79~(4) (2000) 339--425.

\bibitem{MarMer2001}
Y.~Martel, F.~Merle, Instability of solitons for the critical generalized
  {K}orteweg-de {V}ris equation, Geom. Funct. Anal. 11 (2001) 74--123.

\bibitem{MarMer2002_2}
Y.~Martel, F.~Merle, Blow up in finite time and dynamics of blow up solutions
  for the {$L^2$}-critical generalized {gKdV} equation, J. Amer. Math. Soc.
  15~(3) (2002) 617--664.

\bibitem{MarMer2002_3}
Y.~Martel, F.~Merle, Nonexistence of of blow-up solution with minimal
  $l^2$-mass for the critical {gKdV} equation, Duke Math J. 115~(2) (2002)
  385--408.

\bibitem{MarMer2002_1}
Y.~Martel, F.~Merle, Stability of blow-up profile and lower bounds for blow-up
  rate for the critical generalized {KdV} equation, Ann. Math., Sec. Series
  155~(1) (2002) 235--280.

\bibitem{MMR2012_I}
Y.~Martel, F.~Merle, P.~Rapha{\"e}l, Blow up for the critical {gKdV} equation
  {I}: {D}ynamics near the solition, Preprint available at:
  \href{http://arxiv.org/abs/1204.4625/}{\texttt{arXiv:1204.4625}}.

\bibitem{MMR2012_II}
Y.~Martel, F.~Merle, P.~Rapha{\"e}l, Blow up for the critical {gKdV} equation
  {II}: {M}inimal mass dynamics, Preprint available at:
  \href{http://arxiv.org/abs/1204.4624/}{\texttt{arXiv:1204.4624}}.

\bibitem{MMR2012_III}
Y.~Martel, F.~Merle, P.~Rapha{\"e}l, Blow up for the critical {gKdV} equation
  {III}: {E}xotic regimes, Preprint available at:
  \href{http://arxiv.org/abs/1209.2510/}{\texttt{arXiv:1209.2510}}.

\bibitem{McLPSS1986}
D.~W. Mc{L}aughlin, G.~C. Papanicolaou, C.~Sulem, P.~L. Sulem, Focusing
  singularity of the cubic {S}chr{\"o}dinger equation, Phys. Rev. A 34~(2)
  (1986) 1200--1210.

\bibitem{Merle2001}
F.~Merle, Existence of blow-up solutions in the energy space for the critical
  generalized {KdV} equation, J. Amer. Math. Soc. 14~(3) (2001) 555--578.

\bibitem{PSSW1993}
G.~Papanicolaou, C.~Sulem, P.~L. Sulem, X.~P. Wang, Dynamic rescaling for
  tracking point singularities: Application to nonlinear {S}chr{\"o}dinger
  equation and related problems, in: R.~E. Caflisch, G.~C. Papanicolaou (eds.),
  Singularities in fluids, plasmas and optics, vol. 404 of NATO science series
  C, Kluwer, 1993.

\bibitem{Schme}
T.~Schmelzer, The fast evaluation of matrix functions for exponential
  integrators, Ph.D. thesis, Oxford University (2007).

\bibitem{SulemSulem1999}
C.~Sulem, P.-L. Sulem, The nonlinear {S}chr{\"o}dinger equation ---
  {S}elf-focusing and wave collapse, vol. 139 of Applied Mathematical Sciences,
  $1^{\text{st}}$ ed., Springer, 1999.

\bibitem{weinstein}
M.~I. Weinstein, Nonlinear {S}chr{\"o}dinger equations and sharp interpolation
  estimates, Comm. Math. Phys. 87 (1983) 567--576.

\bibitem{ZakharovShvets1988}
V.~E. Zakharov, V.~F. Shvets, Nature of wave collapse in the critical case,
  JETP Lett. 47~(4) (1988) 275--278.

\end{thebibliography}

\end{document}